\documentclass{aa}
\usepackage{graphicx}
\usepackage{natbib}
\usepackage{txfonts}
\bibpunct{(}{)}{;}{a}{}{,}
\begin{document}
\newcommand{\cmtri}{$\rm cm^{-3}$}
\newcommand{\cmdva}{$\rm cm^{-2}$}
\newcommand{\kms}{km\,s$^{-1}$}
\newcommand{\pc}{{\rm pc}}
\newcommand{\kpc}{{\rm kpc}}
\newcommand{\myr}{$M_{\sun}\,{\rm yr}^{-1}$}
\newcommand{\ecs}{$\rm\,erg\,cm^{-2}\,s^{-1}$}
\newcommand{\ecsa}{$\rm\,erg\,cm^{-2}\,s^{-1}\,\AA^{-1}$}
\def\V{\rm {\small V}}
\def\I{\rm {\small I}}
\title{Disentangling the composite continuum of symbiotic binaries}
\subtitle{I. S-type systems}

\author{A. Skopal}

\institute{Astronomical Institute, Slovak Academy of Sciences,
           059\,60 Tatransk\'a Lomnica, Slovakia, 
           \email{skopal@ta3.sk}
          }

\date{Received; accepted}

\abstract{
We describe a method of disentangling the composite, 
0.12 -- 5\,$\mu$m continuum of symbiotic binaries. 
The observed SED is determined by the IUE/HST archival 
spectra and flux-points corresponding to the optical $UBVRI$ 
and infrared $JHKLM$ photometric measurements. The modeled 
SED is given by superposition of fluxes from the cool 
giant, hot stellar source and nebula including the effect 
of the Rayleigh scattering process and considering 
influence of the iron curtain absorptions. 
We applied this method to 21 S-type symbiotic stars during 
quiescence, activity and eclipses. We isolated four main 
components of radiation and determined their properties. 
(i) {\em Stellar radiation from the giant} corresponds to 
a unique luminosity class -- normal giants. Characteristic 
luminosities are $1\,600 \pm 200$ and $290 \pm 30\,L_{\sun}$ 
for red and yellow giants, respectively in our sample 
of objects. 
(ii) {\em Hot object radiation during quiescence} consists 
of the nebular and stellar component. The former radiates at 
a mean electron temperature of 19\,000\,K and its amount of 
emission suggests a mass-loss rate from giants via the wind 
at $\dot M_{\rm W}$ = a few $\times\,10^{-7}$\,\myr. Radiation 
of the latter conforms well with that of a black-body 
photosphere at a characteristic temperature of 105\,000\,K. 
The corresponding effective radii are a factor of $\sim$\,10 
larger than those of white dwarfs, which thus precludes observing 
the accretor's surface. Extreme cases of AX\,Per and V443\,Her, 
for which the hot star temperature from the fit is not capable 
of producing the nebular emission, 
signal a disk-like structure of the hot stellar source even 
during quiescence. 
(iii) {\em Hot object radiation during activity} consists 
of three components -- the stellar and the low- and 
high-temperature nebular radiation. 
The stellar radiation satisfies that of a black-body 
photosphere at a low characteristic temperature of 
$\sim 22\,000$\,K (we call it the 1st type of outbursts) or 
at a very high characteristic temperature of 
$\approx 165\,000$\,K (2nd type of outbursts). 
All the active objects with a high orbital inclination show
features of the 1st-type of outbursts (here Z\,And, AE\,Ara, 
CD-43$^{\circ}$14304, TX\,CVn, BF\,Cyg, CH\,Cyg, CI\,Cyg, 
AR\,Pav, AX\,Per), while AG\,Dra represents the 2nd-type. 
The presence of a two-temperature type of UV spectrum 
and an enlargement of effective radii of the stellar source 
by a factor of $\sim$\,10 with respect to the quiescent values 
during the 1st-type of outburst 
suggest an expansion of an optically thick medium at 
the orbital plane in the form of a disk. 
The low-temperature nebula radiates at a mean electron 
temperature of 14\,000\,K and is subject to eclipses, while 
the high-temperature nebula, which is seen during eclipses 
as the only component, is characterized by 
$T_{\rm e} > 30\,000$\,K. \\
\hspace*{5mm}Radiative and geometric properties of the main 
sources of radiation allowed us to reconstruct a basic 
structure of the hot object during the 1st-type of outburst. 
There is an edge-on disk around the accretor. Its outer 
flared rim represents a warm pseudophotosphere of the hot 
stellar source, whose radiation is Rayleigh attenuated and 
affected by the iron curtain absorptions in the neutral gas 
concentrated at the orbital plane. The low-temperature nebula 
is placed just above/below the disk with a concentration at 
its edge as to be subject to eclipses and to 'see' well 
the central ionizing source. High above/below the orbital 
plane, there is a hot nebular emitting region. 
         }
\maketitle
%
%
\section{Introduction}

Symbiotic stars are interacting binary systems consisting of 
a cool giant and a hot compact star. Typical orbital periods 
are between 1 and 3 years, but can be significantly larger. 
The mass loss from the giant represents the primary condition 
for interaction between the binary components. A part of the 
material lost by the giant is transferred to the more compact 
companion via accretion from the stellar wind or Roche-lobe
overflow. This process generates a very hot
($T_{\rm h} \approx 10^5$\,K)
and luminous
($L_{\rm h} \approx 10^2 - 10^4\,L_{\sun}$)
source of radiation. 
How the generated energy is liberated determines two stages 
of a symbiotic binary. 
During {\em quiescent phases} the hot component releases its
energy approximately at a constant rate and spectral distribution.
The hot radiation ionizes a fraction of the neutral giant's wind, 
which gives rise to nebular emission comprising numerous lines 
of high excitation/ionization and the continuum. As a result 
the spectrum consists 
of three components of radiation -- two stellar and one nebular.
For eclipsing systems, the hot star radiation can be attenuated 
by Rayleigh scattering at the position where the mass losing 
giant is in front \citep{inv89}. 
During {\em active phases} the hot component radiation changes 
significantly, at least in its spectral distribution, which 
leads to the 2-3\,mag brightening of the object in the optical. 
A common feature of active phases is a high-velocity mass outflow, 
which can significantly disrupt the ionization structure of 
symbiotic nebulae. In some cases the nebular emission disappears 
entirely at the optical maximum (e.g. BF\,Cyg), in others 
dominates the UV/optical spectrum (e.g. AG\,Dra). 

Many particular aspects of this general view have been 
recently discussed and summarized with outlined problems 
by \cite{cmm03}. 
Current pivotal problems in the symbiotic stars research 
are connected mainly with understanding the nature of 
outbursts and the observed accompanied phenomena 
\citep{so+02,bode03,sk03c,lee00}. 
In this respect the nature of hot objects in symbiotic stars 
play a crucial role. A way to determine their fundamental 
parameters is by modeling the ultraviolet spectrum. 
A pioneering work in this direction has been made by 
Kenyon \& Webbink (1984,hereafter -- KW), who elaborated a grid 
of synthetic spectra for various binary models. By comparing 
the calculated and observed ultraviolet continuum colours 
they suggested the type of the binary model (MS/WD accretor 
or a single hot star) with the corresponding accretion rate 
and the hot star temperature. 
M\"urset at al. (1991, hereafter -- MNSV) refined their method 
by developing a more 
sophisticated model of the nebular continuum accounting for 
fractional rates of photons capable of ionizing He$^{+}$ and H. 
This approach requires including a geometrical structure 
of He$^{++}$ and H$^{+}$ ionized zones in the binary. 
They fitted the observed IUE spectra scaling the hot star 
contribution, its temperature (calculated as the Zanstra 
temperature) and a parameter $X$, which defines the geometry 
of the nebula in the sense of the STB ionization model 
\citep[][hereafter - STB]{stb}. 
Both groups assumed the electron temperature to be constant 
throughout the nebula, but kept its quantity at 
10\,000\,K and/or 20\,000\,K (KW) or approximated it by 
values in a rather narrow interval of 12\,800 -- 15\,000\,K 
for the H$^{+}$ region (MNSV). 

The aim of this paper is to introduce a simple method of 
disentangling the composite SED in the UV/optical/IR 
continuum of symbiotic binaries (Sect.~3). In our approach 
we fit observations by a model with 6 basic free 
parameters -- temperatures of the three main emitting 
sources and scalings of their contributions. 
In addition we consider effects modifying 
the continuum -- the Rayleigh scattering process and veiling 
the continuum by a forest of blended \ion{Fe}{ii} 
absorptions. 
In the second part of this paper (Sect.~4) we apply the method 
to S-type symbiotic binaries during quiescence, activity and 
eclipses. Particularly we discuss cases in which the modeled 
parameters lead to a conflict with the observed properties. 
In Sect.~5 we summarize common characteristics of the isolated 
components of radiation and their sources. 

\section{Observations}

We apply our method (Sect.~3) to well observed S-type symbiotic 
stars, for which the low-resolution ultraviolet spectroscopy 
and the broad-band optical and infrared $UBVRI$ and $JHKLM$ 
photometry are available. 
The most important data source we used to cover the ultraviolet 
domain is the Final archive of the IUE satellite. In the case 
of AG\,Peg we analyzed also the spectra taken by the FOS on the 
board of the HST satellite. We selected spectra with well exposed 
continuum, taken with the large aperture and the short- and 
long-wavelength prime during one shift. We used the spectra 
taken during quiescence (if available, at dates of different 
conjunctions of the binary components), activity and optical 
eclipses. 
Main sources of the data covering the infrared region are 
represented by broad-band multicolour photometric observations 
published by \cite{mu+92}, \cite{ka99}, \cite{ta00}, \cite{gw73} 
and \cite{frc77}. Measurements in the $I_{\rm C}$ and 
$R_{\rm C}$ bands of the Cousins system were converted into 
the Johnson system according to \cite{bessel}. 
In addition, we summarized appropriate parts of light curves 
(LC) to distinguish stages of quiescence and activity 
at the time of spectroscopic observations. 
The main sources of the data here are represented by $UBVR$ 
photometric observations collected during the last 15 years 
of our campaign for long-term monitoring of symbiotic 
stars \citep[][and references therein]{sk+04}. 
%
%
Other data for individual objects are referred in relevant 
sections.
Stellar magnitudes were converted to fluxes according to 
the calibration of \cite{lena99}. 
We dereddened observations with appropriate value of 
$E_{\rm B-V}$ using the extinction curve of \cite{c+89}. 

In Table~1 we list the objects, ephemerides, colour excesses 
and distances we use in this paper. 

\section{Disentangling the composite spectrum}

\subsection{Model of the continuum}

According to the three-component model of radiation (Sect.~1) 
we can express the observed continuum flux, $F(\lambda)$, of 
symbiotic stars as a superposition of two stellar components, 
$F_{\rm h}(\lambda)$ and $F_{\rm g}(\lambda)$, from the hot and 
cool giant star, respectively, and the nebular contribution, 
$F_{\rm N}(\lambda)$, as
%
%
\begin{equation}
  F(\lambda) =  F_{\rm g}(\lambda) + F_{\rm h}(\lambda) +
                F_{\rm N}(\lambda) 
\end{equation}
%
in units of \ecsa. 
We approximate radiation from the giant by an appropriate 
synthetic spectrum, 
  $F^{\rm synth.}_{\lambda}(T_{\rm eff})$, 
scaled to flux points given by the broad-band photometry. 
Thus we can write 
%
%
\begin{equation}
 F_{\rm g}(\lambda) = F^{\rm synth.}_{\lambda}(T_{\rm eff}),
\end{equation}
%
%
%
The effective temperature of the giant, $T_{\rm eff}$, and 
the scaling factor of its spectrum determine the bolometric 
flux at the Earth's surface as 
%
%
\begin{equation}
 F_{\rm g}^{\rm obs} = \int_{\lambda}\!
         F^{\rm synth.}_{\lambda}(T_{\rm eff})\,{\rm d}\lambda  = 
    k_{\rm g}\!\!\int_{\lambda}\!
         \pi B_{\lambda}(T_{\rm eff})\,{\rm d}\lambda  = 
         k_{\rm g} \sigma T_{\rm eff}^{4}
\end{equation}
%
in units of \ecs. In terms of the giant's luminosity 
$L_{\rm g} = 4\pi R_{\rm g}^{2} \sigma T_{\rm eff}^{4}$, 
the scaling factor, $k_{\rm g}$, can be expressed as 
%
%
\begin{equation}
 k_{\rm g} = \frac{L_{\rm g}}{4\pi d^2 \sigma T_{\rm eff}^4}
 = \Bigl(\frac{R_{\rm g}}{d}\Bigr)^2 \equiv \theta_{\rm g}^2, 
\end{equation}
%
where $\theta_{\rm g}$ is the angular radius of the giant. 
We approximate the hot star continuum by a black-body 
radiation at a temperature $T_{\rm h}$. In addition, we consider 
that it can be attenuated by neutral atoms of hydrogen due 
to the Rayleigh scattering process. Thus, we express the second 
term on the right side of Eq.~(1) as 
%
%
\begin{equation}
F_{\rm h}(\lambda) = k_{\rm h}\times 
   \pi B_{\lambda}(T_{\rm h})e^{-n_{\rm H}\sigma_{\lambda}^{\rm R}}, 
\end{equation}
%
where $k_{\rm h} = F_{\rm h}(\lambda)/\pi B_{\lambda}(T_{\rm h})$
is a dimensionless scaling factor, which can be expressed 
in the same form as that for the giant's radiation, i.e. 
%
%
\begin{equation}
 k_{\rm h} = \frac{L_{\rm h}}{4\pi d^2 \sigma T_{\rm h}^4}
 = \Bigl(\frac{R_{\rm h}^{\rm eff}}{d}\Bigr)^2
   \equiv \theta_{\rm h}^2 ,
\end{equation}
%
where $L_{\rm h}$, $T_{\rm h}$, $R_{\rm h}^{\rm eff}$ and
$\theta_{\rm h}$ are the luminosity, temperature, effective
and angular radius of the hot stellar source, respectively. 
Further in Eq.~(5), $n_{\rm H}$ [\cmdva] is the column density 
of H atoms and $\sigma_{\lambda}^{\rm R}$ [cm$^{2}$] is 
the Rayleigh scattering cross-section for atomic hydrogen. 
According to \cite{nsv89} we calculate the scattering 
cross-sections as 
%
%
\begin{equation}
\sigma_{\lambda}^{\rm R} = \sigma_{\rm e}\biggl[\sum_{k}\frac{f_{1k}}
                           {(\lambda/\lambda_{1k})^2-1}\biggr]^2,
\end{equation}
%
where $\sigma_{\rm e}$ = 6.65E-25\,\cmdva\ is the Thomson 
cross-section, $f_{1k}$ are oscillator strengths of the hydrogen 
Lyman lines and $\lambda_{1k}$ are the corresponding wavelengths. 
An interesting feature (from the astrophysical point of view) of the 
$\sigma_{\lambda}^{\rm R}$ function is its rapid decrease from 
Ly$\alpha$ to $\lambda$1055\,\AA\ \citep[see Fig.~2 of][]{nsv89}. 
This effect should be seen as a rapid increase of the far-UV 
continuum beyond the Ly$\alpha$ line to its scattering free values 
at $\lambda$1055\,\AA. For example, the SWP spectra from 
the IUE archive begin at $\sim$1140\,\AA, which means that 
the effect could be detected only for rather low values of 
$n_{\rm H} \la 10^{23}$\,\cmdva\ and on the well exposed 
spectra. 
%
%

The last term on the right side of Eq.~(1) represents flux 
in the continuum emitted by the symbiotic nebula. For the purpose 
of this paper we consider nebular contributions from hydrogen 
and doubly ionized helium generated by processes of recombination 
(f-b transitions) and thermal bremsstrahlung (f-f transitions). 
We neglect contributions from singly ionized helium, because its 
emission coefficient is comparable, in both the profile and 
the quantity, with that of hydrogen, but it is much less abundant 
in the ionized medium (to about 10\%). This similarity also makes 
it difficult to identify any signatures of the emission from 
the singly ionized helium in the observed spectrum. Therefore, 
and also for simplicity, we express the nebular flux as 
%
%
\begin{eqnarray}
F_{\rm N}(\lambda) = 
         \frac{1}{4\pi d^2}\int_{V}\,\Bigl[\,
         \varepsilon_{\lambda}({\rm H},T_{\rm e})N_{\rm H^{+}} + 
~~~~~~~~~~~~~
\nonumber \\
              \varepsilon_{\lambda}({\rm He^{+}},T_{\rm e})
              N_{\rm He^{++}}\Bigr]N_e {\rm d}V,
\end{eqnarray}
%
where $d$ is the distance to the object and 
$\varepsilon_{\lambda}$ (erg\,cm$^{3}$\,s$^{-1}$\,$\AA^{-1}$) 
is the volume emission coefficient per electron and per ion under 
consideration, which includes all acts of f-b and f-f transitions. 
$N_{\rm H^{+}}$, $N_{\rm He^{++}}$ and $N_{\rm e}$ are concentrations 
of the ionized hydrogen (protons), doubly ionized helium and 
the electrons, respectively, and $T_{\rm e}$ is the electron 
temperature. Further we use the following two assumptions for 
the symbiotic nebula: 
(i) Both the H$^{+}$ and the He$^{++}$ regions occupy the same 
volume (in contrast to the usual case, for which the volume 
$V_{\rm H^{+}} > V_{\rm He^{++}}$). Then the ions in such 
nebula are in the pool of electrons with 
%
%
\begin{equation}
 N_e = (1 + 2\tilde{a})\,N_{\rm H^{+}}, 
\end{equation}
%
where 
%
%
\begin{equation}
 \tilde{a} = N_{\rm He^{++}}/N_{\rm H^{+}} 
\end{equation}
%
represents an average abundance of the doubly ionized helium 
throughout the nebula. 
(ii) The nebular radiation is characterized by a uniform 
electron temperature, $T_{\rm e}$, which is constant within 
the nebula. 
The assumption (i) is based on that the spectra include contributions 
from all the present ions irrespective of their displacement 
within the nebula. Then the average abundance $\tilde{a}$ represents 
a lower limit of that in the real ${\rm He^{++}}$ region. 
Specially, in the case that 
$V_{\rm H^{+}} = V_{\rm He^{++}}$, 
also 
$\tilde{a} = a({\rm He^{++}})$. 
Contrarily, if 
$V_{\rm H^{+}} \gg V_{\rm He^{++}}$ 
then 
$\tilde{a} \ll a({\rm He^{++}})$, which implies that the 
contribution from the ${\rm He^{++}}$ region is very small 
with respect to that from ${\rm H^{+}}$ region. These limiting 
cases can be distinguished from observations. For example, a large 
quantity of $\tilde{a}$ derived from observations (to say around 0.1, 
which can be considered as a maximum given by the helium abundance 
$a({\rm He})$) means that the helium is doubly ionized everywhere in 
the nebula, i.e. $V_{\rm He^{++}} \approx V_{\rm H^{+}}$. 
The assumption (ii) is verified by good fits of the continuum 
at spectral regions, where the ${\rm He^{++}}$ and ${\rm H^{+}}$ 
nebular continuum dominate the spectrum (see Figs.~2 -- 22). 
So, with the aid of Eqs.~(9, 10) and the assumption of uniform 
$T_{\rm e}$ in the nebula, the expression (8) takes the form 
%
%
\begin{equation}
F_{\rm N}(\lambda) = \frac{EM}{4\pi d^2} \times 
  \hat{\varepsilon}_{\lambda}({\rm H, He^{+}},T_{\rm e},\tilde{a}), 
\end{equation}
where the emission measure $EM = \int_V N_{\rm H^{+}}^2 {\rm d}V$ 
and the emission coefficient 
%
%
\begin{eqnarray}
\hat{\varepsilon}_{\lambda}({\rm H, He^{+}},T_{\rm e},\tilde{a}) = 
  (1 + 2\tilde{a})\varepsilon_{\lambda}({\rm H},T_{\rm e}) + ~~~
\nonumber \\
   \tilde{a}(1 + 2\tilde{a})
   \varepsilon_{\lambda}({\rm He^{+}},T_{\rm e}). 
\end{eqnarray}
%
For small values of $\tilde{a}$, to say $\tilde{a} < 0.05$, 
$\hat{\varepsilon}_{\lambda} \dot = 
 \varepsilon_{\lambda}({\rm H},T_{\rm e}) + \tilde{a}
 \varepsilon_{\lambda}({\rm He^{+}},T_{\rm e})$, which can be 
applied, for example, to symbiotic nebulae during quiescent 
phases that are characterized by 
$V_{\rm H^{+}} > V_{\rm He^{++}}$. 
Finally, according to relations (2), (5) and (11), we can express 
Eq.~(1) in the form 
%
%
\begin{eqnarray}
 F_{\lambda} = F^{\rm synth.}_{\lambda}(T_{\rm eff}) + 
~~~~~~~~~~~~~~~~~~
\nonumber \\
k_{\rm h}\times \pi B_{\lambda}(T_{\rm h})
       e^{-n_{\rm H}\sigma_{\lambda}^{\rm R}} + 
~~~
\nonumber \\
k_{\rm N}\times 
       \hat{\varepsilon}_{\lambda}({\rm H, He^{+}},T_{\rm e},\tilde{a}),
\end{eqnarray}
%
where $k_{\rm N}$ is the scaling factor for the nebular component 
of radiation. According to Eq.~(11) 
%
%
\begin{equation}
k_{\rm N} = \frac{EM}{4\pi d^{2}} ~~~ [\rm cm^{-5}] .
\end{equation}
%

\subsubsection{Influence of the line blanketing effect}

This effect represents an additional source of the absorbing 
spectrum due mostly to \ion{Fe}{ii} transitions. 
This so-called "iron curtain" was first identified by 
\cite{sa93} as the origin of 
the anomalies in the emission-line fluxes resulting from the 
differential absorption by the environment as the path length 
along the line-of-sight changes. Generally, the intervening 
absorption originates in a slab of cool gas that veils the source 
of the hot radiation. In symbiotic binaries the veiling gas can 
be represented by the neutral wind of the giant, the effect 
of which is largest at/around the giant's inferior conjunction 
\citep[e.g.][]{d+99}. In this study we find that such a slab 
of absorbing gas can temporarily be created at the orbital 
plane during the Z\,And type of outbursts. 

The line blanketing effect results in a complex profile of 
the observed continuum. On the low-resolution spectra broad 
absorption bands at 1500 to 1800\,\AA\ and 2\,300 to 
2\,800\,\AA\ accompanied by spurious emissions are 
characteristic features. To estimate the level of 
the continuum at wavelengths with the lowest absorption, we 
have chosen three representative low-resolution IUE spectra 
of MWC\,560 to obtain a template for a low, modest and 
strong influence of the continuum by the iron curtain 
(Fig.~1). With the help of these observations and considering 
theoretical calculations given by \cite{sa93}, \cite{ho+94} 
and \cite{d+99} we select points of the continuum 
at $\sim$\,1280, $\sim$\,1340, $\sim$\,1450, $\sim$\,1500, 
($\sim$\,1580-1600), around 1800, $\sim$\,1950-2150, 
($\sim$\,2450), $\sim$\,2640 and beyond 2800\,\AA\ depending 
on the amount of absorption. 

The resulting effect of the absorbing gas in the spectrum of 
symbiotic binaries depends also on its relative position with 
respect to the location of different emitting regions in the 
system. Generally, radiation from the hot stellar source is 
absorbed most, while the nebular emission -- having its origin 
in a region located well away from the hot star -- is less 
affected. An additional complication is the superposition of 
the strong emission lines on the broad absorption bands of 
the iron curtain. For example, emission of \ion{He}{ii}\,1640, 
\ion{O}{i}\,1641] and \ion{O}{iii}\,1664\,\AA\ lines fills up 
in part the underlying absorption bands, which results in 
an artificial absorption located at/around 1700\,\AA. 
Its depth reflects the magnitude of the iron curtain effect. 
This part of the spectrum cannot be used to estimate 
the level of the continuum. 

%
%
\begin{figure}
\vspace{2mm}
\centering
\begin{center}
\resizebox{\hsize}{!}{\includegraphics[angle=-90]{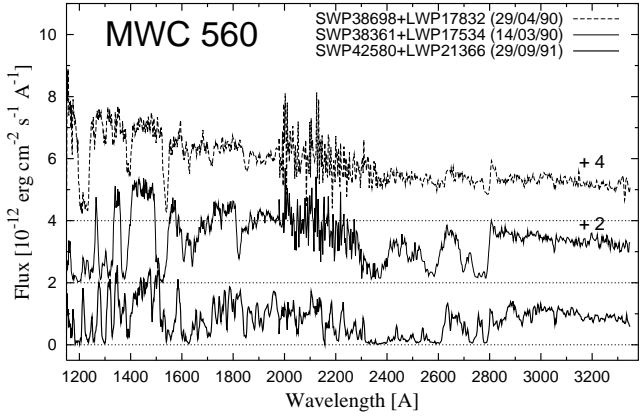}}
\caption[]{IUE spectra of MWC\,560 obtained on 29/04/90, 
14/03/90 and 29/09/91 with different amounts of absorption 
by the veiling gas. 
          }
\end{center}
\end{figure}

\subsection{The model}

To determine the giant's continuum we scaled a synthetic 
spectrum to the broad-band $[BVRI]JKLM$ photometric fluxes 
(depending on the giant's contribution in the optical). 
This approach is applicable for S-type symbiotic stars 
(i.e. with a stellar type of the spectrum; no significant 
dust emission is present in the IR wavelengths). 
An appropriate synthetic spectrum, which best matches 
the photometric flux-points, was selected from a grid of 
models made by \cite{h+99}. We found it satisfactory 
to use only a ($T_{\rm eff},\,\log(g)$) grid and to smooth 
the spectrum within 20\,\AA\ bins to keep a basic profile 
of the giant's continuum. The resulting parameters 
are $T_{\rm eff}$ and $\theta_{\rm g}$, which define the 
observed bolometric flux, $F_{\rm g}^{\rm obs}$ (Sect.~3.1). 
For most of the objects we selected $\log(g) \equiv 0.5$, 
for CH\,Cyg and T\,CrB we put $\log(g) \equiv 0$ and for 
yellow symbiotics, 1.5 -- 2, in accordance with their detailed 
atmospheric analysis \citep[e.g.][]{smith+97}. 

Having determined the radiation from the giant we subtracted 
its contribution from the selected flux-points of the 
UV-continuum, $F_{\rm obs}(\lambda_{\rm i})$, and fitted 
them by the synthetic continuum expressed by Eq.~(13), 
for which the function 
%
%
\begin{equation}
 \chi^{2} = \frac{1}{N}\sum_{\rm i=1}^{\rm N}
    [(F_{\rm obs}(\lambda_{\rm i}) - F_{\rm g}(\lambda_{\rm i})) - 
             (F(\lambda_{\rm i}) - F_{\rm g}(\lambda)_{\rm i})]^{2}
\end{equation}
%
reaches a minimum. In this way we determined the other 6 free 
parameters, $k_{\rm h}$, $T_{\rm h}$, $n_{\rm H}$, $k_{\rm N}$, 
$T_{\rm e}$ and $\tilde{a}$, which define the individual
components of radiation. In Eq.~(15), $F(\lambda_{\rm i})$ 
are the theoretical fluxes given by Eq.~(13) at selected 
wavelengths $\lambda_{\rm i}$. $N$ is their number. 
We estimated fluxes $F_{\rm obs}(\lambda_{\rm i})$ at about 
10 to 20 wavelengths between 1\,250 and 3\,300\,\AA\ depending on 
quality of the used spectrum. Because of noise, numerous emission 
lines and absorption features we estimated the continuum fluxes 
by eye at the wavelengths specified in section 3.1.1. Then, if 
possible and applicable, we subtracted the contribution from 
the hot temperature nebula (see Sect.~5.3.1). 
Second, we prepared a grid of models for reasonable ranges 
of the fitting parameters and selected that corresponding 
to a minimum of the $\chi^{2}$ function. We found the initial 
ranges for $k_{\rm h}$, $T_{\rm h}$ and $n_{\rm H}$ parameters 
by comparing the Planck function for different temperatures 
(15\,000 -- 40\,000\,K for some active phases) attenuated 
with $n_{\rm H} \approx 10^{23}$\,\cmdva\ (if applicable) 
to the far-UV spectrum between about 1\,250 and 1\,600\,\AA. 
For the spectra from quiescent phases we adopted in some cases 
their Zanstra temperature according to MNSV and/or determined 
their lower limits as introduced in Sect.~4.1. 
Section~5.2.2 gives more details. 
The range for $T_{\rm e}$ can be estimated according to 
the slope of the near-UV continuum from about 2\,400\,\AA\ 
to the Balmer jump, where the nebular emission represents 
a dominant contribution. The average abundance of the doubly 
ionized helium, $\tilde{a}$, can be estimated from the difference 
of the continuum level in the range of about 1\,600\,\AA\ to 
1\,980\,\AA\ and about 2\,300\,\AA\ to $\sim$3\,200\,\AA. 
If we use the IUE spectra, these parts are also well defined. 
Unfortunately, the part from the beginning of the LWP(R) spectra 
to about 2\,300\,\AA\ is very noisy and the following region 
in between $\sim 2\,300$ and $\sim 2\,700$\,\AA\ is often 
affected by the iron curtain absorptions, which makes it 
difficult to recognize signatures from the He$^{++}$ emission. 
Finally, the scaling factor $k_{\rm N}$ results from 
the fitting procedure. 
If the initial grid of models (one model = one combination 
of the fitting parameters) included the minimum of the 
$\chi^{2}$ function, then a close grid could be determined 
around the minimum. Repeating this approach we selected models 
fitting the observed fluxes within their uncertainties. 

\subsection{Uncertainties of the fitting parameters}

To estimate uncertainties of the resulting fitting 
parameters, we determined a $\chi_{\rm max}$ quantity, 
which separates the 
reliable models from those given by all combinations of the 
grid parameters. The models with $\chi < \chi_{\rm max}$ then 
reconstruct the observed fluxes within their uncertainties. 
According to evaluation of the well exposed spectra from 
the IUE archive, they are of about 10\% of the measured 
values. By preparing many trials we found an approximative 
relation for $\chi_{\rm max}$ as 
%
%
\begin{equation}
  \chi_{\rm max} \approx 0.05\times
                 \tilde F_{\rm obs}, 
\end{equation}
%
where $\tilde F_{\rm obs}$ is the average of the 
$F_{\rm obs}(\lambda_{\rm i})$ fluxes. 
%
%
The $\chi_{\rm max}$ value then bounds the ranges for the 
fitting parameters in the $\chi (n_{\rm H}, T_{\rm h}, 
k_{\rm h}, \tilde{a}, T_{\rm e}, k_{\rm N})$ diagrams. 
Note that a change of the parameter under consideration within 
the range of its possible values requires a different 
combination of the other fitting parameters to keep the $\chi$ 
value of the corresponding model under the $\chi_{\rm max}$ 
limit. 
The primary source of uncertainties is the accuracy of 
the measured continuum. The ranges of possible values of 
the fitting parameters are also given by dependencies 
of the synthetic continuum on individual parameters. 
The profile of the far-UV continuum depends strongly on the 
$n_{\rm H}$ parameter (Eq.~5). Any other parameter can 
exceed its influence on the continuum at these wavelengths. 
For the cases with a significantly Rayleigh-attenuated 
far-UV continuum, the $\chi(n_{\rm H})$ function has a rather 
sharp minimum. A complication here is a lower temperature of 
the stellar source, $T_{\rm h} \approx 20\,000 - 30\,000$\,K 
measured during some active 
phases, because the maximum of the corresponding Planck's 
function lies around the far-UV region or just beyond it 
at shorter wavelengths. In this case a higher temperature 
requires also a slightly higher $n_{\rm H}$ to keep the fit 
under $\chi_{\rm max}$. This dependence mainly enlarges 
the range of possible values of $T_{\rm h}$. 
The profile of the nebular continuum from about 2\,000\,\AA\ 
to the long-wavelength end of the spectrum depends 
significantly on the parameters $\tilde a$ and $T_{\rm e}$. 
For the spectra strongly affected by the iron curtain 
absorption it is not possible to determine unambiguously 
the $\tilde a$ parameter. In addition, our models 
(Sects.~4.2 -- 4.22) suggest it to be very small. 
In conclusion, ranges of the fitting parameters 
$T_{\rm h}$, $n_{\rm H}$, $T_{\rm e}$, $k_{\rm h}$ and 
$k_{\rm N}$ were estimated to be 20\% to 50\% of their 
best values. 

Uncertainties for the stellar component of radiation from 
the cool giant are given by those in $T_{\rm eff}$ and 
the scaling. The former is proximately $\pm 100$\,K 
\citep[see also][]{sk00} and the latter is given by that of 
photometric measurements, which we assumed to be less than 
10\%. Then the relative mean square error in the observed 
bolometric flux from the giant is about of 15\%. 
%
%
%
%
%
\begin{table}
\begin{center}
\caption{Ephemerides, reddening and distances} 
\begin{tabular}{clclll}
\hline
\hline
Object                   & 
${\rm Sp.~conj.}^{a}$    &
$P_{\rm orb}$            &
$E_{\rm B-V}$            &
$d$                      &
Ref.                     \\
                         &
[MJD]                    &
[day]                    &
[mag]                    &
[kpc]                    &
                         \\
\hline
EG\,And     & 50\,683.2 & 482.57   & 0.05     &0.59$^b$ & 1,2  \\
Z\,And      & 14\,625.2 & 757.5~~  & 0.30     &1.5      & 3,4  \\
AE\,Ara     & 50\,217   & 812~~~~  & 0.25$^b$ &3.5$^b$  & 5    \\ 
CD-43$^{c}$ & 45\,567   &1\,448~~~~~~~& 0.2   &2.1$^b$  & 6    \\
T\,CrB      & 47\,861.7 & 227.57   & 0.15     &0.96     & 1,7,8\\
TX\,CVn     & 45\,130.5 & 198~~~~  & 0.0      &1.0$^b$  & 9    \\
BF\,Cyg     & 45\,395.1 & 757.2~~  & 0.35     &3.8$^b$  & 10   \\
CH\,Cyg     & 45\,888   & 756~~~~  & 0.0      &0.27     & 11,12\\
CI\,Cyg     & 11\,902   & 855.25   & 0.35     &2.0$^b$  & 13,14\\
V1329\,Cyg  & 27\,687   & 958.0$^d$& 0.37     &4.2$^b$  & 15   \\
LT\,Del     & 45\,910   & 478.5~~  & 0.20$^b$ &3.9$^b$  & 31   \\
AG\,Dra     & 43\,629.2 & 549.73   & 0.08     &1.1$^b$  & 16,17\\
CQ\,Dra     & 42\,932   &1\,703~~~~~~~ & 0.10 &0.178    & 18,19\\
V443\,Her   & 43\,660$^e$&594~~~~  & 0.25$^b$ &2.2$^b$  & 20   \\
YY\,Her     & 40\,637   & 592.8~~  & 0.20     &6.3$^b$  & 3,21 \\
RW\,Hya     & 49\,512   & 370.4~~  & 0.10     &0.82$^b$ & 22,2 \\
SY\,Mus     & 50\,176   & 625.0~~  & 0.35$^b$ &1.0$^b$  & 23   \\
AR\,Pav     & 11\,266.1$^f$& 604.45& 0.26     &4.9      & 24--6\\
AG\,Peg     & 27\,664.2 & 812.6~~  & 0.10     &0.80     & 3,27 \\
AX\,Per     & 36\,673.3 & 679.9~~  & 0.27     &1.73     & 28-30\\
IV\,Vir$^{g}$ & 49\,016.9 & 281.6~~  & 0.20$^b$ &1.3$^b$& 32   \\
%
%
\hline
\end{tabular}
\end{center}
$a$ -- inferior conjunction of the cool component \\
$b$ -- this paper \\
$c$ -- CD-43$^{\circ}$14304 \\
$d$ -- pre-outburst ephemeris \\
$e$ -- $JD_{\rm Min(U)}$ \\
$f$ -- a mean linear ephemeris of eclipses \\
$g$ -- BD-21$^{\circ}$3873 \\
References: 1 -- \cite{f+00a}, 2 -- MNSV, 3 -- \cite{sk98}, 
            4 -- \cite{mk96}, 5 -- \cite{mika+03}, 
            6 -- \cite{schn93}, 7 -- \cite{sell+92}, 
            8 -- \cite{bm98}, 9 -- \cite{kg89}, 
           10 -- \cite{f+01}, 11 -- \cite{hinkle+93}, 
           12 -- \cite{viotti+97}, 13 -- \cite{a54}, 
           14 -- \cite{k+91}, 15 -- \cite{ss97}, 
           16 -- \cite{ga+99}, 17 -- \cite{bi+00},
           18 -- \cite{rei+88}, 19 -- \cite{per+97},
           20 -- \cite{kmy95}, 21 -- \cite{mu+97a},
           22 -- \cite{sch+96}, 23 -- \cite{d+99},
           24 -- \cite{sk+01b}, 25 -- KW,
           26 -- \cite{sch+01}, 27 -- \cite{k+93},
           28 -- \cite{sk91}, 29 -- \cite{mk92a},
           30 -- \cite{sk00}, 31 -- \cite{arch+95},
           32 -- \cite{smith+97}
\end{table}

\section{Application to S-type symbiotic stars}

\subsection{Fitting and derived parameters}

Fitting parameters, $k_{\rm g}$, $k_{\rm h}$, $k_{\rm N}$ 
and corresponding temperatures determine stellar luminosities 
and emission measure of the nebula (Eqs.~4, 6, 14). 
A useful parameter is the ratio of the hot star and the cool 
giant luminosity, which is independent of the distance. 
According to our notation this can be expressed as 
%
%
\begin{equation}
  \frac{L_{\rm h}}{L_{\rm g}} = \frac{k_{\rm h}}{k_{\rm g}}
           \Big(\frac{T_{\rm h}}{T_{\rm eff}}\Big)^4 .
\end{equation}
At the wavelengths, where the nebular emission dominates the 
spectrum, one can estimate the upper limit of $EM$ directly 
from observations by assuming that $F_{\rm N}(\lambda) = 
F_{\rm obs}(\lambda)$, which is usually fulfilled for fluxes 
in the near-UV region. Then, according to Eq.~(11) the emission 
measure is limited by 
%
%
\begin{equation}
 EM = 4\pi d^2 \frac{F_{\rm obs}(\lambda)}
     {\hat{\varepsilon}_{\lambda}({\rm H, He^{+}},T_{\rm e},\tilde{a})}.
\end{equation}
%
We calculate the luminosity of the nebula from Eq.~(11) as 
%
%
\begin{equation}
L_{\rm N} = EM \int_{912}^{\infty} 
       \hat{\varepsilon}_{\lambda}({\rm H, He^{+}},T_{\rm e},\tilde{a})
       {\rm d}\lambda ,
\end{equation}
%
which is valid for an ionised medium that is optically thick in 
the Lyman continuum. 
Finally, we introduce a parameter 
%
%
\begin{equation}
 \delta = \frac{EM_{\rm obs}}{EM_{\rm B}}, 
\end{equation}
%
where $EM_{\rm obs}$ is the emission measure derived from 
the model (Eq.~14) and $EM_{\rm B} = L_{\rm ph}/\alpha_{\rm B}$ 
results from the equilibrium equation between the rate of 
ionizing photons, $L_{\rm ph}$ (photons\,s$^{-1}$), and the rate 
of ionization/recombination acts within the ionized region; 
$\alpha_{\rm B}$ ($\rm cm^{3}\,s^{-1}$) stands for the total 
hydrogen recombination coefficient (STB). 
Thus $EM_{\rm B}$ represents a maximum of the nebular emission, 
which can be produced by the ionizing source under conditions 
of case $B$ in the hydrogen plasma. This implies that 
$\delta \le 1$. Values close to 1 indicate that all ionizing 
photons are converted by the circumbinary medium and are not 
lost. Small values of $\delta$ indicate the opposite case. 
Using the relation for $L_{\rm ph}$ from \cite{sk01a} 
the $\delta$ parameter can be expressed as 
%
%
\begin{equation}
 \delta = \frac{k_{\rm N}}{k_{\rm h}} \alpha_{\rm B}/f(T_{\rm h}), 
\end{equation}
%
where the function 
%
%
\begin{equation}
f(T_{\rm h}) = \frac{\pi}{hc}\int^{912}_{0}\!\!\!
                \lambda\, B_{\lambda}(T_{\rm h})\,\rm d\lambda. 
\end{equation}
%
For example, $f(8\,10^{4}\,\rm K)$ = 4.6E+25, 
             $f(1\,10^{5}\,\rm K)$ = 1.1E+26, 
             $f(1.3\,10^{5}\,\rm K)$ = 2.7E+26 and 
             $f(1.5\,10^{5}\,\rm K)$ = 4.3E+26\,
                                       $\rm cm^{-2}\,s^{-1}$. 
The limiting case, $\delta = 1$, determines a lower limit of 
the temperature of the ionizing source, $T_{\rm h}^{\rm min}$, 
required to produce the observed emission measure. 
Thus, the number of ionizing photons produced by the 
hot object radiating at $T_{\rm h}^{\rm min}$ just balances 
the number of recombinations in the continuum. 
In modeling the ultraviolet continuum by Eq.~(13) we often used 
$T_{\rm h} = T_{\rm Zanstra}$ (from MNSV) for 
a first-model-fit. If the ionizing capacity of the hot stellar 
source was not consistent with the observed nebular emission 
($\delta > 1$) we searched for $T_{\rm h}^{\rm min}$. As 
a temperature change in the range of $T_{\rm h} > 10^{5}$\,K 
practically does not influences the resulting fit, we 
can get $T_{\rm h} = T_{\rm h}^{\rm min}$ by solving equation 
%
%
\begin{equation}
  \frac{k_{\rm N}}{k_{\rm h}(T_{\rm h})} \alpha_{\rm B} - f(T_{\rm h}) = 0
\end{equation}
for parameters $k_{\rm N}$ and $T_{\rm e}$ from the first-model-fit. 

In the following sections we give results for individual 
objects with special attention to common properties 
of their SEDs during quiescent and active phases as well as 
eclipses. The resulting parameters of the reconstructed 
continuum are in Tables~2, 3 and 4. Corresponding synthetic 
continua together with observations are shown in Figs.~2 -- 22. 

\subsection{EG~Andromedae} 
%
EG\,And is a quiet symbiotic star -- no active phase has been 
recorded to date. $UBV$ LCs display a periodic orbitally-related 
variation of a double-wave in the profile (Fig.~2). 
Occasionally, 0.3 - 0.5\,mag flares are observed in the $U$-LC. 
The infrared photometry was summarized by \cite{sk00}. 
At positions of the inferior conjunction of the giant, 
the ultraviolet continuum is Rayleigh scattered by H\,\I\ 
atoms of the giant's wind, which suggests a high 
inclination of the orbit \citep{v91}. 
To show properties of the ultraviolet continuum we selected 
IUE observations taken close to the superior conjunction of 
the giant 
(SWP42347 + LWP21103, 28/08/91, $\varphi$ = 0.47 and 
 SWP55897 + LWP31444, 10/09/95, $\varphi$ = 0.52) 
and at the opposite position 
(SWP18993 + LWR15045, 13/01/83, $\varphi$ = 0.95). 

{\em Radiation from the giant}.  
Fluxes of the infrared photometry can be compared well with 
the synthetic spectrum calculated for 
  $T_{\rm eff} = 3\,500$\,K. 
The scaling factor, 
  $k_{\rm g} = 2.23\,10^{-17}$ 
%
%
yields the integrated flux 
  $F_{\rm g}^{\rm obs} = 1.9\,10^{-7}$\,\ecs. 
The timing of the eclipse at $\lambda$1320\,\AA\ corresponds to 
  $R_{\rm g}/A = 0.32 \pm 0.05$ \citep[see Appendix A of][]{sk01a}. 
For $P_{\rm orb}$ = 482 days and the total mass of the binary, 
  $M_{\rm T} \equiv 2.5\,M_{\sun}$ 
\citep[$M_{\rm h} \equiv 0.6\,M_{\sun}$ and elements of][]{f+00a} 
we get the separation of the stars 
  $A = 386\,R_{\sun}$, 
which yields 
  $R_{\rm g} = 124\,\pm\,19\,R_{\sun}$. 
This quantity can be converted by $\theta_{\rm g}$ to the distance 
  $d = 590\,\pm\,92$\,pc, 
which is well within the range of 400 to 1960\,pc measured 
by Hipparcos. 

{\em Radiation from the ultraviolet}. 
The last low-resolution spectrum of EG\,And made by 
IUE (10/09/95) is characterized by a strong nebular 
emission. Its parameters of our solution 
  ($T_{\rm e} = 25\,000\,\pm\,3\,000$\,K, 
   $k_{\rm N} = 1.25\,10^{15}\,\rm cm^{-5}$) 
constrain a minimum temperature of the ionizing source, 
   $T_{\rm h}^{\rm min}$ = 95\,000\,K 
   ($k_{\rm h} = 1.53\,10^{-24}$, $\delta \sim 1$, Eq.~23), 
which makes it just capable of producing the observed emission 
measure, 
  $EM_{\rm obs} = 5.2\,10^{58}(d/590\,\pc)^2$\,\cmtri. 
These parameters imply 
  $L_{\rm h} = 77(d/590\,\pc)^2\,L_{\sun}$, 
  $\theta_{\rm h} = 1.24\,10^{-12}$ 
and 
   $R_{\rm h} = 0.032(d/590\,\pc)\,R_{\sun}$.
We adopted this temperature to analyze the remaining two 
spectra. 
Observations from 28/08/91 show a significantly smaller 
amount of nebular emission 
  ($k_{\rm N} = 2.5\,10^{14}\,\rm cm^{-5}$) 
radiating at a low temperature of 
   $T_{\rm e} = 13\,500\,\pm\,1\,000$\,K, 
even though it was taken at similar orbital phase as 
the previous spectrum. 
However, the luminosity of the hot stellar source 
was comparable: 
  $k_{\rm h} = 1.27\,10^{-24}$ 
 ($\theta_{\rm h} = 1.1\,10^{-12}$), which 
yields 
  $R_{\rm h} = 0.029(d/590\,\pc)\,R_{\sun}$
and 
  $L_{\rm h} = 60(d/590\,\pc)^2\,L_{\sun}$.
The spectrum taken close to the inferior conjunction 
of the giant (13/01/83) shows a strongly attenuated 
far-UV continuum by the Rayleigh scattering process 
  ($n_{\rm H} = 9.4\,10^{22}\,\rm cm^{-2}$) 
and features of the iron curtain absorptions pronounced at
$\sim \lambda$1400\,\AA\ and $\sim \lambda$1700\,\AA. 
However, the stellar radiation can be scaled with 
approximately the same quantity of $k_{\rm h}$ as on 
28/08/91 and 10/09/95, i.e. if one removes the effect of 
the Rayleigh scattering, the hot star luminosity remains 
practically unchanged in all spectra.  
The nebular component of radiation is fainter at positions 
near to the giant's occultation. 
\cite{sk01a} showed that the variation in the $EM$ follows 
well that observed in the LCs. Here the amplitude 
$\Delta m_{\rm EM} = -2.5\log(EM_{\rm min}/EM_{\rm max})$ 
                   = 0.36\,mag 
is comparable with that measured photometrically in the $U$ 
band (Fig.~2). 
%
%
%
\begin{figure}
\centering
\begin{center}
\resizebox{\hsize}{!}{\includegraphics[angle=-90]{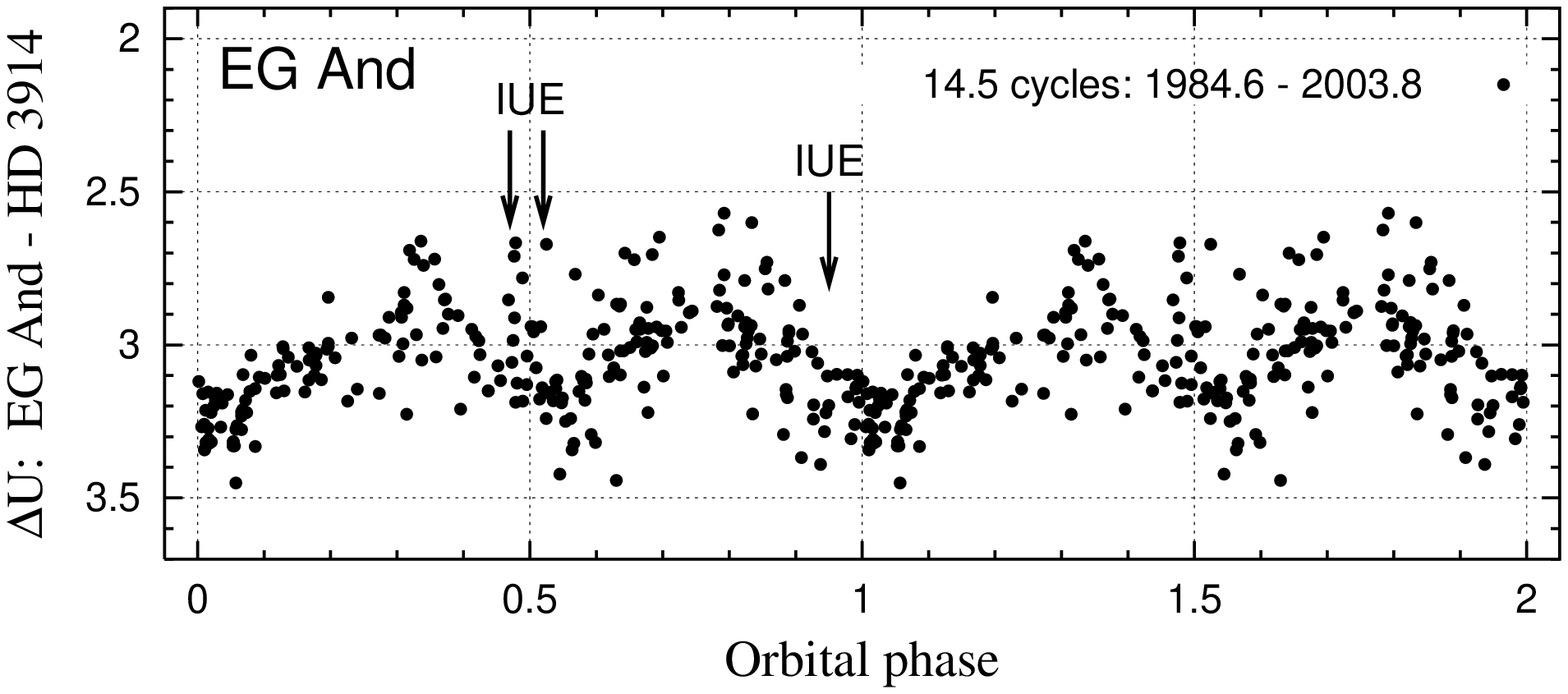}}


\resizebox{\hsize}{!}{\includegraphics[angle=-90]{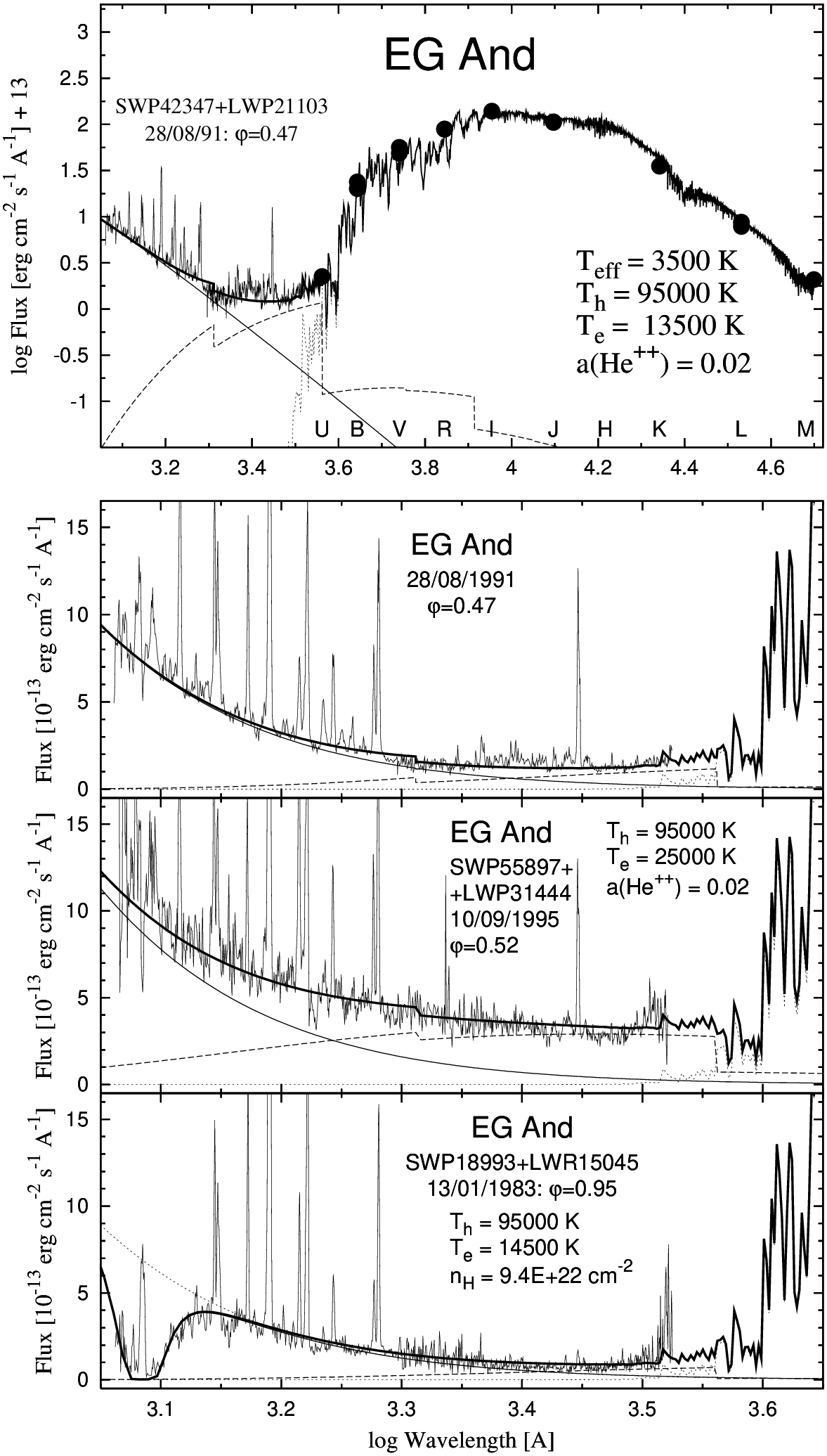}}
\caption[]{Reconstructed SED in the continuum of EG\,And between 
           0.12 and 5\,$\mu$m. The solid thin and dashed lines 
           represent the hot stellar and nebular components of 
           radiation. The solid thick line is the resulting 
           modeled continuum. Top panel shows the phase diagram 
           of $U$ magnitudes. Observations are marked by arrows. 
          }
\end{center}
\end{figure}
Parameters of the hot star in EG\,And suggest that the observed 
hot star luminosity is balanced by accretion processes at 
a rate of 
$\dot M_{\rm acc} = {\rm a~ few} \times 10^{-8}$\,\myr\ 
\citep[see][for details]{sk05}. 
Then the observed increase in the nebular emission on 
10/09/95 could be caused by a transient increase in the 
mass-loss rate -- an injection of a surplus of emitters 
into an open \ion{H}{ii} zone ($\delta < 1$ on 28/08/91) 
increases the rate of ionization/recombination acts. 
From this point of view, the 0.3 -- 0.5\,mag flares in $U$ 
result from a variable mass-loss rate from the giant. 
%
%
\begin{figure}
\centering
\begin{center}
\resizebox{\hsize}{!}{\includegraphics[angle=-90]{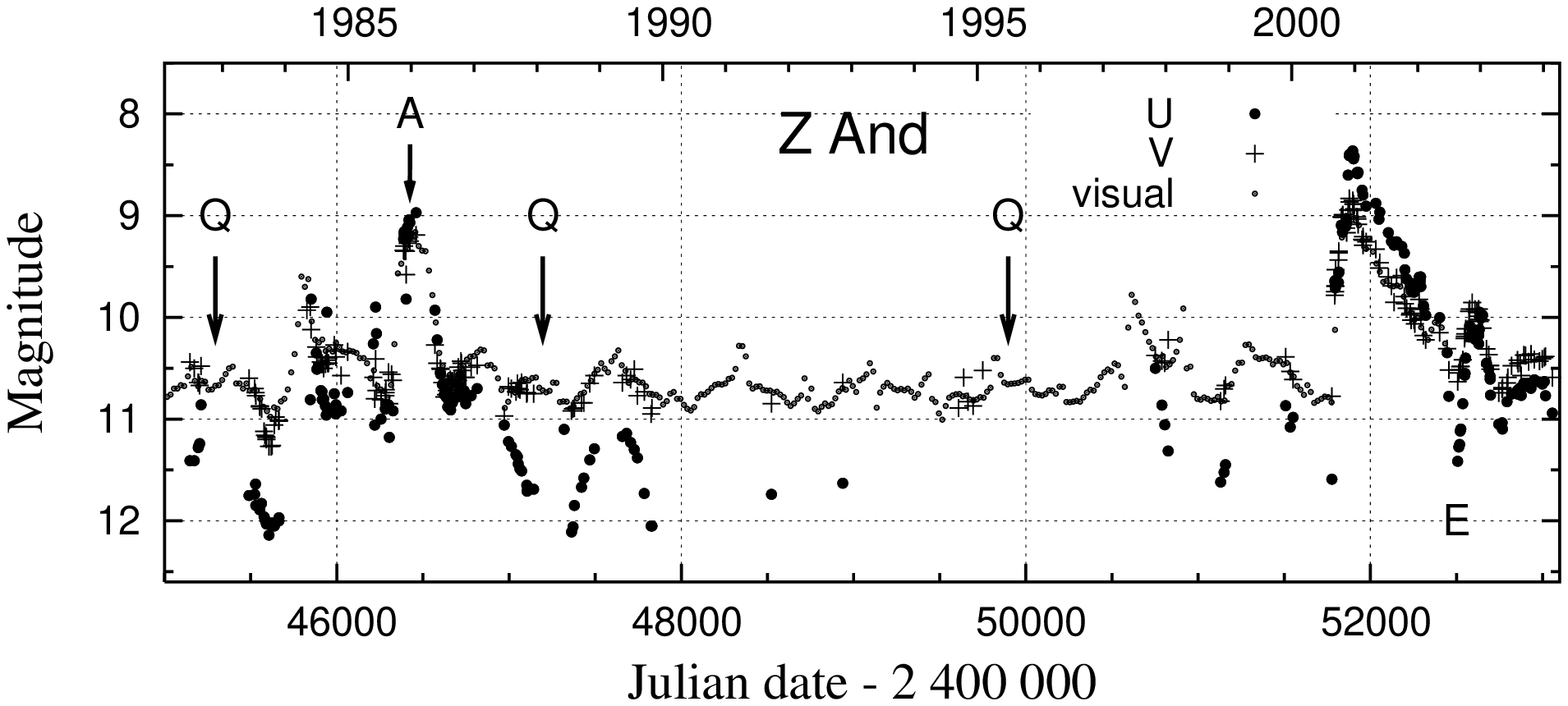}}

\vspace*{2mm}

\resizebox{\hsize}{!}{\includegraphics[angle=-90]{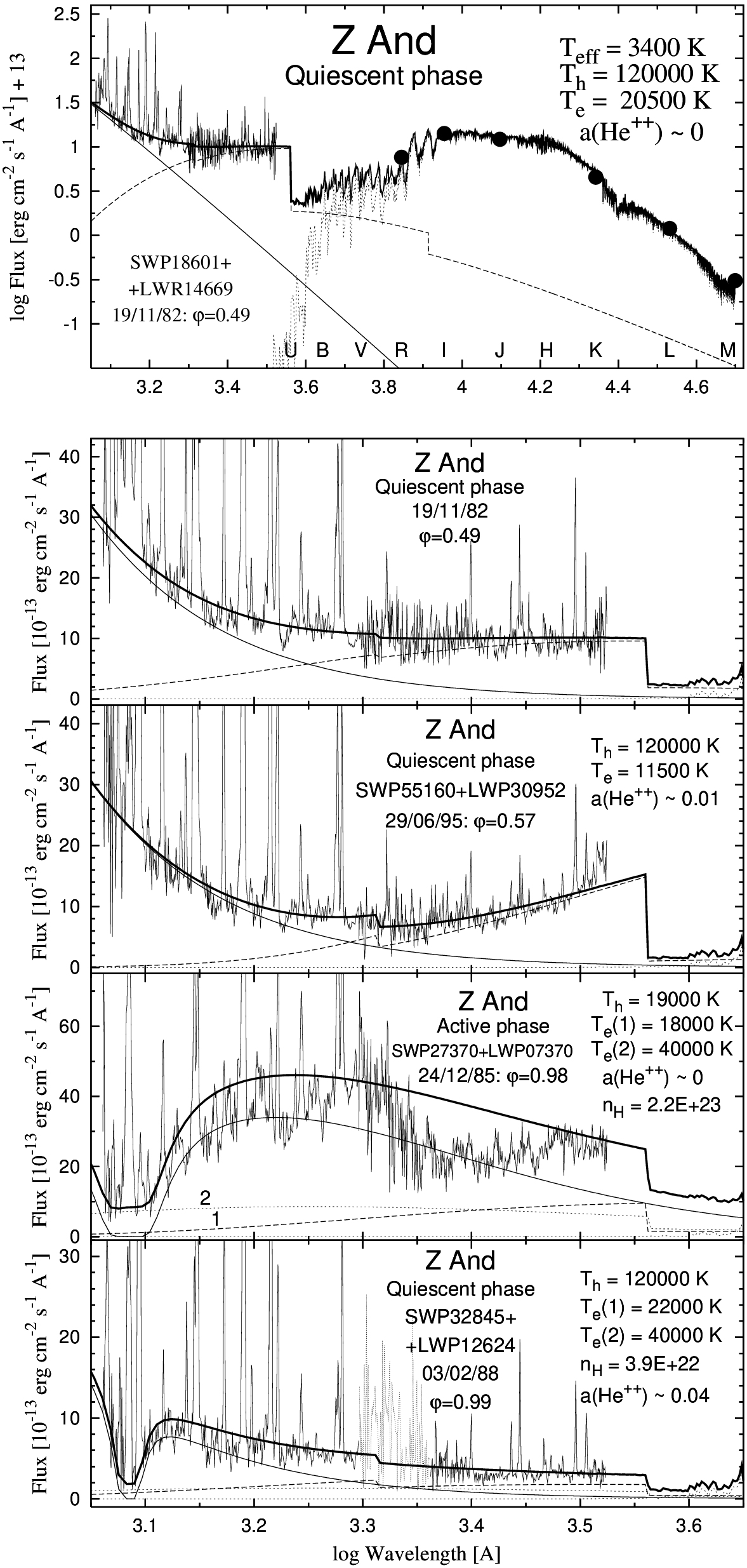}}
\caption[]{
As in Fig.~2, but for Z\,And during quiescent and active 
phases. Dates of IUE observations are marked in the top 
panel with the LC. 
$\sf Q$, $\sf A$ and $\sf E$ denote quiescence, activity and 
the eclipse effect, respectively. 
          }
\end{center}
\end{figure}

\subsection{Z~Andromedae}

Z\,And is considered as a prototype symbiotic star. The LC 
is characterized by phases of activity with up to 2-3\,mag 
brightenings, alternating with periods of quiescence (Fig.~3). 
Recently \cite{sk03a} revealed a high inclination of 
the Z\,And orbit. 
To show properties of the ultraviolet continuum during 
quiescence we selected IUE spectra taken at the position 
when the hot star was in front 
  (SWP18601 + LWR14669, 19/11/82, $\varphi$ = 0.49; 
   SWP55160 + LWP30952, 19/11/82, $\varphi$ = 0.57) 
and at the opposite position 
  (SWP32845 + LWP12624, 03/02/88, $\varphi$ = 0.99). 
For the active phase we selected spectra taken at 
the 1985 maximum 
  (SWP27370 + LWP07370, 24/12/85, $\varphi$ = 0.98). 

{\em Radiation from the giant}.  
We matched the IR fluxes by a synthetic spectrum with 
  $T_{\rm eff} = 3\,400$\,K 
and scaling 
  $k_{\rm g} = 2.66\,10^{-18}$, 
which corresponds to the bolometric flux, 
  $F_{\rm g}^{\rm obs} = 2.0\,10^{-8}$\,\ecs. 
Our value of $\theta_{\rm g}$ is in good agreement with that 
derived from the surface brightness relation for M-giants 
\citep{ds98}, $\theta_{\rm g} = 1.7\,10^{-9}$, given by 
the reddening free magnitudes $K$ = 4.85 and $J$ = 6.09\,mag.
Our value of $\theta_{\rm g}$ then gives
the radius of the giant 
  $R_{\rm g} = 106\,(d/1.5\,\kpc)\,R_{\sun}$ 
and the luminosity
  $L_{\rm g} \sim 1\,400\,(d/1.5\,\kpc)^2\,L_{\sun}$. 

{\em Radiation from the ultraviolet: Quiescent phase}. 
Features of the iron curtain can be recognized mainly in 
the short-wavelength part of the spectrum. 
Our solution for the spectra at $\varphi \sim 0.5$ 
corresponds to 
  $T_{\rm h} = 120\,000$\,K = $T_{\rm h}^{\rm min}$ 
(see discussion below) and very different electron temperatures, 
  $T_{\rm e}(19/11/82) = 20\,500\,\pm\,2\,000$\,K 
and 
  $T_{\rm e}(29/06/95) = 11\,500\,\pm\,1\,000$\,K. 
The significant change in $T_{\rm e}$ by about 10\,000\,K 
is observationally given by the change of the slope of 
the near-UV continuum profile (Fig.~3). 
The stellar component of radiation was scaled with 
  $k_{\rm h} = 2.73\,10^{-24}$, 
which yields 
  $\theta_{\rm h} = 1.65\,10^{-12}$, 
  $R_{\rm h} = 0.11(d/1.5\,\kpc)\,R_{\sun}$
and 
  $L_{\rm h} = 2\,300(d/1.5\,\kpc)^2\,L_{\sun}$. 
The nebular component of the radiation was scaled with 
  $k_{\rm N}(19/11/82) = 3.6\,10^{15}\,{\rm cm^{-5}}$  
and
  $k_{\rm N}(29/06/95) = 2.7\,10^{15}\,{\rm cm^{-5}}$. 
These fitting parameters give $\delta$ = 0.92 and 1.1 on 
19/11/82 and 29/06/95, respectively. The case of 
$\delta \sim 1$ corresponds to a lower limit of the hot 
star temperature, 
  $T_{\rm h}^{\rm min} \sim 120\,000$\,K, 
under which the hot star is not able to produce the observed 
amount of nebular emission (Eq.~23). 
However, parameters of Z\,And suggest a very open \ion{H}{ii} 
zone \citep[e.g. Fig.~7 of][]{fc+88}, which means that a certain 
fraction of the ionizing photons escapes the system without 
being converted to nebular radiation (i.e. $\delta \ll 1$). 
This situation requires even higher $T_{\rm h}$ at the same 
scaling to balance the measured $EM$. Therefore the real values 
of $L_{\rm h}$ and $R_{\rm h}^{\rm eff}$  are larger/smaller 
than we derived from our model for the $T_{\rm h}^{\rm min}$ 
temperature. 
The spectrum from 03/02/88 ($\varphi \sim 0$) is characterized 
by the Rayleigh attenuated far-UV continuum with 
  $n_{\rm H} \sim 3.9\,10^{22}$\,\cmdva\ 
and features of the iron curtain that are more pronounced than 
at the opposite binary position. This 
can be interpreted in terms of the atmospheric eclipse due 
to Rayleigh scattering of the far-UV photons on the hydrogen 
atoms of the neutral giant's wind \citep{inv89}. During 
quiescent phases this effect is a strong signature of a highly 
inclined orbit of the binary. 
A further distinct difference between the spectra taken at 
$\varphi\,\sim\,0.5$ and $\varphi\,\sim$\,0 is an attenuation 
of the UV continuum at all wavelengths by a factor of about 2. 
This effect is present in all IUE spectra of Z\,And in 
the range of phases $\varphi\,\approx\,0\,\pm\,0.15$. 
As a result such an extinction process results in lower values 
of the derived parameters, 
  $\theta_{\rm h} < 1.0\,10^{-12}$, 
  $R_{\rm h}^{\rm eff} < 0.068(d/1.5\,\kpc)\,R_{\sun}$
and 
  $L_{\rm h} > 860(d/1.5\,\kpc)^2\,L_{\sun}$. 
Also the amount of the nebular emission decreased by a factor 
of about 5 
and the parameter $\delta$ = 0.2 for $k_{\rm h} = 1.0\,10^{-24}$, 
which reflects a significant attenuation/loss of the nebular 
photons in the direction of the observer at $\varphi\,\sim\,0$. 

{\em Radiation from the ultraviolet: Active phase}. 
The selected spectrum is characterized by a strong 
Rayleigh attenuation of the far-UV continuum and the effect 
of the iron curtain absorption, which drastically 
depressed the level of the continuum in between 2\,150 and 
3\,000\,\AA\ (Fig.~3). 
Note that during activity these effects are also significant 
out of the occultation by the giant 
(e.g. on SWP22684 + LWP03099, $\varphi$ = 0.15). 
The temperature of the stellar component of the radiation 
significantly decreased to $T_{\rm h} \sim 19\,000$\,K and 
is transferred throughout the $\sim\,2\,10^{23}$\,\cmdva\ 
hydrogen atoms on the line of sight. Below we call this 
stellar component of radiation during active phases 
the 'hot stellar source' (HSS). 
There are two components of nebular radiation: 
 (i) A low-temperature nebula (LTN) with typical properties 
     of quiescent phases and 
(ii) a high-temperature nebula (HTN) corresponding to 
     a high electron temperature 
     ($T_{\rm e} \sim 40\,000$\,K). 
The latter is required to fit the far-UV continuum. 
With analogy to AR\,Pav \citep[see.][]{sk03b}, we scaled 
a 40\,000\,K hot nebular radiation of hydrogen to 
the non-zero continuum around the Ly${\alpha}$ line. 
Ignoring the effect of the iron curtain leads to 
an artificially high He$^{++}$ abundance and too low electron 
temperature of the LTN \citep[see Fig.~3 of][]{sk03a}. 
%
%
\begin{figure}
\centering
\begin{center}
\resizebox{\hsize}{!}{\includegraphics[angle=-90]{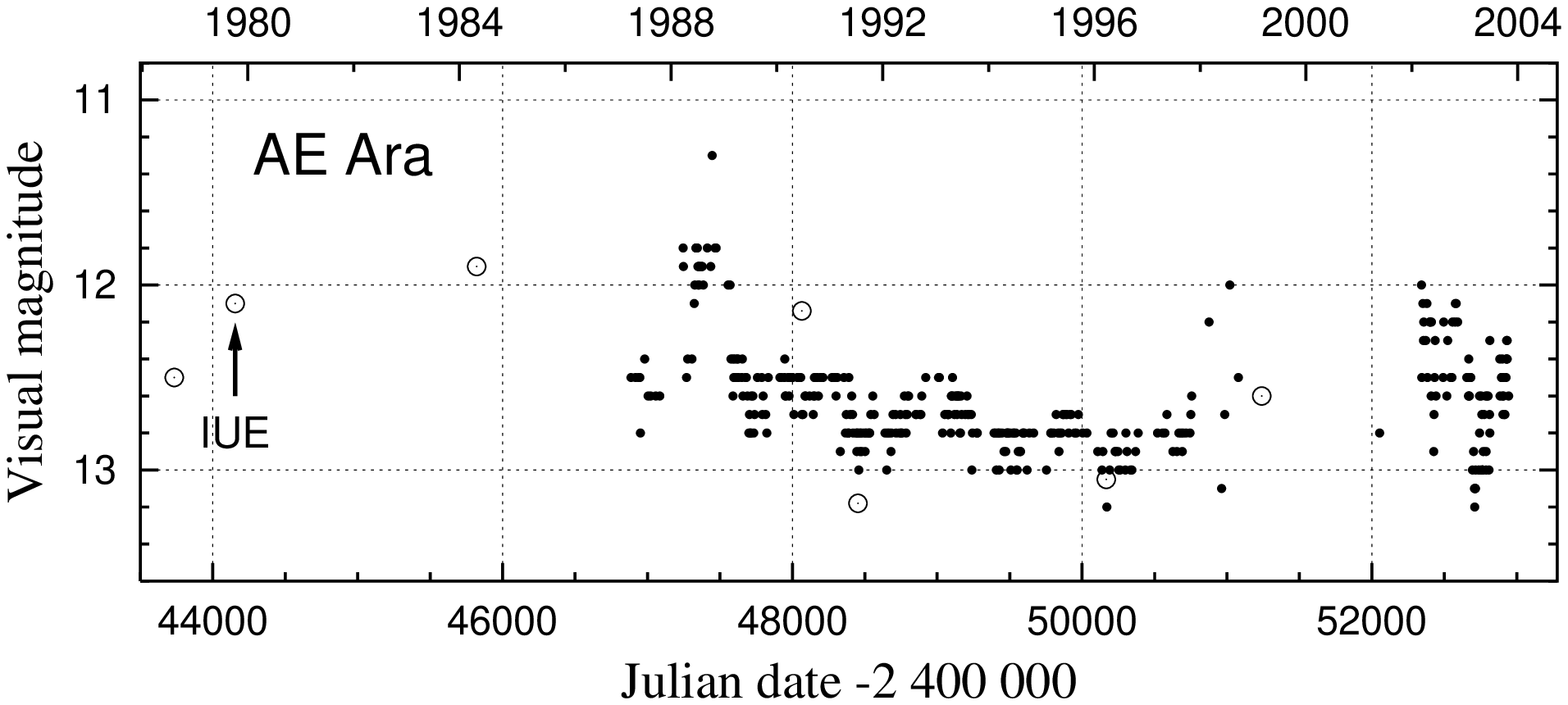}}

\vspace*{2mm}

\resizebox{\hsize}{!}{\includegraphics[angle=-90]{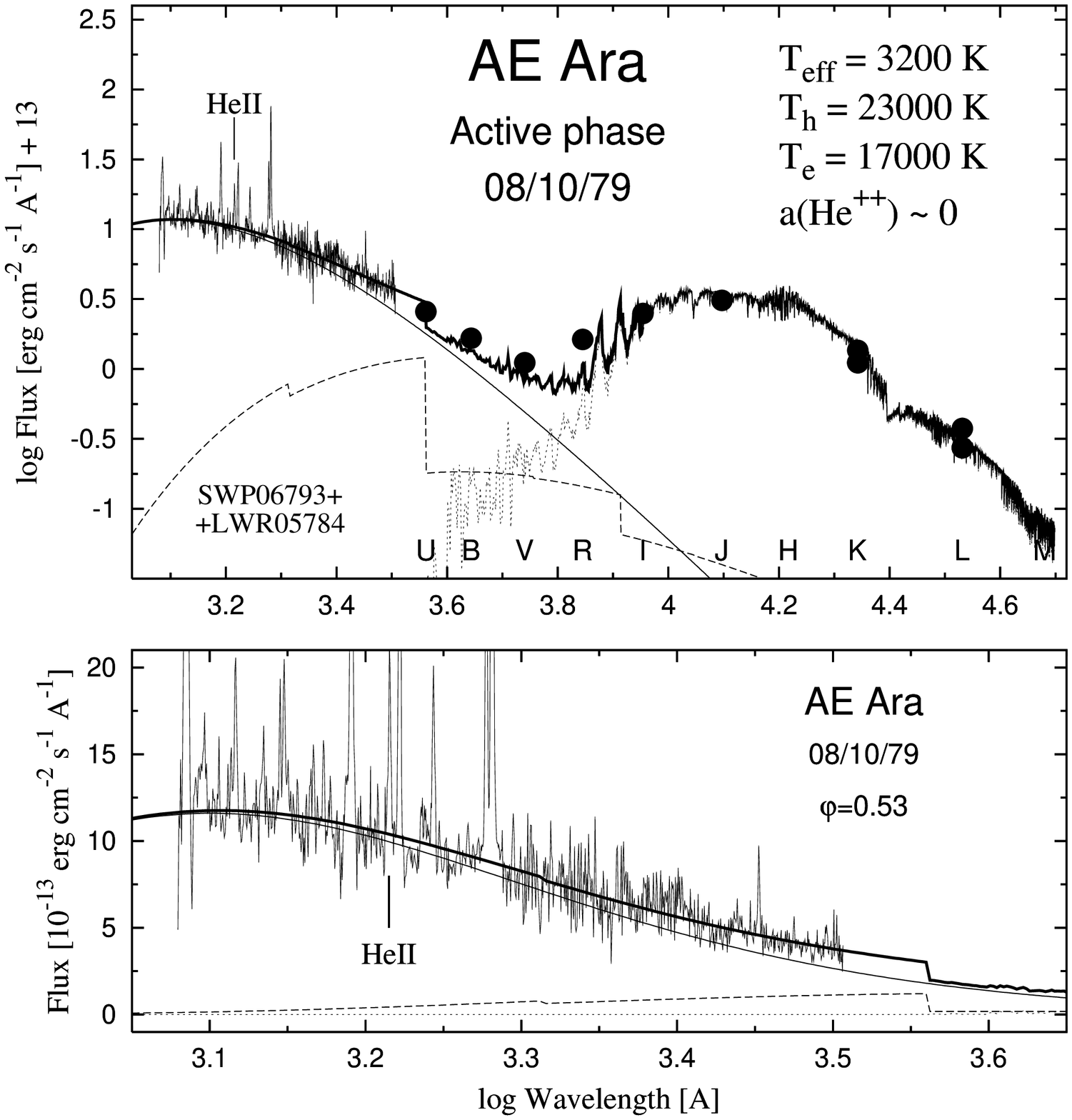}}
\caption[]{
Top: Visual light curve of AR\,Ara. At the time of IUE 
observations the system was at a higher level of its activity. 
Data are from A.~Jones ($\bullet$, private communication) 
and \cite{mika+03} ($\circ$). Middle and bottom: 
Reconstructed SED shows typical characteristics of an 
active phase of the system. 
          }
\end{center}
\end{figure}

\subsection{AE~Ara}
%

Visual LC of AE\,Ara (1987 -- 2004) displays some bright 
stages alternating with periods of quiescence (Fig.~4). 
For the purpose of this paper we used the only well 
exposed low-resolution spectrum from the IUE archive
  (SWP06793+LWR05784, 08/10/79, $\varphi = 0.53$). 
To flatten its pronounced 2\,200\,\AA\ feature we determined 
the colour excess, $E_{\rm B-V}\, =\, 0.25\,\pm 0.05$. 

{\em Radiation from the giant.}   
A synthetic spectrum with $T_{\rm eff} = 3\,200$\,K matches 
best the IR fluxes from $IJKL$ photometry \citep{ag74,mu+92}. 
Its scaling, 
 $k_{\rm g} = 8.1\,10^{-19}$, 
yields 
  $F_{\rm g}^{\rm obs} = 4.8\,10^{-9}$\,\ecs. 
If we adopt the radius of the giant, 
   $R_{\rm g} = 130 - 150\,R_{\sun}$, 
in agreement with the empirical relations between the spectral 
type and $T_{\rm eff}$ \citep[e.g.][]{bel+99}, then 
$\theta_{\rm g}$ is transformed to $d = 3.5\,\pm\,0.3$\,\kpc, 
where the uncertainty is given only by the adopted range 
of the giant's radius. This result is in excellent agreement 
with that derived from the surface brightness relation 
for M-giants \citep{ds98}. 
The giant's luminosity 
 $L_{\rm g} = 1\,800\,\pm\,300\,(d/3.5\,\kpc)^2\,L_{\sun}$ 
(Eq.~4). 

{\em Radiation from the ultraviolet.}  
The hot star in AE\,Ara during the time of the IUE observation 
was in an active stage with a significant contribution in the 
optical. A moderate influence of the iron curtain absorptions 
can be also recognized. 
Our model identified a warm pseudophotosphere (the HSS) 
radiating at 
  $T_{\rm h} = 23\,000\,+ 10\,000/ - 3\,000$\,K. 
The scaling factor, 
  $k_{\rm h} = 1.40\,10^{-22}$, 
implies its effective radius, 
  $R_{\rm h} = 1.8(d/3.5\,\kpc)\,R_{\sun}$ (Eq.~6). 
To obtain a better fit of the long-wavelength end of the 
spectrum, a nebular radiation determined by 
  $T_{\rm e} = 17\,000\,\pm\,3\,000$\,K 
and the scaling, 
  $k_{\rm N} = 3.6\,10^{14}\,{\rm cm^{-5}}$, 
was required. 
There is no direct signature for the presence of the HTN 
component in the spectrum. 
The active stage of AE\,Ara is signaled by its 
two-temperature UV spectrum, which develops during active 
phases of all the objects with a high orbital 
inclination (Sect.~5.3.4). 
%
%
\begin{figure}
\centering
\begin{center}
\resizebox{\hsize}{!}{\includegraphics[angle=-90]{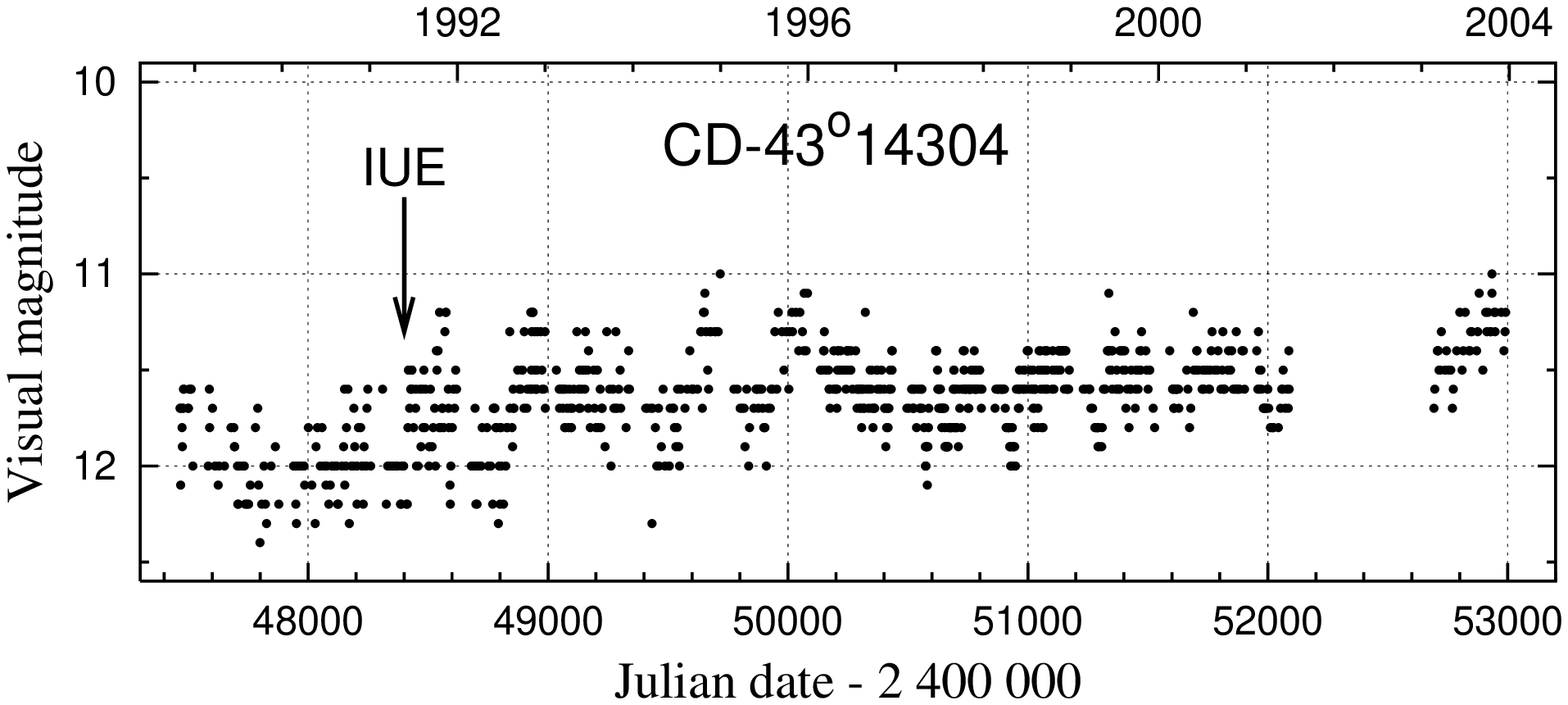}}

\vspace*{2mm}

\resizebox{\hsize}{!}{\includegraphics[angle=-90]{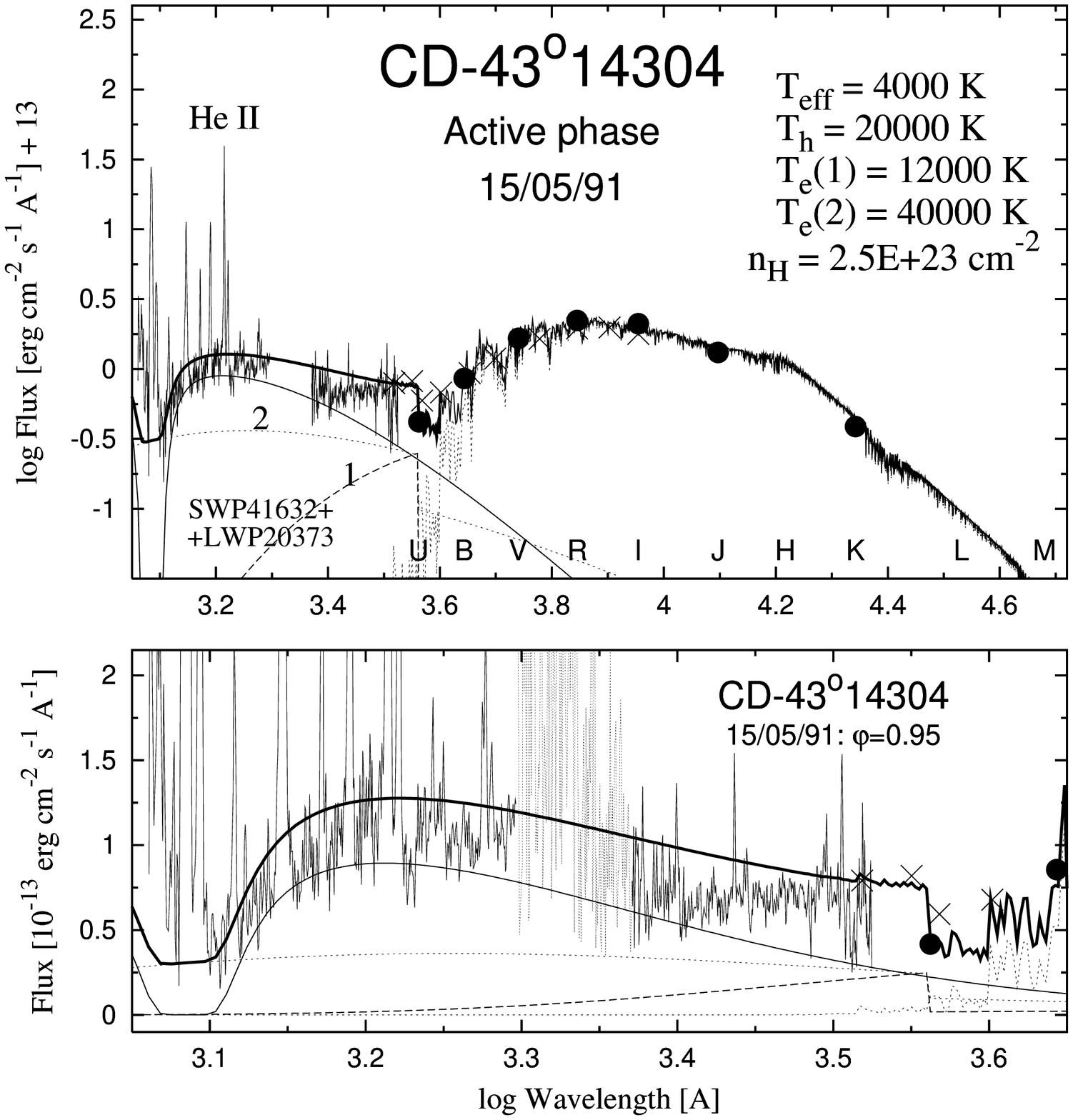}}
\caption[]{The SED for CD\,43$^{\circ}$14304. It is very 
           similar to that from active phases of other well 
           known systems. Crosses ($\times$) represent fluxes 
           in the continuum of the optical spectroscopy 
           carried out in July 1991 \citep{gmc99}. 
           At the time of IUE observations the system 
           just entered an active phase (top panel). 
          }
\end{center}
\end{figure}

\subsection{CD-43$^{\circ}$14304}
%

Visual estimates of this system were regularly carried out 
by Albert Jones (Fig.~5, private communication). 
The LC shows periods of a lower and higher level 
of activity differing by about 1\,mag. 
Only a few multicolour photometric observations were made 
by \cite{mu+92} and \cite{gmc99} in the optical $UBVRI$ and 
infrared $JHK$ passbands. 
\cite{schn93} found that the light of CD-43$^{\circ}$14304 
is reddened with $E_{\rm B-V}\, <\,$0.2. The observed 
index $V-K$ = 4.0 and that corresponding to the 4\,000\,K 
giant, $(V-K)_{\rm g}\,\dot =\, 3.6$ \citep{bel+99}, yields 
$E_{\rm B-V}\,=\,$0.15. This value represents a lower limit, 
because of a contribution from other sources in the $V$ band. 
Therefore we adopted $E_{\rm B-V}\,=\,$0.2. 
There is only one spectrum in the final IUE archive 
  (SWP41632 + LWP20373, 15/05/91, $\varphi = 0.95$) 
applicable to our modeling. The spectrum taken by 
the long wavelength prime was poorly exposed (Fig.~5). 

{\em Radiation from the giant.}   
The $(B)VRIJK$ photometric measurements can be compared well 
with a synthetic spectrum of $T_{\rm eff} = 4\,000$\,K 
scaled with $k_{\rm g} = 1.9\,10^{-19}$, which gives 
  $F_{\rm g}^{\rm obs} = 2.7\,10^{-9}$\,\ecs. 
If we adopt the giant's radius to be 
  $R_{\rm g} \equiv 40\,R_{\sun}$ 
as a typical value for the spectral type K7\,I\I\I\ 
\citep[][]{ms99,bel+99} then the angular radius yields 
  $d = 2.1\,$\,kpc. 
Accordingly, the giant's luminosity 
 $L_{\rm g} = 380\,(d/2.1\,\kpc)^2\,L_{\sun}$. 

{\em Radiation from the ultraviolet}  
The two-temperature type of the ultraviolet spectrum 
signals an active stage of CD-43$^{\circ}$14304. 
Independently, \cite{gmc99} found an increase of 
the intensity shortward of 5\,000\,\AA\ in their optical 
spectra from 1991 June and July 
(i.e. taken near to the IUE observations) with respect to 
the spectrum from 1987 by a factor of 3 at 4\,000\,\AA. 
They ascribed this optical brightening to a mild outburst. 
Our solution indicates a cool shell (i.e. the HSS) re-radiating 
the hot star photons in the direction of the observer at 
  $T_{\rm h} = 20\,000\,+\,7\,000/-\,3\,000$\,K 
and scaling of 
  $k_{\rm h} = 2.4\,10^{-23}$, 
which results in a significant contribution in the optical. 
The spectrum shows the far-UV continuum Rayleigh attenuated 
by 2.5\,10$^{23}$\,\cmdva\ \ion{H}{i} atoms and pronounced 
features of the iron curtain absorption. 
The position of the binary at the time of observation 
($\varphi$ = 0.95) enhances both effects. 
The effective radius of the shell is 
  $R_{\rm h}^{\rm eff} = 0.46(d/2.1\,\kpc)\,R_{\sun}$ 
and its luminosity 
  $L_{\rm h} = 30\,(d/2.1\,\kpc)^2\,L_{\sun}$ (Eq.~6). 
A non-zero level of the Rayleigh attenuated far-UV continuum 
at/around $Ly{\alpha}$ indicates a contribution from the HTN 
with $k_{\rm N} = 1.9\,10^{14}\,\rm cm^{-5}$ 
for $T_{\rm e} \equiv 40\,000$\,K. 
The continuum profile at the end of the LWP spectrum 
and the following part containing the Balmer jump reflects 
the presence of a faint radiation from the LTN: 
  $T_{\rm e} \sim 12\,000$\,K, 
  $k_{\rm N} \sim 0.5\,10^{14}\,\rm cm^{-5}$. 
Evolution in the H$\alpha$
line profile along the orbital motion \citep{sch+98} 
is very similar to that observed in eclipsing systems. 
For example, 
EG\,And \citep[][]{oliv+85}, 
V1329\,Cyg \citep[][]{it00}, 
AX\,Per \citep[][]{iijima88}, 
RW\,Hya \citep[][]{sch+96} and 
SY\,Mus \citep[][]{schmutz+94}. 
This suggests that the orbital plane of CD-43$^{\circ}$14304 
is highly inclined to the observer, which is consistent with 
the presence of the two-temperature ultraviolet spectrum. 
%
%
%
\begin{figure}
\centering
\begin{center}
\resizebox{\hsize}{!}{\includegraphics[angle=-90]{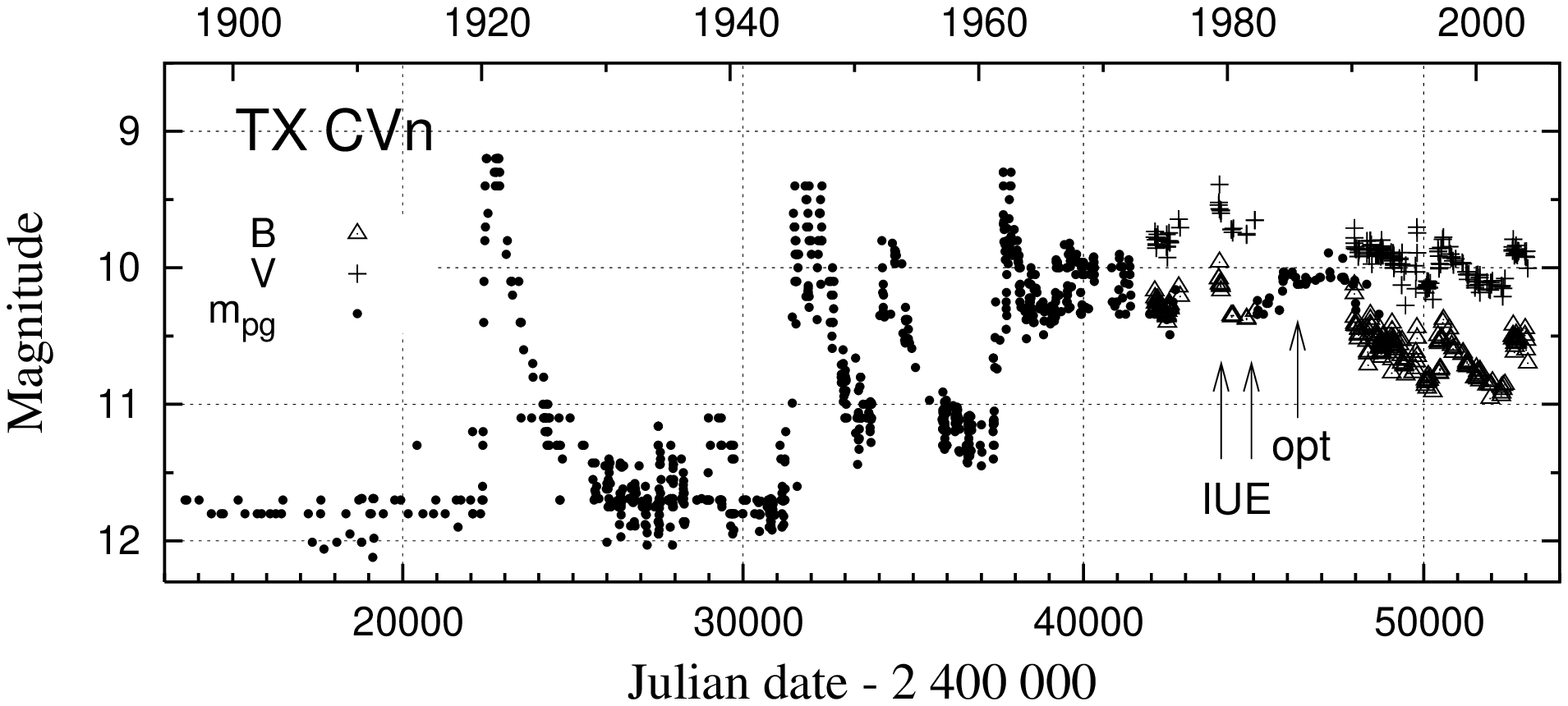}}

\vspace*{2mm}

\resizebox{\hsize}{!}{\includegraphics[angle=-90]{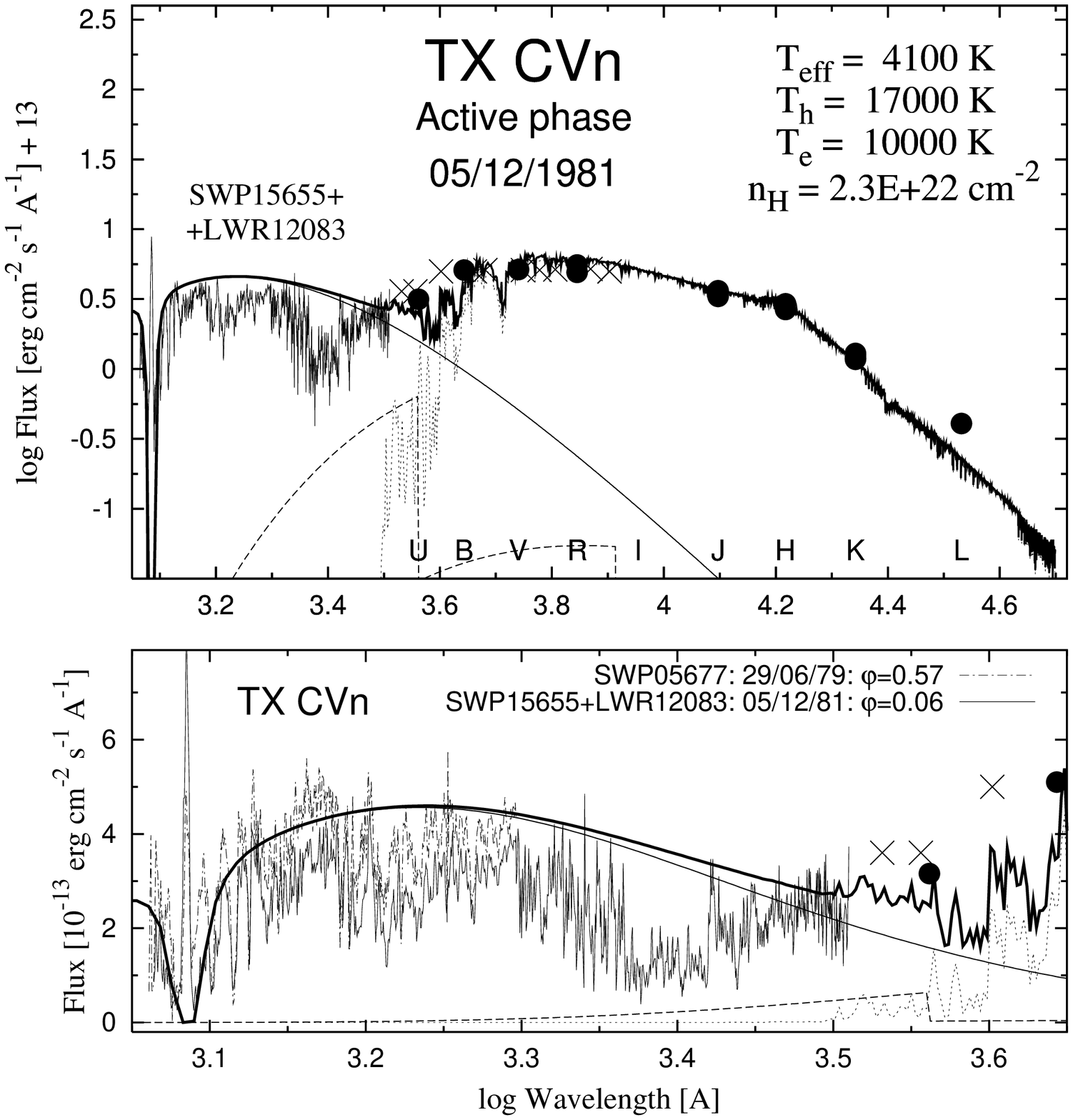}}
%
%
\caption[]{
The observed and modeled SED of TX\,CVn. The ultraviolet 
continuum is drastically affected by the iron curtain 
absorptions and the far-UV part is moderately 
attenuated by the Rayleigh scattering -- typical features of 
active phases of symbiotic stars. 
Flux-points of the optical spectroscopy ($\times$) 
are from \cite{kg89}. 
Dates of spectroscopic observations are marked 
by arrows at the top panel. 
          }
\end{center}
\end{figure}

\subsection{TX~Canum Venaticorum}

The historical LC shows the eruptive character of TX\,CVn. 
The present active phase is maintained from the last major 
outburst in 1962 (Fig.~6). Recently, \cite{sk+04} revealed 
two minima in the $U$-LC, whose timing coincides with that 
of the inferior conjunction of the giant \citep[][]{kg89}. 
This suggests a high orbital inclination of TX\,CVn. 
Here we used well-exposed IUE spectra taken close 
to the inferior and superior occultation of the giant
(SWP15655 + LWR12083, 05/12/81, $\varphi = 0.06$ and 
 SWP05677, 29/06/79, $\varphi = 0.57$). 

{\em Radiation from the giant.}  
A synthetic spectrum of 
  $T_{\rm eff} = 4\,100$\,K 
matches well the $VRJHKL$ photometric measurements. 
Its scaling, 
 $k_{\rm g} = 4.5\,10^{-19}$, 
corresponds to 
  $F_{\rm g}^{\rm obs} = 7.1\,10^{-9}$\,\ecs. 
According to the empirical relations between the spectral type, 
$V-K$ index and $T_{\rm eff}$ \citep[e.g.][]{bel+99}, 
the giant's radius 
  $R_{\rm g} = 30\pm 10\,R_{\sun}$, which then transforms 
the $\theta_{\rm g}$ parameter to the distance 
  $d = 1.0\,\pm\,0.3$\,kpc. 
The corresponding luminosity 
 $L_{\rm g} = 230\,(d/1.0\,\kpc)^2\,L_{\sun}$. 

{\em Radiation from the ultraviolet.}   
The profile of the ultraviolet continuum displays 
characteristics of an active phase in agreement with evolution 
in the LC (Fig.~6). The effect of Rayleigh scattering and 
absorptions of the veiling gas are present at very different 
orbital phases ($\varphi$ = 0.06 and 0.57), which suggests 
its distribution around the central star. 
However it is not possible to determine its geometry more 
accurately, because 
the spectrum does not contain emission lines -- the shell 
probably encompasses the whole star core 
\citep[see also the discussion on the nature of the hot component 
by][]{kg89}. 
Our solution for the 05/12/81 spectrum suggests a black-body 
radiation at 
   $T_{\rm h} \sim 17\,000$\,K
  ($k_{\rm h} = 2.6\,10^{-22}$) 
as a dominant component from the hot object. 
Only a very small nebular contribution is required 
to fit better the long-wavelength end of the spectrum. 
However, the line blanketing affects drastically the ultraviolet 
continuum, which does not allow us to determine reliably 
the continuum fluxes and thus to distinguish unambiguously 
the presence of a faint LTN in the spectrum. 
%
%
%
\begin{figure}
\centering
\begin{center}
\resizebox{\hsize}{!}{\includegraphics[angle=-90]{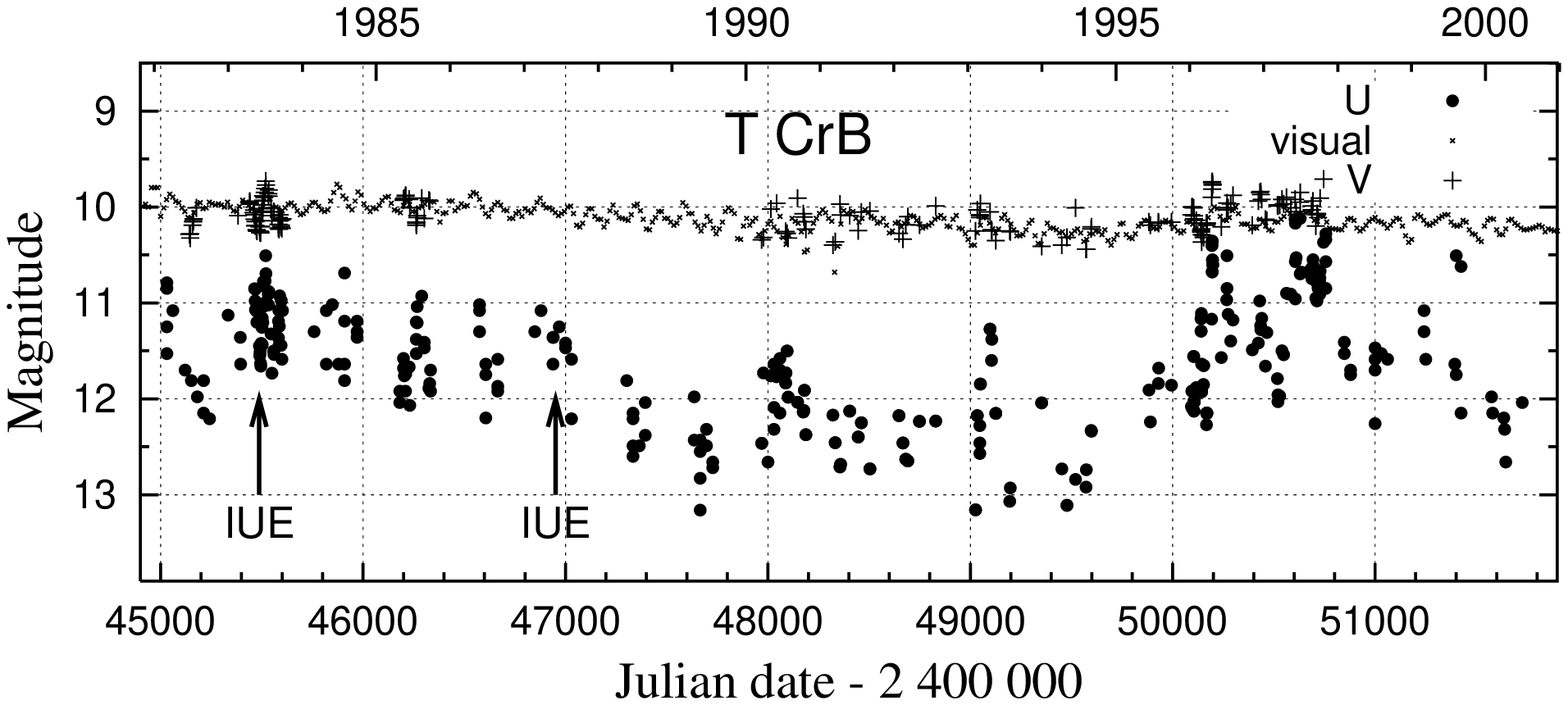}}

\vspace*{2mm}

\resizebox{\hsize}{!}{\includegraphics[angle=-90]{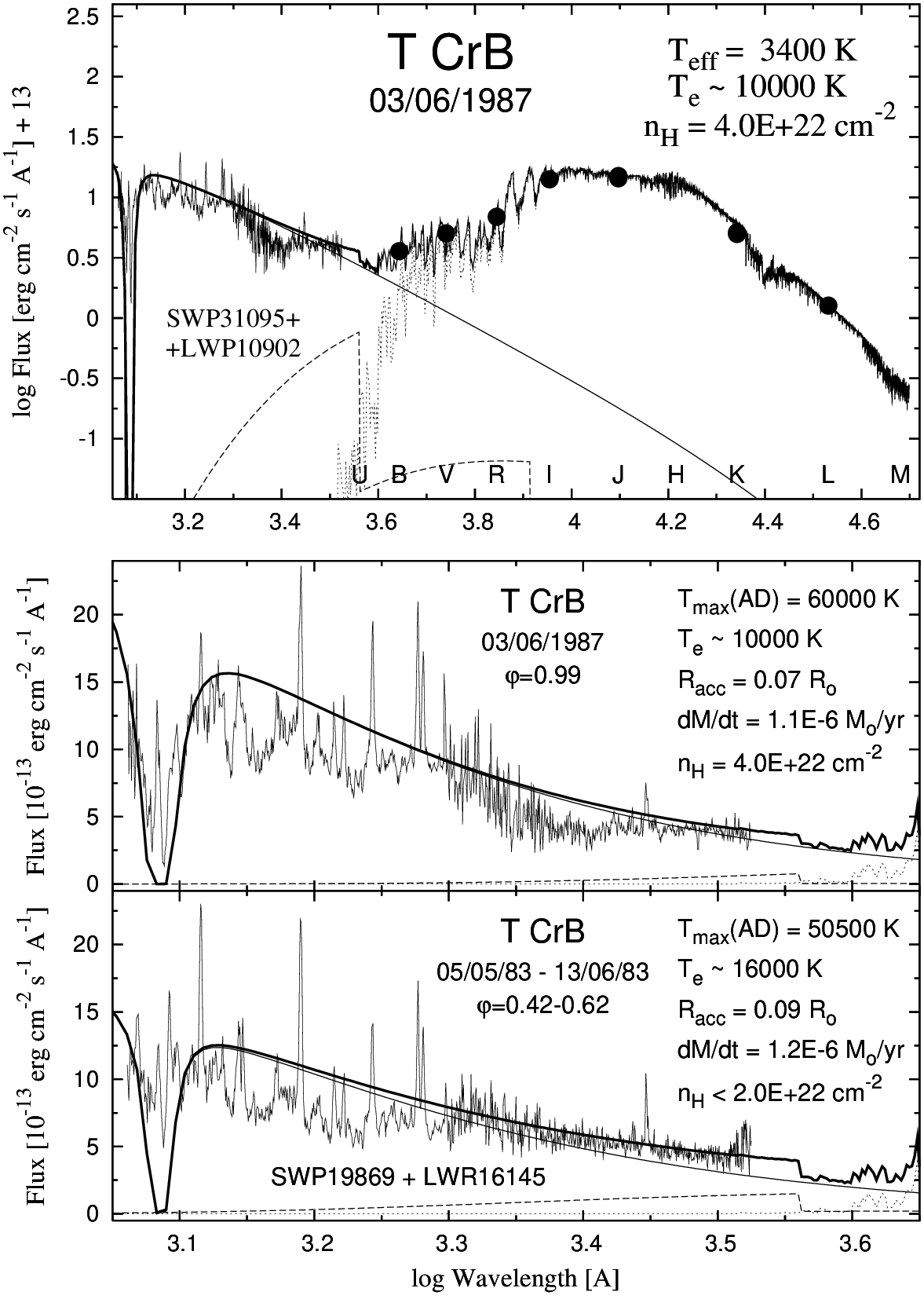}}
\caption[]{
The UV/optical/IR SED for T\,CrB. A black-body accretion disk 
dominates the UV domain with a significant contribution in 
$UB$ passbands. The Rayleigh scattering and absorptions of 
the iron curtain affect the UV continuum. The contribution from 
the nebula is very small. Dates of observations are marked 
in the LC at the top panel. 
          }
\end{center}
\end{figure}

\subsection{T\,Coronae Borealis}
%
T\,CrB is a symbiotic recurrent nova with two major outbursts,
recorded in 1866 and 1946. Recent evolution in the LC shows
periods of higher and lower states of activity 
\citep[e.g.][Fig.~7]{s+04}. 
Currently, it is believed that the hot star luminosity is 
powered by an accretion process triggered by a mass transfer 
from the lobe-filling giant. 
\cite{sell+92} suggested a white dwarf as the accretor 
to satisfy mainly the ultraviolet IUE observations, while 
\cite{ck92} tested accretion of a gaseous torus onto 
a main-sequence star to explain the secondary outburst of 
the 1946 eruption. 
To model the ultraviolet continuum we selected observations 
from the giant's inferior conjunction 
 (SWP31095 + LWP10902, 03/06/87, $\varphi$\,=\,0.99) 
and from the opposite site 
(SWP19869, 01/05/83, $\varphi$\,=\,0.42 and LWR16145, 13/06/83, 
 $\varphi$\,=\,0.62 -- no simultaneous SWP/LWR exposures are 
 available at this position). 

{\em Radiation from the giant}   
The red giant's fluxes can be matched by a synthetic 
spectrum of $T_{\rm eff} = 3\,400$\,K and the scaling, 
  $k_{\rm g} = 3.13\,10^{-18}$, 
which corresponds to the observed bolometric flux, 
  $F_{\rm g}^{\rm obs} = 2.37\,10^{-8}$\,\ecs\ 
and the angular radius, 
  $\theta_{\rm g} = 1.77\,10^{-9}$. 
These parameters imply the giant's radius,
  $R_{\rm g} = 75\,\pm\,12\,(d/960\,\pc)\,R_{\sun}$
and the luminosity,
  $L_{\rm g} = 680\,\pm\,230\,(d/960\,\pc)^2\,L_{\sun}$ 
(Table~1). 

{\em Radiation from the ultraviolet.}  
Both spectra display the Rayleigh attenuated far-UV 
continuum 
%
%
and pronounced features of the iron curtain. This constrains 
the veiling material surrounding the hot source, being 
located at/behind the outer rim of the disk, extended 
vertically from the disk plane, so as to cause the observed 
absorptions. 
At $\varphi \sim 0$ the effect of the absorbing slab is 
enhanced by the cool giant's wind. 
We estimated flux-points of the continuum with the aid 
of calculated spectra including the blanketing effect 
\citep{sa93}. 
We performed modeling of the continuum by substituting the single 
Planck function in Eq.~(13) by a component of radiation from 
an accretion disk. Then Eq.~(13) in the ultraviolet is 
%
%
\begin{equation}
 F(\lambda) = k_{\rm AD}\times F_{\lambda}({\rm AD})\,
              e^{-n_{\rm H}\sigma_{\lambda}^{\rm R}} +
              k_{\rm N}\times
\hat{\varepsilon}_{\lambda}({\rm H, He^{+}},T_{\rm e},\tilde{a}).
\end{equation}
%
The function $F_{\lambda}({\rm AD})$ represents a flux distribution 
from an optically thick accretion disk that radiates locally 
like a black body. 
Here we adopted expressions according to the textbook of 
\cite{w95}, recently used also by \cite{sk03b}. 
A fundamental parameter in models with an accretion disk 
is the disk temperature $T_{\star} = 2\,T_{\rm max}$, which 
determines the slope of the UV continuum. Observations from 
03/06/87 suggest $T_{\star} \approx 110\,000$\,K at all 
wavelengths. We found that 
it is possible to match the selected points of the continuum 
by very different sets of model parameters. Below we summarize 
all possible solutions to select ranges of fitting parameters 
that are physically plausible. In all calculations we adopted 
the core mass of the accretor to be 1.37\,$M_{\sun}$ and 
the orbital inclination $i\,=\,67^{\circ}$ \citep{s+04}. 

1. Solutions with the accretion disk only: 

(a) If the radius $R_{\rm h} \sim R_{\rm WD}$ 
  (we adopted $R_{\rm h} = 0.01\,R_{\sun}$) then the temperature 
   required by the slope of the continuum, 
   $T_{\star}\sim 110\,000$\,K, corresponds to the mass 
   accretion rate $\dot M_{\rm acc} \sim$ a few $\times
   10^{-9}M_{\sun}\,\rm yr^{-1}$, which yields 
   $L_{\rm acc}\sim$ a few $\times\, L_{\sun}$. These values 
   are much lower than those given by observations, 
  $L_{\rm obs}(AD) = \int_{\lambda} F_{\lambda}(AD)\,{\rm d}\lambda\,
  /\cos(i)\sim 300\,L_{\sun}$. 

(b) If the radius $R_{\rm h} > R_{\rm WD}$ (we tested a range of 0.10 
  to 0.05\,$R_{\sun}$), the $T_{\star}$ temperature can be satisfied 
  by $\dot M_{\rm acc}\sim 10^{-6}M_{\sun}\,\rm yr^{-1}$ 
  corresponding to $L_{\rm acc} \sim L_{\rm obs}(AD)$. 

2. Solutions with the accretion disk and nebula. These models 
are not unambiguous. Models with a higher 
$T_{\star}$ express better only the far-UV region, which then 
requires a strong nebular contribution to fit the near-UV 
part of the spectrum. 
%

(a) The case of $R_{\rm h} \sim R_{\rm WD}$ and a large nebular 
  contribution leads to $T_{\star} \gg 110\,000$\,K, 
  $L_{\rm acc} \sim 10^{3}\,L_{\sun}$ and 
  $\dot M_{\rm acc}\,\sim\,1-10\,10^{-7}M_{\sun}\,\rm yr^{-1}$. 
  These models are not consistent with observations 
  in too high $L_{\rm acc}$ and the nebular emission. 
  Models with small contributions from the nebula converge 
  to the cases of point 1\,(a) above. 

(b) Solutions for $R_{\rm h} > R_{\rm WD}$ correspond to 
  $T_{\star}\sim 1.1 - 1.3\,10^{5}$\,K, 
  $\dot M_{\rm acc}\sim 1 - 1.6\times 10^{-6}M_{\sun}\,\rm yr^{-1}$ 
  and 
  $L_{\rm acc} \sim L_{\rm obs}(AD) \sim 300 - 400\,L_{\sun}$. 
  The emission measure is small and it is not possible 
  to determine $T_{\rm e}$ unambiguously. 
%

Our modeling of the SED of T\,CrB shows that a major fraction 
of the UV radiation is produced by an optically thick accretion 
disk. Models whose parameters are consistent with observations 
result from an accretion process at high accretion rates of 
$\dot M_{\rm acc}\sim 1 - 1.6\times 10^{-6}M_{\sun}\,\rm yr^{-1}$
onto the accretor with 
  $R_{\rm h} > R_{\rm WD}$ (Fig.~7). 
These results, however, do not support the presence of 
the white dwarf as the accretor in T\,CrB. 
%
%
%
\begin{figure}
\centering
\begin{center}
\resizebox{\hsize}{!}{\includegraphics[angle=-90]{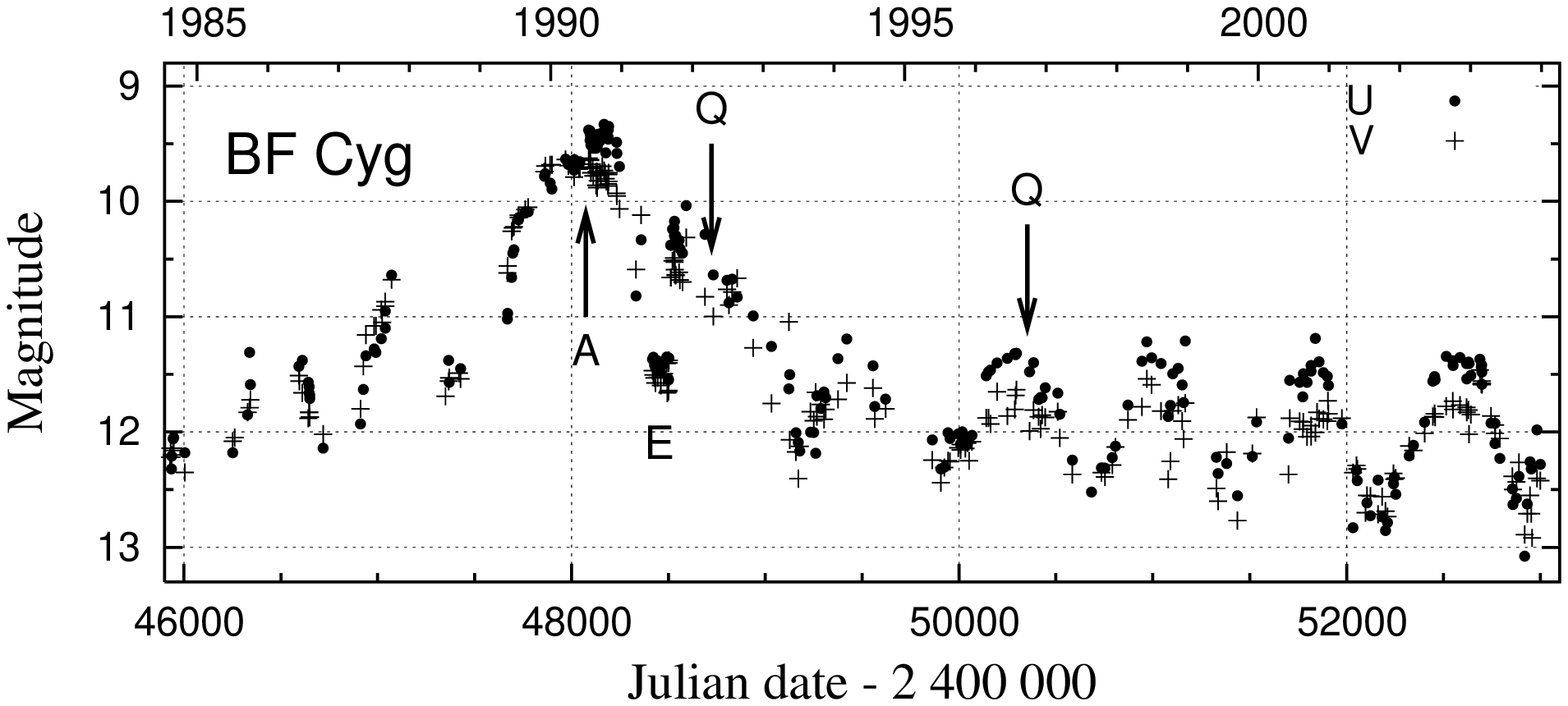}}

\vspace*{2mm}

\resizebox{\hsize}{!}{\includegraphics[angle=-90]{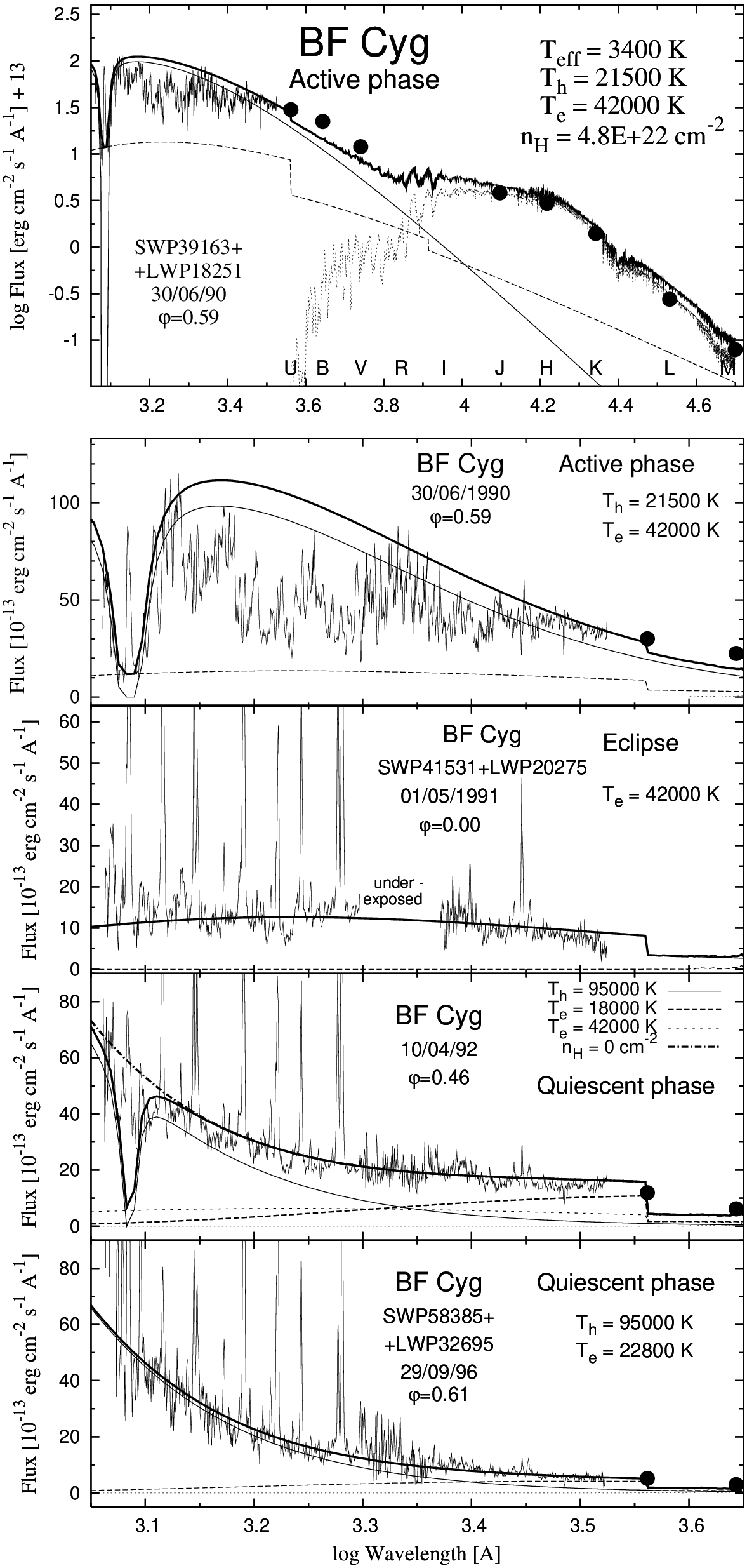}}
%
%
\caption[]{
The SEDs of BF\,Cyg during the maximum of the star's brightness 
(denoted by {\sf A} at the top panel), eclipse ({\sf E}) and 
quiescence ({\sf Q}). Dates of observations are marked 
in the LC at the top panel. 
          }
\end{center}
\end{figure}

\subsection{BF\,Cygni}
%

BF\,Cyg is an eclipsing symbiotic star 
\citep[][Fig.~8 here]{sk92}. Its historical LC 
shows three basic types of active phases -- nova-like and Z\,And 
type of outbursts and short-term flares \citep[Fig.~1 of][]{sk+97}. 
Here we selec\-ted observations from 
the optical maximum 
(SWP39163 + LWP18251, 30/06/90, $\varphi = 0.59$), 
the eclipse effect 
(SWP41531 + LWP20275, 01/05/91, $\varphi = 0.0$), 
a transition to quiescence 
(SWP44370 + LWP22778, 10/04/92, $\varphi = 0.46$) 
and a quiescence 
(SWP58385 + LWP32695, 29/09/96, $\varphi = 0.61$). \\
\hspace*{5mm}{\em Radiation from the giant.} 
Fluxes of the photometric measurements can be compared with 
a synthetic spectrum of 
  $T_{\rm eff} = 3\,400$\,K (Figs.~8). 
Its scaling, 
  $k_{\rm g} = 4.5\,10^{-19}$, 
corresponds to the bolometric flux, 
  $F_{\rm g}^{\rm obs} = 5.9\,10^{-9}$\,\ecs. 
Empirical relations between the spectral type and 
$T_{\rm eff}$ suggest the giant's radius 
  $R_{\rm g} \sim 150\,R_{\sun}$, 
which yields the distance
  $d \sim 3.8$\,kpc. 
On the other hand the contact times of the 1991 eclipse 
give $R_{\rm g}/A = 0.54\,\pm\,0.02$, which suggests a very 
large radius of the giant \citep{sk+97}. A solution for 
the spectroscopic orbit \citep{f+01} yields the Roche 
lobe radius $R_{\rm L}/A = 0.49$. These results imply 
the lobe-filling giant in BF\,Cyg. 
However, the broad eclipse profile may be partly caused 
by attenuation of the optical light in the extended dense 
atmosphere of the giant. Note that the SWP42356 spectrum 
from 29/08/91 shows a steep far-UV profile, which signals 
the out-of-eclipse position, while the brightness in the 
optical remains at the level of the total eclipse. Therefore 
here we adopted $R_{\rm g} = 150\,R_{\sun}$. \\
%
%
\hspace*{5mm}{\em Active phase}.
The continuum is 
drastically depressed by the iron curtain absorptions and 
the far-UV continuum profile indicates a hydrogen column 
density with $n_{\rm H} \sim 5\,10^{22}\rm cm^{-2}$ and 
a contribution from the HTN. Other spectra from the maximum 
are of the same type, i.e. its profile does not depend on the 
orbital phase. Only very faint emissions of \ion{Si}{iv}\,1403, 
\ion{He}{ii}\,1640 and \ion{O}{iii}\,1664\,\AA\ can be still 
recognized. 
Our solution suggests the presence of a strong stellar component 
  ($T_{\rm h} \sim 21\,500$\,K, 
   $k_{\rm h} \sim 1.8\,10^{-21}$) 
superposed on the HTN emission 
  ($T_{\rm e} \sim 42\,000$\,K, 
   $k_{\rm N} \sim 6.9\,10^{15}\,{\rm cm^{-5}}$ -- as derived 
from the following eclipse). 
Both components dominate the optical and in part influence also 
the IR wavelengths. 
This composition explains colour indices during the totality 
of the 1991-eclipse, which also reflect a combined
spectrum \citep[][]{sk92}. \\
\hspace*{5mm}{\em Eclipse}.
The ultraviolet continuum from the 1991-eclipse is flat in 
profile radiating at a level of $\sim 10^{-12}$\,\ecsa. 
We ascribe this component to the emission from the HTN 
  ($T_{\rm e} \sim 42\,000$\,K, 
   $EM_{\rm HTN} \sim 1.1\,10^{61}(d/3.8\,\kpc)^2$\,\cmtri\ 
and
   $L_{\rm HTN} \sim 2\,100\,(d/3.8\,\kpc)^2\,L_{\sun}$). \\
\hspace*{5mm}{\em Transition to quiescence}. 
First we subtracted the HTN contribution and fitted the rest 
fluxes adopting the hot star temperature 
$T_{\rm h}$ = 95\,000\,K = $T_{\rm h}^{\rm min}$ 
($\delta = 1$, Eq.~23). 
The presence of the HTN emission is given by observations 
from the preceding eclipse. 
A high $T_{\rm h}$ here is required by a large emission 
measure 
 $EM = 1.2\,10^{61}(d/3.8\,\kpc)^2$\,\cmtri\
in total (Table~3). 
We note that a fraction of the nebular emission can be 
produced by collisions as suggested by the presence of 
the HTN (Sect.~5.3.1). 
This would allow the ionizing source to be cooler than 
$T_{\rm h}^{\rm min}$. 
Iron curtain features are relatively faint and the far-UV 
spectrum is attenuated with only 
  $n_{\rm H} \le \,10^{22}\rm cm^{-2}$. 
If one ignores influence of the iron curtain a formal good 
fit with $T_{\rm h} = 60\,000$\,K is also possible 
\citep[][]{f-c+90}. \\
\hspace*{5mm}{\em Quiescent phase}.
The stellar component 
has practically the same properties as that from the transition 
period, but without signatures of more pronounced absorptions 
from the veiling neutral gas. However, the nebular component 
  ($T_{\rm e} = 22\,800$\,K) 
decreased by a factor of about 4, to 
 $EM = 3.1\,10^{60}(d/3.8\,\kpc)^2$\,\cmtri, 
despite of that both the spectra were taken at the same 
orbital phase ($\varphi \approx 0.5$). A decrease of the emission 
measure at the same ionizing capacity of the hot star implies 
that a fraction of the ionizing photons escapes the system 
without being converted into the nebular emission, 
i.e. the parameter $\delta $ decreased and the nebula became 
more open than during the transition period. 
A constant mass-loss rate from the giant would mean that 
the hot star wind, which developed during the activity, 
caused a surplus of ionizations. In part it is itself 
a subject of ionization and in part it creates a fraction 
of the HTN by collisions. Otherwise the mass-loss rate from 
the giant had to significantly decreased. 
%
%
\begin{figure}
\centering
\begin{center}
\resizebox{\hsize}{!}{\includegraphics[angle=-90]{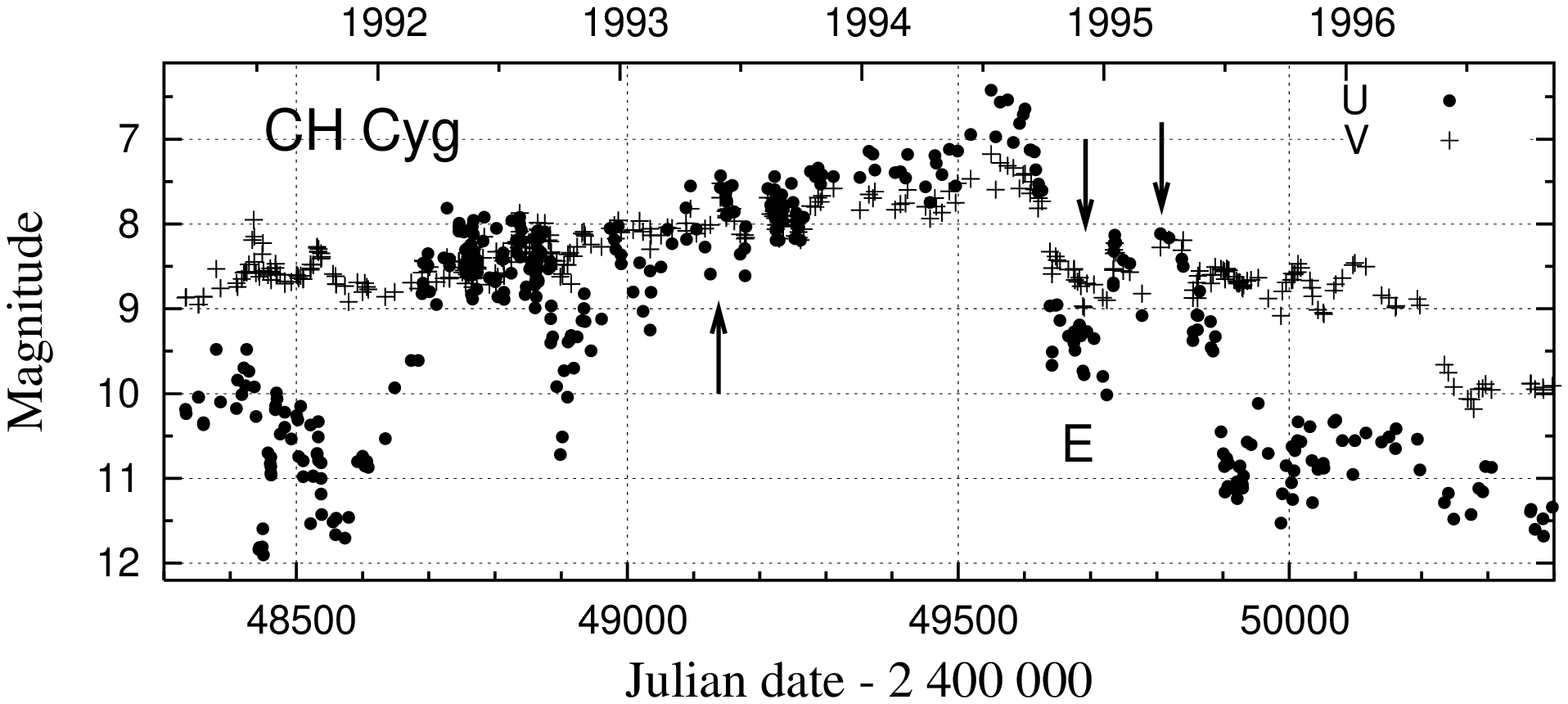}}

\vspace*{2mm}

\resizebox{\hsize}{!}{\includegraphics[angle=-90]{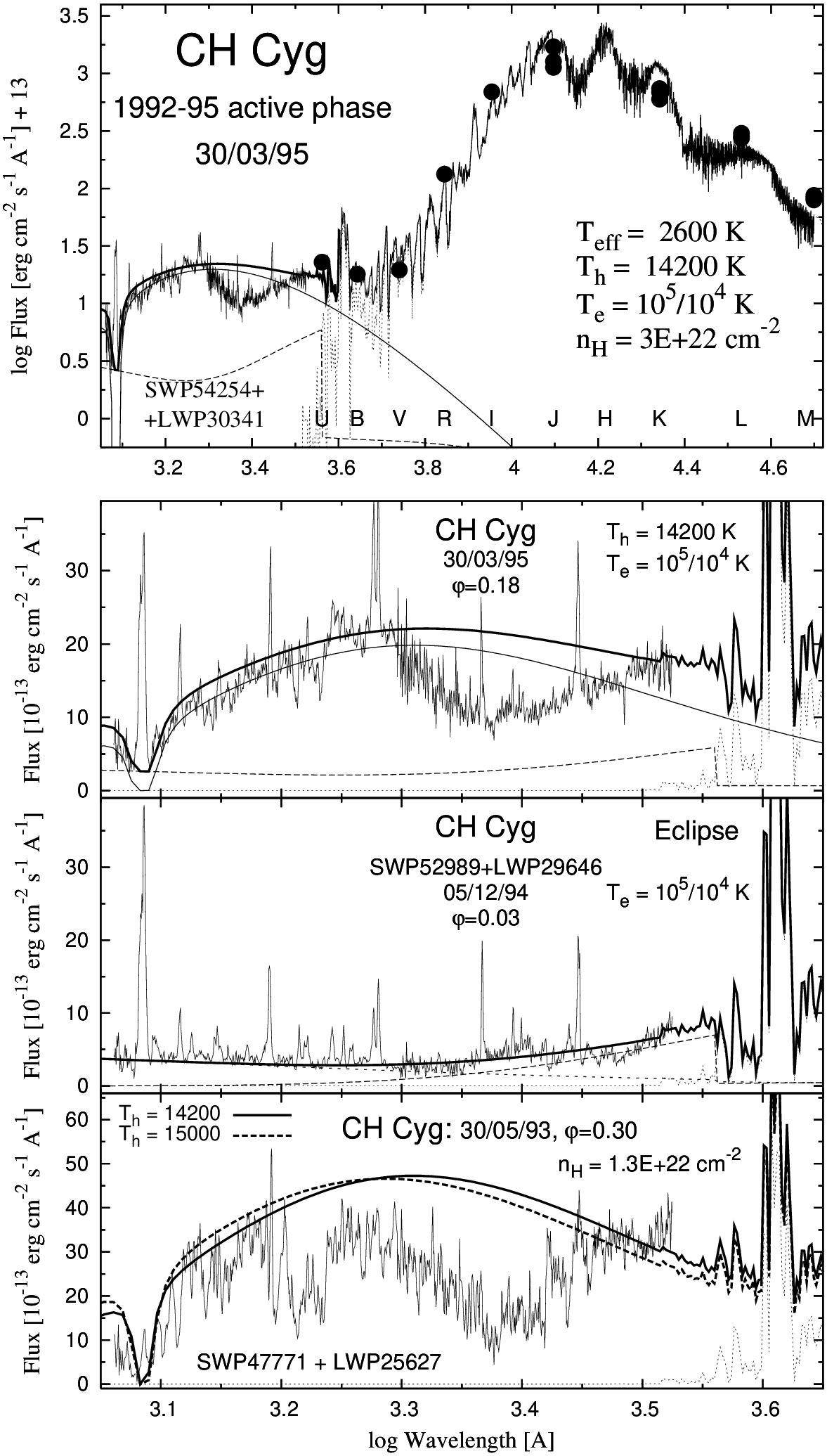}}
\caption[]{
Example of the SED for CH\,Cyg during its 1992-95 active phase. 
Spectra prior to, in and after the 1994 eclipse were selected. 
Corresponding dates are marked at the LC on the top panel. 
          }
\end{center}
\end{figure}

\subsection{CH\,Cygni}

CH\,Cyg is the brightest symbiotic star in the infrared 
passbands ($J \sim 0.9$, $K \sim -0.6$\,mag). It is at 
$270\,\pm\,66$\,pc \citep{viotti+97}. 
We consider CH\,Cyg to be a triple-star system consisting 
of the inner, 756-day period symbiotic binary, which is 
moving in a common 5300-day period outer orbit with another 
red giant \citep{hinkle+93,sk+96}. 
To demonstrate evolution in the SED during active phases 
of CH\,Cyg we selected observations from its 1992-95 activity: 
Before the 1994 eclipse in the inner binary 
(SWP47771 + LWP25627, 30/05/93, $\varphi = 0.30$), 
during the totality 
(SWP52989 + LWP29646, 05/12/94, $\varphi = 0.03$) 
and after the eclipse 
(SWP54254 + LWP30341, 30/03/95, $\varphi = 0.18$). \\
\hspace*{5mm}{\em Radiation from the giant.} 
The slope of the $VRIJ$ fluxes suggests a very low temperature 
of the giant(s) in the system. We matched 
the $UBV$ fluxes from the giant and the observed fluxes in 
the $RIJKLM$ bands by a synthetic spectrum of 
   $T_{\rm eff}$ = 2\,600\,K 
\citep[see Appendix B of][for details]{sk+02a}. 
Corresponding parameters, 
  $F_{\rm g}^{\rm obs} = 2.4\,10^{-6}\,\rm erg\,cm^{-2}\,s^{-1}$ 
and 
  $\theta_{\rm g} = 3.1\,10^{-8}$ 
yield the giant's radius 
  $R_{\rm g} \sim 370\,(d/270\,\pc)\,R_{\sun}$ 
and the luminosity 
 $L_{\rm g} \sim 5\,600\,(d/270\,\pc)^2\,L_{\sun}$. 
These parameters probably belong in major part to the giant 
at the outer orbit in the triple-star model. We note that 
the presence of the two red giants in the system has been 
suggested by the timing of photometric eclipses 
\citep[e.g.][]{sk97} and by the COAST interferometric 
measurements at 0.9\,$\mu$m \citep{y+00}. \\
\hspace*{5mm}{\em Eclipse.}  
The spectrum taken during the totality is almost flat in 
profile with a moderate increase of fluxes from about 
2\,000\,\AA\ to the long-wavelength end of the spectrum 
(Fig.~9). On the radio maps, \cite{croc+01} found two 
components of nebular emission. One is of thermal nature located 
around the central bright peak and another one is of non-thermal 
nature associated with the extended regions. 
\cite{eyr+02} suggested that the extended emission has 
electron temperatures of $\sim$100\,000\,K. 
Therefore we compared a $10^5$\,K hot nebular radiation 
to the SWP spectrum and filled in the rest of the IUE 
spectrum by a $10^4$\,K nebular radiation. The resulting 
fit expresses satisfactorily the overall profile. 
The long-wavelength part of the spectrum 
is affected by the iron curtain absorptions, which is 
consistent with the location of the LTN at the central 
bright peak on the HST image. \\
\hspace*{5mm}{\em The out-of-eclipse SED.}  
During active phases the profile of the out-of-eclipse 
continuum is of the same type 
\citep[here Fig.~9 and Figs.~1, 2 of][]{sk+98}. 
A relatively cool pseudophotosphere 
(here $T_{\rm h} \sim 14\,000$\,K) is drastically affected 
by absorption of the veiling gas and the far-UV continuum 
is attenuated by a few $\times 10^{22}\,\rm cm^{-2}$ hydrogen 
atoms. These effects are present at all orbital phases. 
Modeling the continuum of the 30/05/93 spectrum we neglected 
the nebular component of radiation due to a very low level of 
the continuum at Ly$\alpha$. Finally, we note that 
including the influence of the iron curtain in modeling 
the ultraviolet continuum  changes the interpretation 
suggested by \cite{sk+98}. \\
\hspace*{5mm}{\em Quiescent phases}. 
All components of radiation associated with the hot star are 
negligible during quiescent phases \citep[cf. Fig.~1 of][]{sk+98}. 
Such a behaviour represents a significant difference from 
all other symbiotic stars with relevant observations. 
%
%
\begin{figure}
\centering
\begin{center}
\resizebox{\hsize}{!}{\includegraphics[angle=-90]{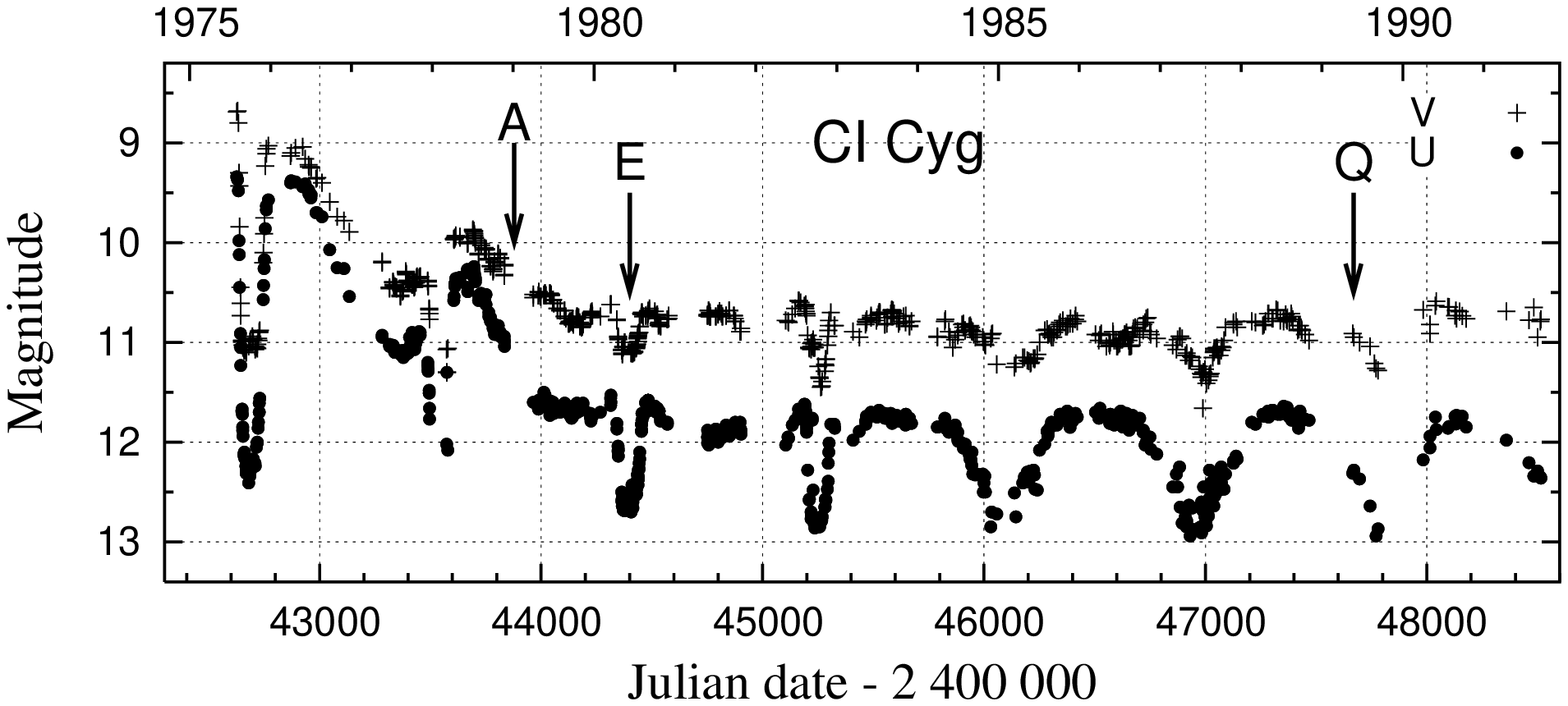}}

\vspace*{2mm}

\resizebox{\hsize}{!}{\includegraphics[angle=-90]{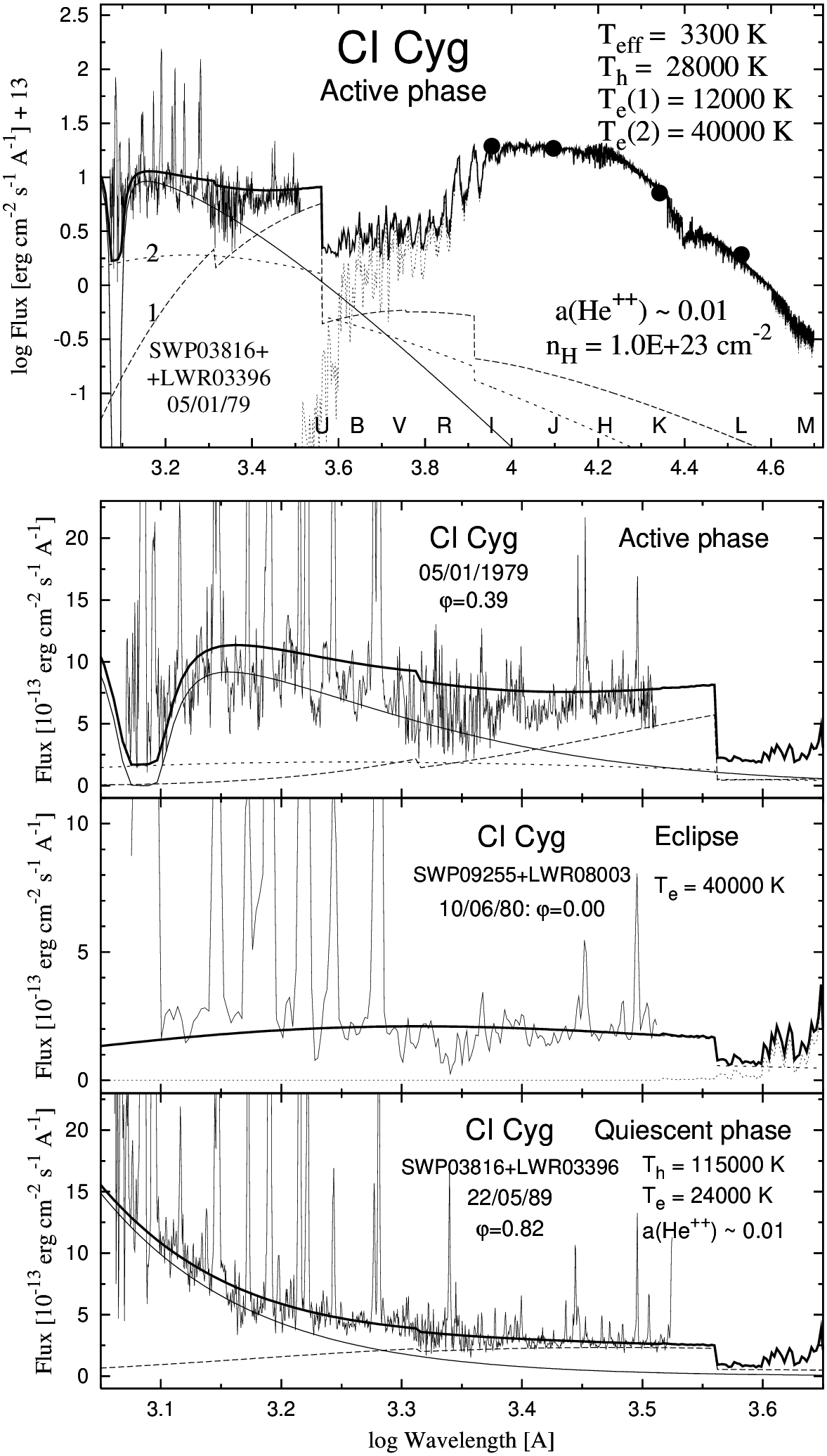}}
\caption[]{
Examples of the SED during the activity ({\sf A}), 
eclipse ({\sf E}) and quiescence ({\sf Q}) for CI\,Cyg. 
Corresponding dates are marked in the LC. 
          }
\end{center}
\end{figure}

\subsection{CI\,Cygni}

The last active phase of CI\,Cyg began in 1975 \citep{bel92}. 
During the first four cycles from the maximum, narrow 
minima -- eclipses -- developed in the LC; a typical feature 
of active phases of symbiotic stars with a high orbital 
inclination. 
From about 1985 the profile of minima became very broad 
indicating a quiescent phase (Fig.~10). 
Accordingly we selected observations from the active phase 
(SWP03816 + LWR03396, 05/01/79, $\varphi = 0.39$),
from the following eclipse 
(SWP09255 + LWR08003, 10/06/80, $\varphi = 0.0$) 
and the quiescence 
(SWP36321 + LWP15571, 22/05/89, $\varphi = 0.82$). \\
\hspace*{5mm}{\em Radiation from the giant}.  
Fluxes corresponding to the $IJKL$ photometry can be matched 
by a synthetic spectrum of 
  $T_{\rm eff} = 3\,300$\,K 
scaled with 
   $k_{\rm g} = 4.0\,10^{-18}$
  ($\theta_{\rm g} = 2.0\,10^{-9}$), 
which yields the integrated flux, 
   $F_{\rm g}^{\rm obs} = 2.8\,10^{-8}$\,\ecs. 
We note that the surface brightness relation for M-giants
gives a similar value of $\theta_{\rm g} = 2.1\,10^{-9}$ 
for the observed reddening-free magnitudes,
$K$ = 4.37 and $J$ = 5.63\,mag. 
Timing of the eclipse profile, which yields 
$R_{\rm g}/A = 0.38 \pm 0.02$, and other fundamental 
parameters derived by \cite{k+91} suggest the giant's radius
  $R_{\rm g} \sim 180\,R_{\sun}$. 
This implies the distance
  $d \sim 2.0$\,kpc 
and the luminosity
 $L_{\rm g} \sim 3\,400\,(d/2.0\,\kpc)^2\,L_{\sun}$. \\
\hspace*{5mm}{\em Radiation from the ultraviolet: Active phase}.  
The profile of the UV continuum during the active phase of CI\,Cyg 
is of the same type as that of other active symbiotics with 
a high orbital inclination -- Rayleigh attenuated 
radiation from a warm HSS, which is affected by the iron 
curtain absorptions (here 
   $T_{\rm h} \sim 28\,000$\,K, 
   $k_{\rm h} \sim 5.7\,10^{-23}$
and 
   $n_{\rm H}\,\sim\,1\,10^{23}\,\rm cm^{-2}$). 
This component represents the main contribution in the UV/U 
spectral region. 
In addition, two components of the nebular radiation, from 
the HTN and LTN, as described for Z\,And, are required to get 
a satisfactory fit. The HTN emission is seen directly during 
the eclipse as the only component and out of the eclipse its 
presence is signaled by the non-zero level of the Rayleigh 
attenuated far-UV continuum (Fig.~10). \\
\hspace*{5mm}{\em Radiation from the ultraviolet: Quiescent phase}.  
In fitting the ultraviolet continuum we fixed the hot star 
temperature to its Zanstra temperature, 
  $T_{\rm h} = 115\,000$\,K 
(MNSV, see Sect.~5.2.2 for details). 
Our solution yields a rather high electron temperature for 
the nebula, $T_{\rm e} = 24\,000\,\pm\,4\,000$\,K, which is 
constrained by a flat profile and a high level of the 
long-wavelength part of the continuum ($\lambda > 2\,000$\,\AA). 
The scaling factor of the nebular component, 
  $k_{\rm N} = 1.0\,10^{15}\,\rm cm^{-5}$, 
gives 
  $EM_{\rm obs} = 4.8\,10^{59}(d/2.0\,\kpc)^2$\,\cmtri. 
%
%
\begin{figure}
\centering
\begin{center}
\resizebox{\hsize}{!}{\includegraphics[angle=-90]{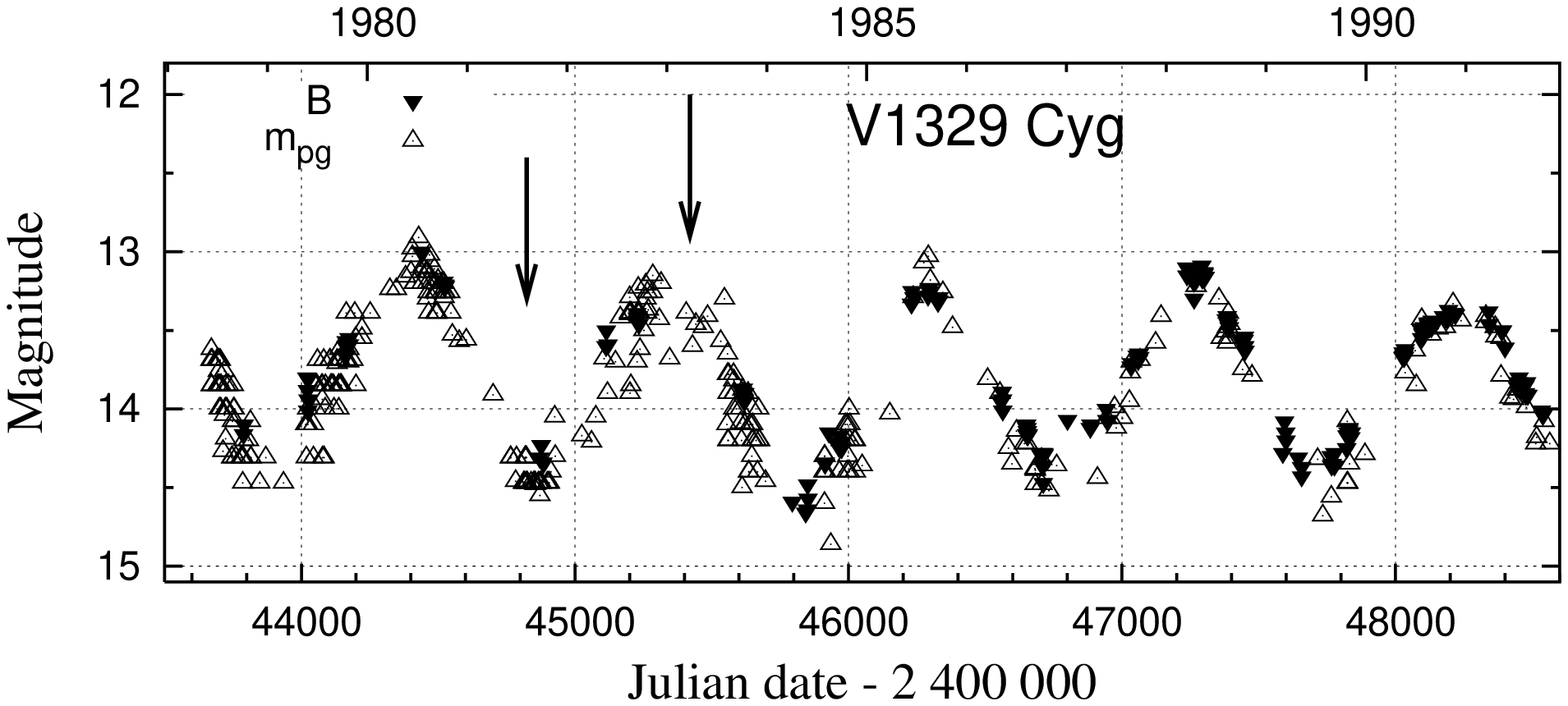}}

\vspace*{2mm}

\resizebox{\hsize}{!}{\includegraphics[angle=-90]{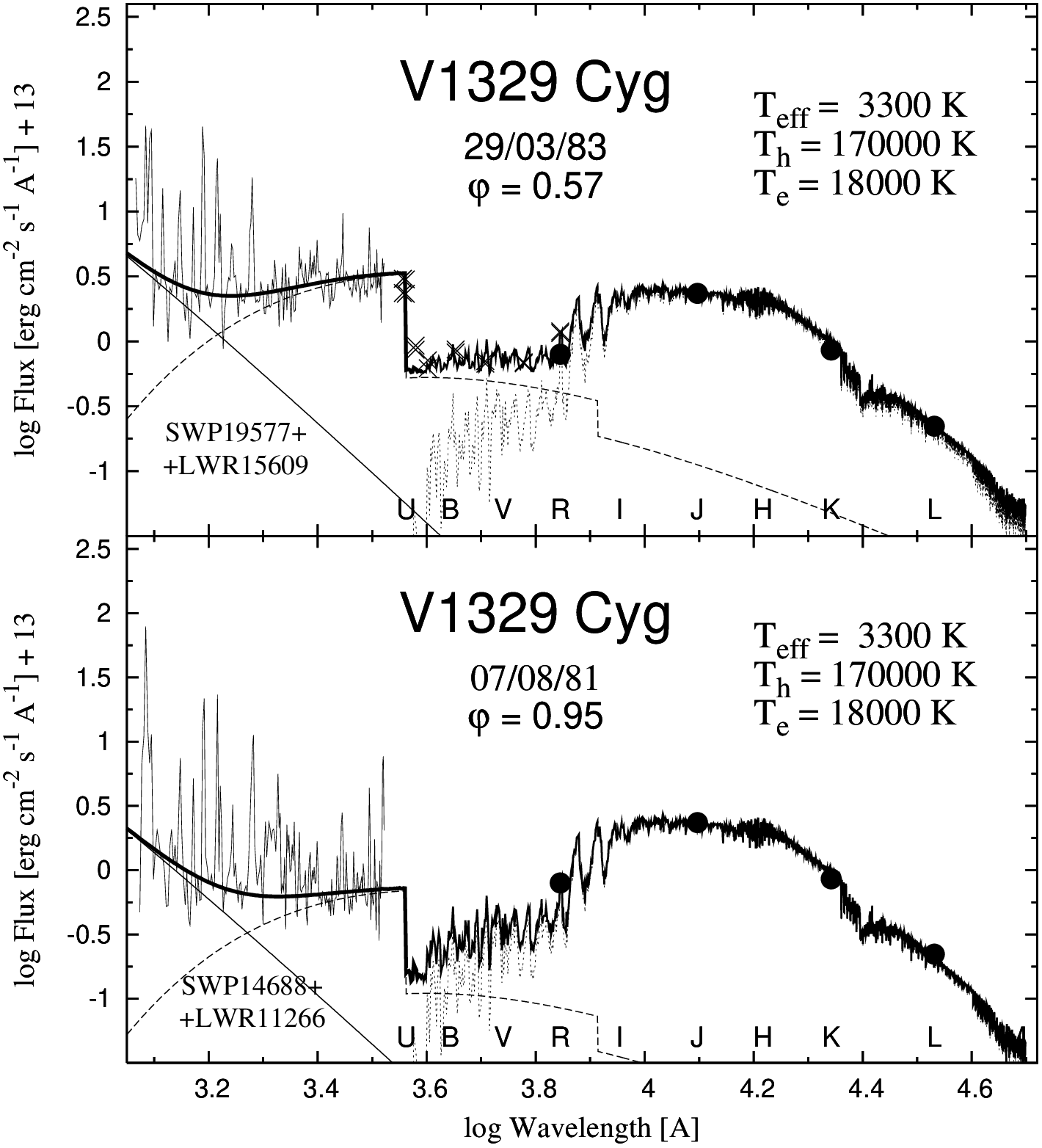}}
\caption[]{
The SED in the continuum of V1329\,Cyg at maximum and 
minimum of the wave-like variation in the LC. 
Crosses ($\times$) at the SED from the maximum represent continuum 
fluxes taken from the optical spectra presented in Figs.~192 and 
193 of \cite{mz02}. 
Dates of the IUE observations are marked in the LC. 
          }
\end{center}
\end{figure}

\subsection{V1329\,Cygni}
V1329\,Cyg is a symbiotic nova which erupted in 1964 
and peaked at m$_{pg}$ = 11.5\,mag in 1966 October. 
Prior to this outbreak, V1329\,Cyg was an inactive star 
of about 15th magnitude displaying $\sim$2\,mag deep eclipses 
\citep[see Fig.~1 of][]{mu+88}. 
After the outburst the symbiotic phenomenon with characteristics 
of a classical symbiotic star developed. 
For the purpose of this paper we selected observations from 
the maximum 
(SWP19577 + LWR15609, 29/03/83, $\varphi = 0.57$)
and minimum 
(SWP14688 + LWR11266, 07/08/81, $\varphi = 0.95$)
of the wave-like variation in the optical LCs (Fig.~11). 
To better visualize their profile we smoothed them by averaging 
the measured points within 10\,\AA\ bins. \\
\hspace*{5mm}{\em Radiation from the giant}.  
Infrared fluxes were derived from average values of $RJKL$ 
measurements published by \cite{t83}, \cite{l+85}, \cite{ty86}, 
\cite{mu+88}, \cite{nv91} and \cite{mu+92}.
They can be matched by a synthetic spectrum of 
  $T_{\rm eff} = 3\,300$\,K 
scaled to 
  $F_{\rm g}^{\rm obs} = 3.4\,10^{-9}\,\rm erg\,cm^{-2}\,s^{-1}$, 
  ($k_{\rm g} = 5.2\,10^{-19}$, 
   $\theta_{\rm g} = 7.2\,10^{-10}$). 
The same value of $\theta_{\rm g}$ can be obtained for the 
M-giant's magnitudes, $K$ = 6.67 and $J$ = 7.88\,mag, by using 
the Dumm \& Schild (1998) relation. 
Combining these parameters with those recently derived by 
\cite{f+01} 
($M_{\rm g} = 2.2\,M_{\sun}$, 
$M_{\rm h} = 0.75\,M_{\sun}$, 
$R_{\rm g} = 132\,\pm\,40\,R_{\sun}$) 
we get the distance 
   $d = 4.2\pm 1.3$\,kpc, 
and the luminosity
   $L_{\rm g} = 1\,900\,(d/4.2\,\kpc)^2\,L_{\sun}$, 
where the uncertainty reflects only that in $R_{\rm g}$. 
On the other hand, the maximum width of the pre-outburst eclipses, 
$t_{4} - t_{1}\, \sim\, 0.13\,P_{\rm orb}$ \citep[Fig.~1 of][]{ss97}, 
yields $R_{\rm g}\,\sim\,228\,R_{\sun}$ assuming the eclipsed 
object to be a point source. However, such a large radius of 
the giant could result from an atmospheric extinction as in 
the case of BF\,Cyg (Sect.~4.8). Therefore we prefer the value 
of 132\,$R_{\sun}$, which was derived under the assumption that 
the giant rotates synchronously with the orbital revolution. \\
\hspace*{5mm}{\em Quiescent phase}.  
Radiation of V1329\,Cyg in the UV/optical domain is characterized 
by a very strong nebular emission 
($T_{\rm e} = 18\,000$\,K, 
 $k_{\rm N} = 1.1\,10^{15}\,{\rm cm^{-5}}$), 
which dominates the spectrum in the range of 
1\,660 -- 7\,000\,\AA\ at the maximum (Fig.~11). 
This constrains the very high temperature of the ionizing source, 
           $T_{\rm h} > 170\,000$\,K = $T_{\rm h}^{\rm min}$ 
scaled to the far-UV continuum with 
           $k_{\rm h} = 2.91\,10^{-25}$, 
to produce the observed emission measure, 
$EM_{\rm obs} = 2.3\,10^{60}(d/4.2\,\kpc)^2$\,\cmtri\ at 
the maximum (i.e. to get the parameter $\delta < 1$, Eq.~23). 
We note that $T_{\rm h}^{\rm min}$ required by our model 
is higher then the Zanstra temperature of 145\,000\,K (MNSV). 
At the minimum the far-UV continuum faded by a factor of 
about 2 as in the case of Z\,And (Sect.~4.3) and the nebular 
emission decreased by a factor of nearly 5 (Table 3). 
The latter is responsible for the large amplitude of 
the periodic variation in the LC. 
%
%
\begin{figure}
\centering
\begin{center}
\resizebox{\hsize}{!}{\includegraphics[angle=-90]{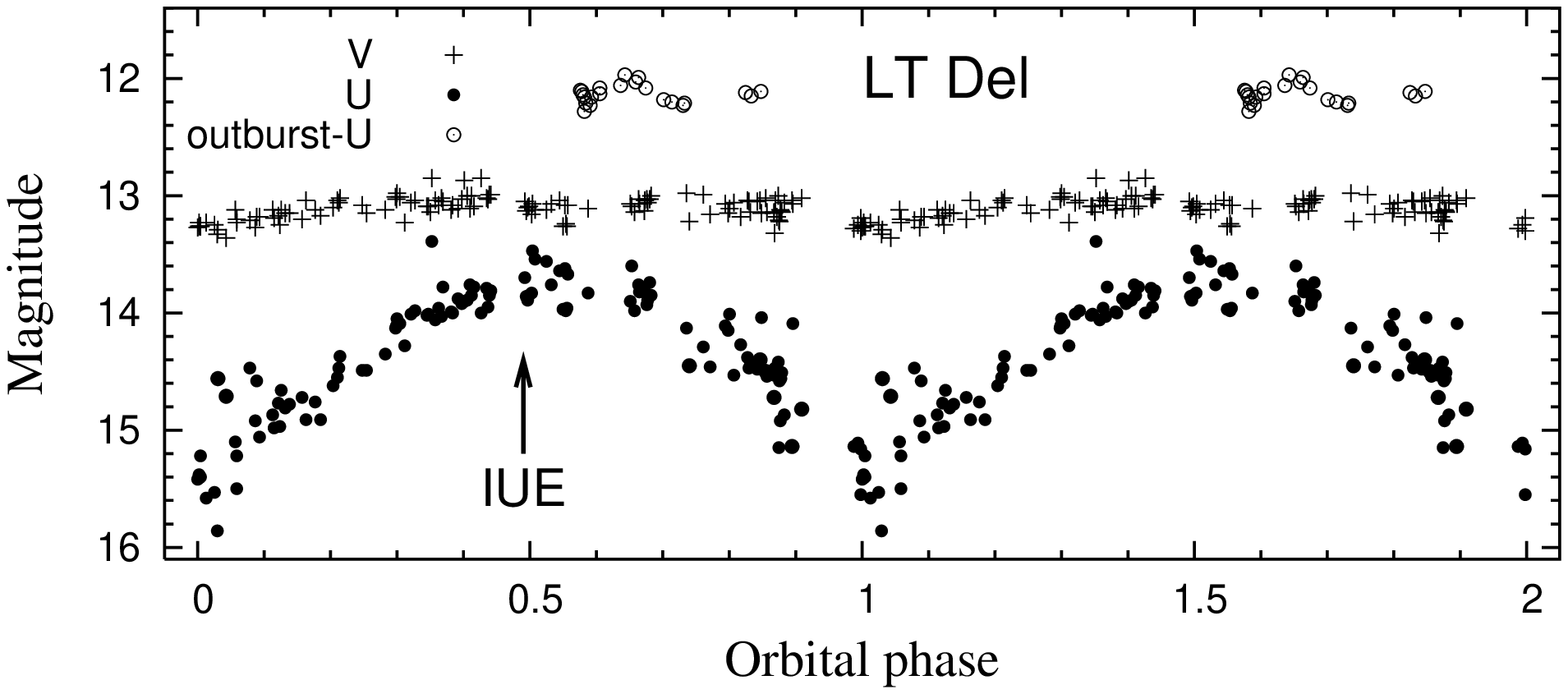}}

\vspace*{2mm}

\resizebox{\hsize}{!}{\includegraphics[angle=-90]{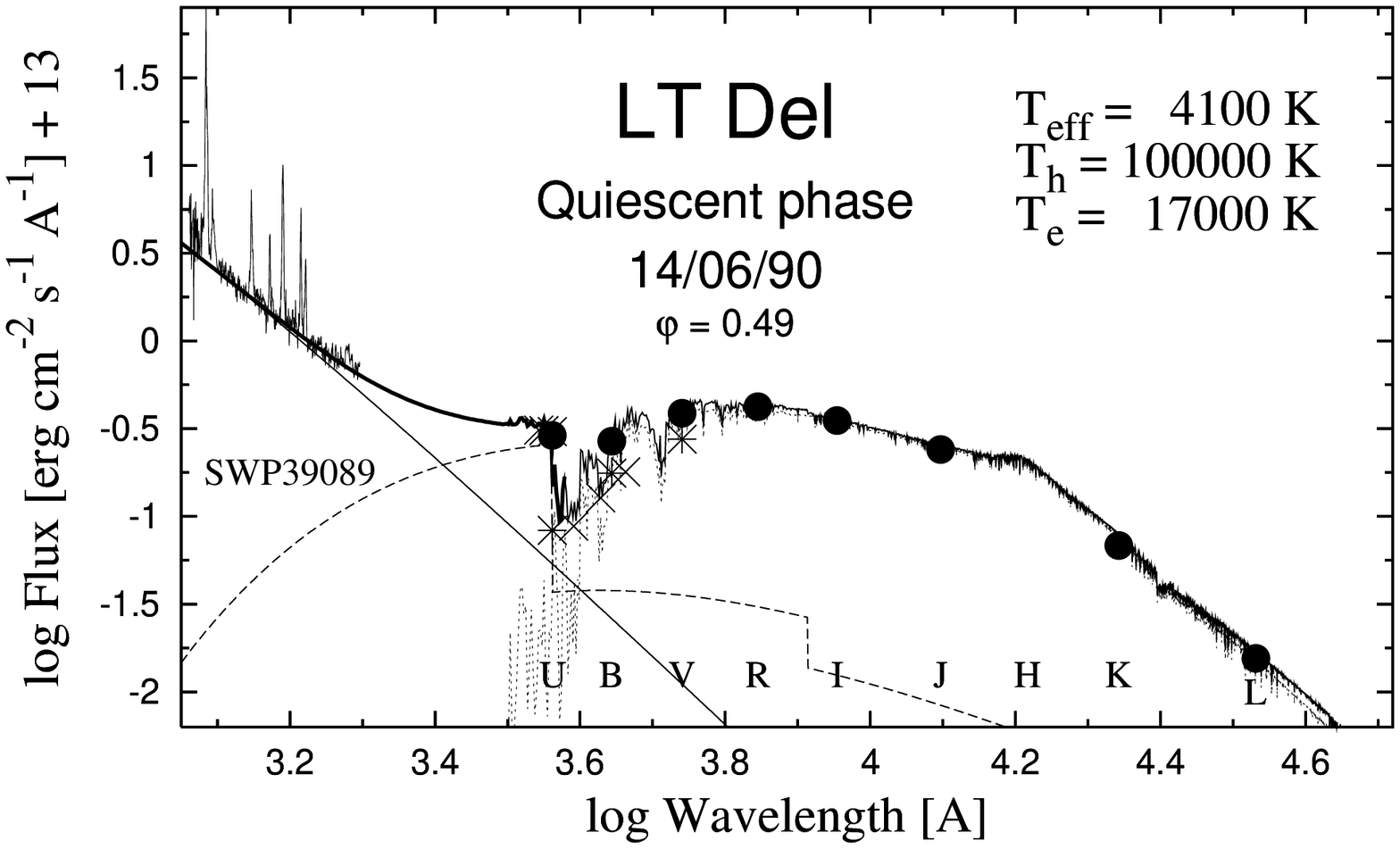}}
\caption[]{
Approximate SED for LT\,Del (LWP spectrum was not available). 
The SWP spectrum was taken at the maximum of the wave-like 
brightness variation in $U$. Full circles in the SED ($\bullet$) 
are photometric fluxes from the maximum, while asterisk ($\ast$) 
from the minimum. Crosses ($\times$) represent continuum fluxes 
from the optical spectrum \citep{m+b92}. 
Small values of the optical/IR fluxes from the giant suggest 
that LT\,Del is a rather distant object. 
          }
\end{center}
\end{figure}

\subsection{LT\,Delphini}
%
%
This yellow symbiotic star shows very pronounced 
orbitally-related brightness variation in the $U$ band. 
This suggests a strong and variable nebular emission in 
the system (Fig.~12). In 1994, \cite{arch+95} detected for 
the first time an optical outburst of this star. 
There is only one spectrum in the IUE archive taken by the 
short-wavelength prime (SWP39089, 14/06/90, $\varphi = 0.49$). 
In spite of this we included this object in our sample to get 
a better representation for S-type yellow symbiotics. 
We determined a new value of $E_{\rm B-V}$ = 0.20 by reddening 
1 -- 2\,10$^{5}$\,K Plank functions to fit the observed continuum 
points of the SWP spectrum. Note that the amount of the interstellar 
extinction at the far-UV is similar to that at 2\,200\,\AA. 
A previous value of $E_{\rm B-V}$ = 0.35 \citep{m+b92} led to 
a strong enhancement of the far-UV fluxes, far beyond of any 
reasonable model. \\
\hspace*{5mm}{\em Radiation from the giant.}  
Fluxes of $VRIJKL$ photometric measurements can be matched 
by a synthetic spectrum of 
  $T_{\rm eff} = 4\,100$\,K 
and scaling of 
  $k_{\rm g} = 3.0\,10^{-20}$, 
which yields the bolometric flux, 
  $F_{\rm g}^{\rm obs} = 4.8\,10^{-10}$\,\ecs. 
Our value of $T_{\rm eff}$ is markedly lower than that 
suggested by the current spectral classification as 
K0 \citep{ms99}. With respect to this disagreement we note 
that spectral classification of yellow symbiotics is based 
on features of G band, \ion{Ca}{i}$\lambda 4227$, 
\ion{Fe}{i}$\lambda 4405$ and absorptions of \ion{Fe}{i}, 
\ion{Sr}{ii} and the CH band. 
\cite{ms99} pointed out that this approach may be
inaccurate for stars with peculiar abundance patterns 
(here AG\,Dra, LT\,Del, BD-21$^{\circ}$3873). 
Our effective temperature from IR fluxes suggests a later 
spectral type, K3 -- K4, of a normal giant \citep{bel+99}. 
Consequently we adopted the corresponding average giant's 
radius of 
  $R_{\rm g} \sim 30\,R_{\sun}$, 
which gives the distance 
  $d = R_{\rm g}/\theta_{\rm g} \sim 3.9\,\kpc$
and the luminosity 
  $L_{\rm g} = 230\,(d/3.9\,\kpc)^2\,L_{\sun}$. 
We note that the giant's radius of $30\,R_{\sun}$ is well inside 
the range of 
  18 to 36\,$(M_{\rm g}/1.5\,M_{\sun})^{1/2}\,R_{\sun}$, 
which can be derived from atmospheric $\log(g) = 1.8 \pm 0.3$ 
\citep{pereira+98}. \\
\hspace*{5mm}{\em Radiation from the ultraviolet.}  
Having only SWP spectrum and a few flux-points from the 
simultaneous optical spectroscopy \citep{m+b92} the resulting 
model is approximate, mainly in $T_{\rm e}$ with an uncertainty 
of $\pm 3\,000$\,K. $T_{\rm h} = 100\,000$\,K was adopted as 
a characteristic temperature. The model is shown in Fig.~12. 

\subsection{AG\,Draconis}
%
%
The LC of AG\,Dra shows numerous brightenings by 2-3 mag in $U$ 
with maxima often separated approximately by 1 year. 
Recently, 
two major eruptions were recorded during 1981-83 and 1994-96 
accompanied by many short-term events lasting from a few weeks 
to a few months. \cite{g-r+99} identified {\em cool} (1981-83 and 
1994-96) and {\em hot} (e.g. 1985-86) outbursts differing in 
their Zanstra temperatures. The quiescent LC is characterized 
by a periodic wave-like variation. 
The top panel of Fig.~13 demonstrates these characteristics. 
The data are from \cite{sk+04} and \cite{l+04}. 
There are no signs either in the optical or far-UV regions 
of eclipses. \cite{ss97ag} derived the orbital inclination 
$i=60\,(\pm 8^{\circ}.2)$. 
To demonstrate the nature of the observed light variation we 
selected two IUE spectra from quiescence taken at different 
spectroscopic conjunctions of the binary components 
(SWP37473 + LWP16675, 27/10/89, $\varphi = 0.62$; 
SWP06650 + LWR05691, 25/09/79, $\varphi = 0.91$) 
and two spectra from active phases; one from the major 1994 
{\em cool} outburst 
(SWP51632 + LWP28752, 28/07/94) 
and one from the {\em hot} 1985 eruption 
(SWP25443 + LWP05513, 13/03/85). \\
\hspace*{5mm}{\em Radiation from the giant.}  
$VRI$ measurements from the light minima during quiescence 
\citep[recently][]{l+04} and the infrared $JHKLM$ photometry 
of AG\,Dra can be matched by synthetic spectra with 
  $T_{\rm eff} = 4\,100 - 4\,300$\,K. 
According to the analysis of \cite{smith+96} we adopted here 
$T_{\rm eff} = 4\,300$\,K. The scaling factor of this spectrum, 
  $k_{\rm g} = 4.9\,10^{-19}$, 
%
%
corresponds to the bolometric flux 
  $F_{\rm g}^{\rm obs} = 9.51\,10^{-9}$\,\ecs. 
From the surface gravity, $\log(g) = 1.6 \pm 0.3$ \citep{smith+96}, 
one can derive the giant's radius as 
%
  $R_{\rm g} = (33 \pm 11)(M_{\rm g}/1.5M_{\sun})^{1/2}\,R_{\sun}$ 
%
and the distance 
%
  $d = R_{\rm g}/\theta_{\rm g} = 
       (1.1 \pm 0.4)(M_{\rm g}/1.5M_{\sun})^{1/2}$\,\rm kpc, 
%
where the giant's mass of 1.5\,$M_{\sun}$ was taken from 
\cite{mika+95}. The giant's luminosity 
  $L_{\rm g} = 360\,(d/1.1\,\kpc)^2\,L_{\sun}$. \\
\hspace*{5mm}{\em Radiation from the ultraviolet: Quiescent phase.}  
The SED of AG\,Dra during quiescent phase is characterized by 
 (i) a dominant giant's radiation in the optical, which results 
in large differences between amplitudes of the orbitally-related 
variation ($\Delta U \gg \Delta B \ga \Delta V$), 
(ii) 
a strong nebula radiating at a high temperature 
($T_{\rm e} \ga 20\,000$\,K) and varying in its $EM$ along 
the orbit, and 
(iii)
a stable hot stellar component of radiation, which is contrary 
to the eclipsing systems. 
Fitting the spectrum from the maximum (27/10/89) we found 
the lower limit of the hot star temperature, 
  $T_{\rm h}^{\rm min} = 110\,000$\,K 
  ($k_{\rm h} = 5.4\,10^{-25}$), 
to satisfy the ionization condition given here through 
the parameter $\delta = 1$ (Sect.~4.1, Eq.~23). 
These parameters determine the lower limit of the hot star's 
ionizing capacity to produce the observed emission measure 
in maximum given by the fitting parameters 
  $k_{\rm N} = 5.5\,10^{14}\,{\rm cm^{-5}}$ 
and 
  $T_{\rm e} = 21\,500$\,K. \\
\hspace*{5mm}{\em Radiation from the ultraviolet: Active phase}. 
The SED during the major 1994 outburst is dominated by the 
nebular radiation at a high electron temperature, 
  $T_{\rm e} = 35\,000\,\pm\,5\,000$\,K 
  ($k_{\rm N} = 2.1\,\pm\,0.4\,10^{16}\,{\rm cm^{-5}}$) 
which determines the continuum profile in the ultraviolet and 
significantly affects the optical with a large contribution 
in the infrared. At 6\,cm radio wavelengths, \cite{ogley+02} 
measured a total emission of $\sim$\,1\,mJy from AG\,Dra 
in 2000 March. They suggested a nebular nature for this emission. 
Also during the smaller eruptions (e.g. 1985, 1986) the nebular 
emission represents a significant component of radiation 
in the UV spectrum (right panels of Fig.~13). Here we present 
the case of the 1985 eruption. We estimated the lower limits 
for the temperature of the ionizing source, 
$T_{\rm h}^{\rm min}$ = 150\,000\,K and 180\,000\,K, for 
the major 1994 'cool' outburst and the 'hot' 1985 eruption, 
respectively. 
%
%
\begin{figure*}
\centering
\begin{center}
\resizebox{\hsize}{!}{\includegraphics[angle=-90]{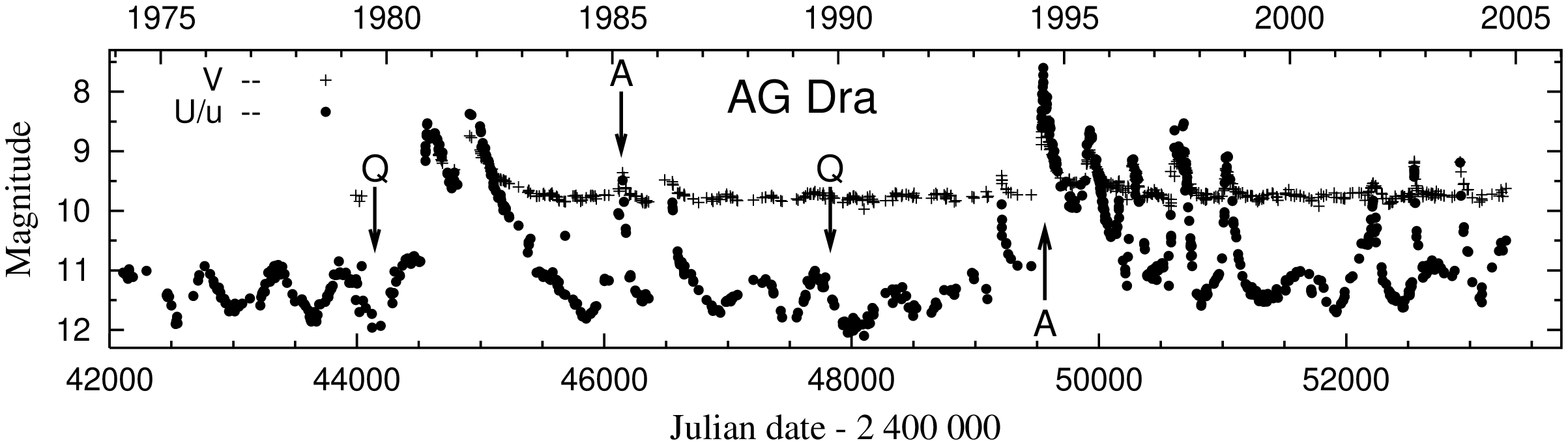}}

\vspace{2mm}

\resizebox{\hsize}{!}{\includegraphics[angle=-90]{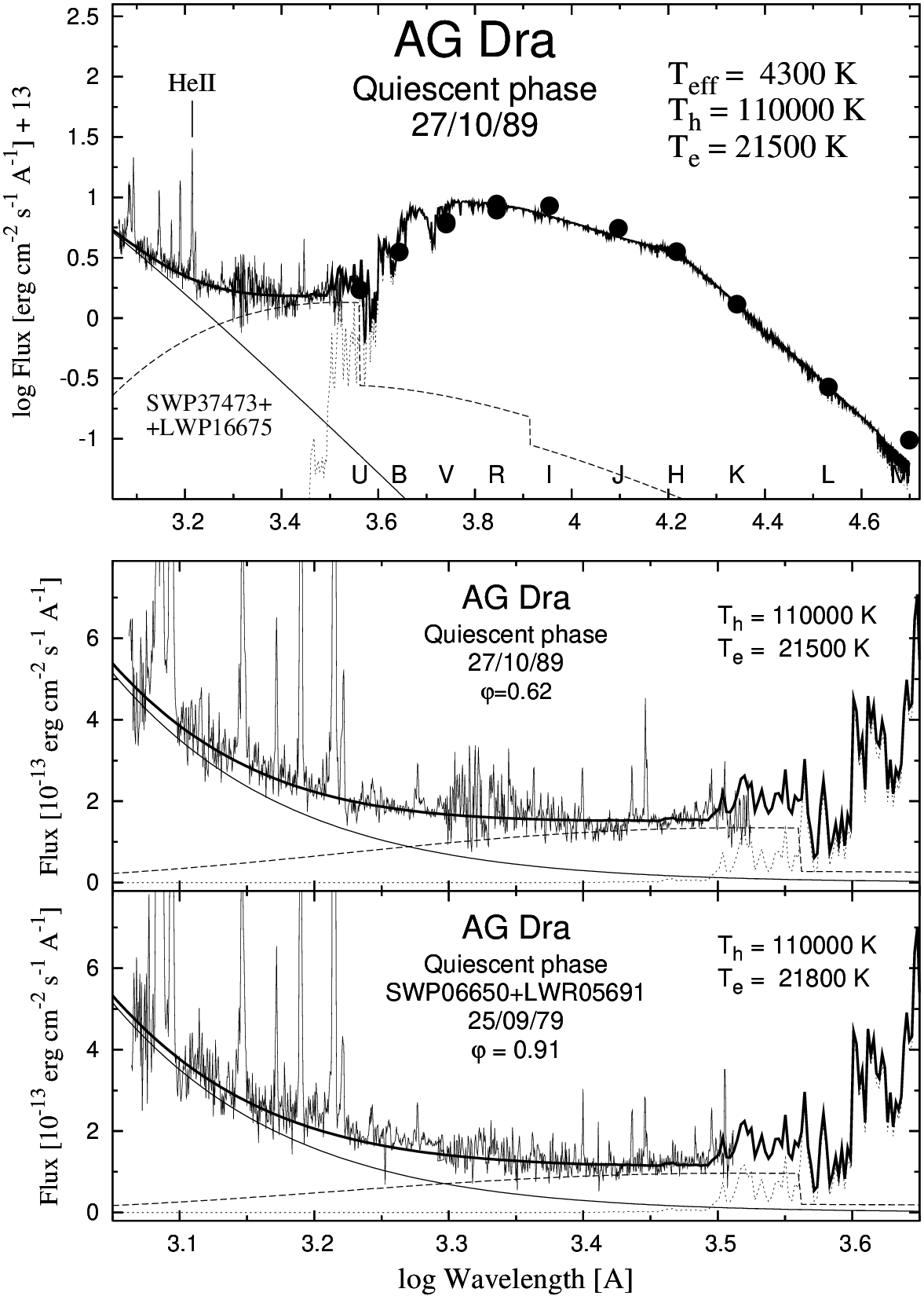}\hspace*{5mm}
                      \includegraphics[angle=-90]{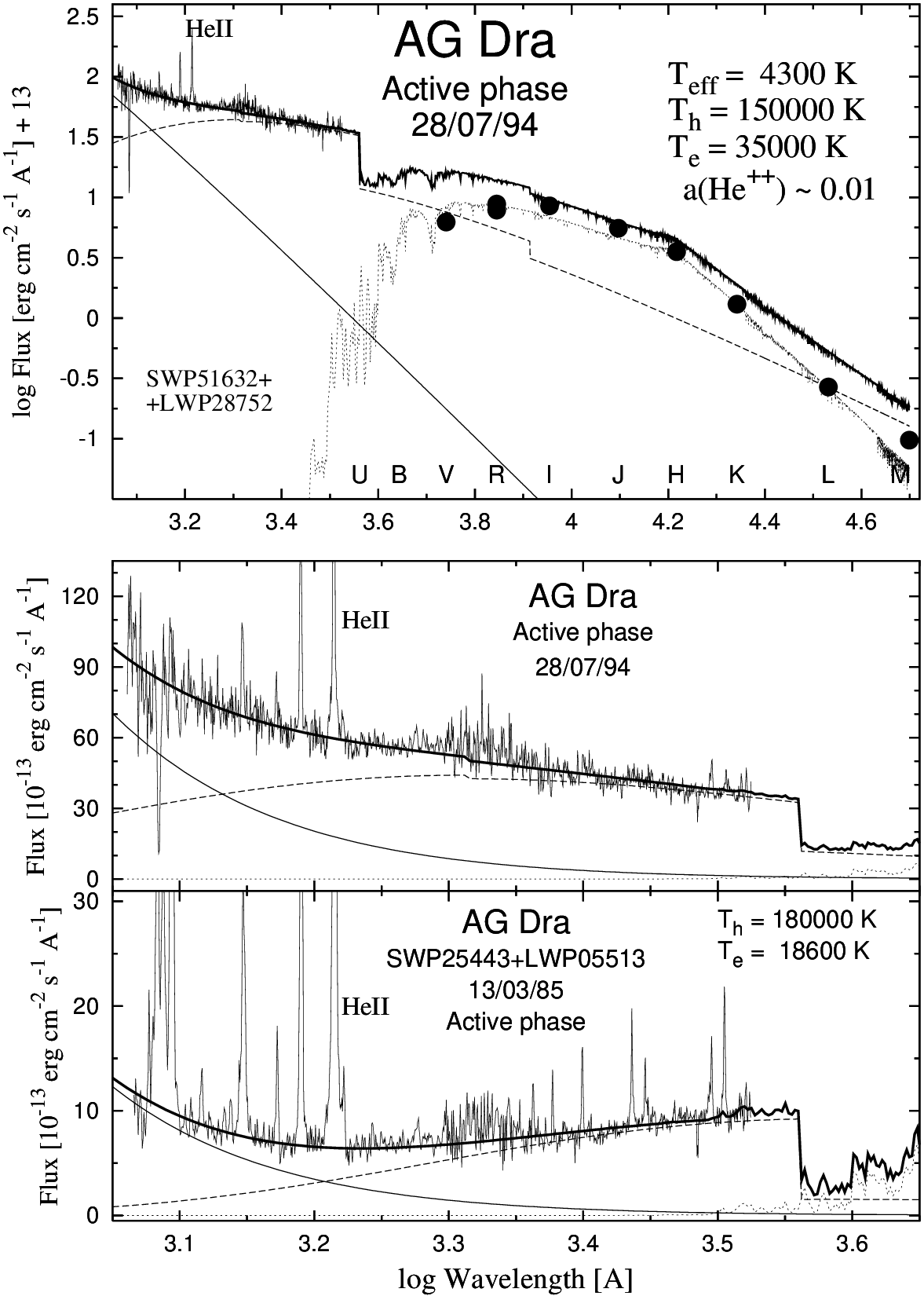}}
\caption[]{
The SED of AG\,Dra during quiescence (left panels) and active 
phases (right). Note the variation of the nebular emission and, 
in contrast, the stability of the hot stellar component with 
the orbital phase during quiescence. The SED during the active 
phases is characterized by a significant increase of the nebular 
emission, which constrains very high temperature of the ionizing 
source. 
Dates of the IUE observations are marked in the U/V LCs (top). 
          }
\end{center}
\end{figure*}
%
%
%
\begin{figure}  
\centering
\begin{center}
%
\resizebox{\hsize}{!}{\includegraphics[angle=-90]{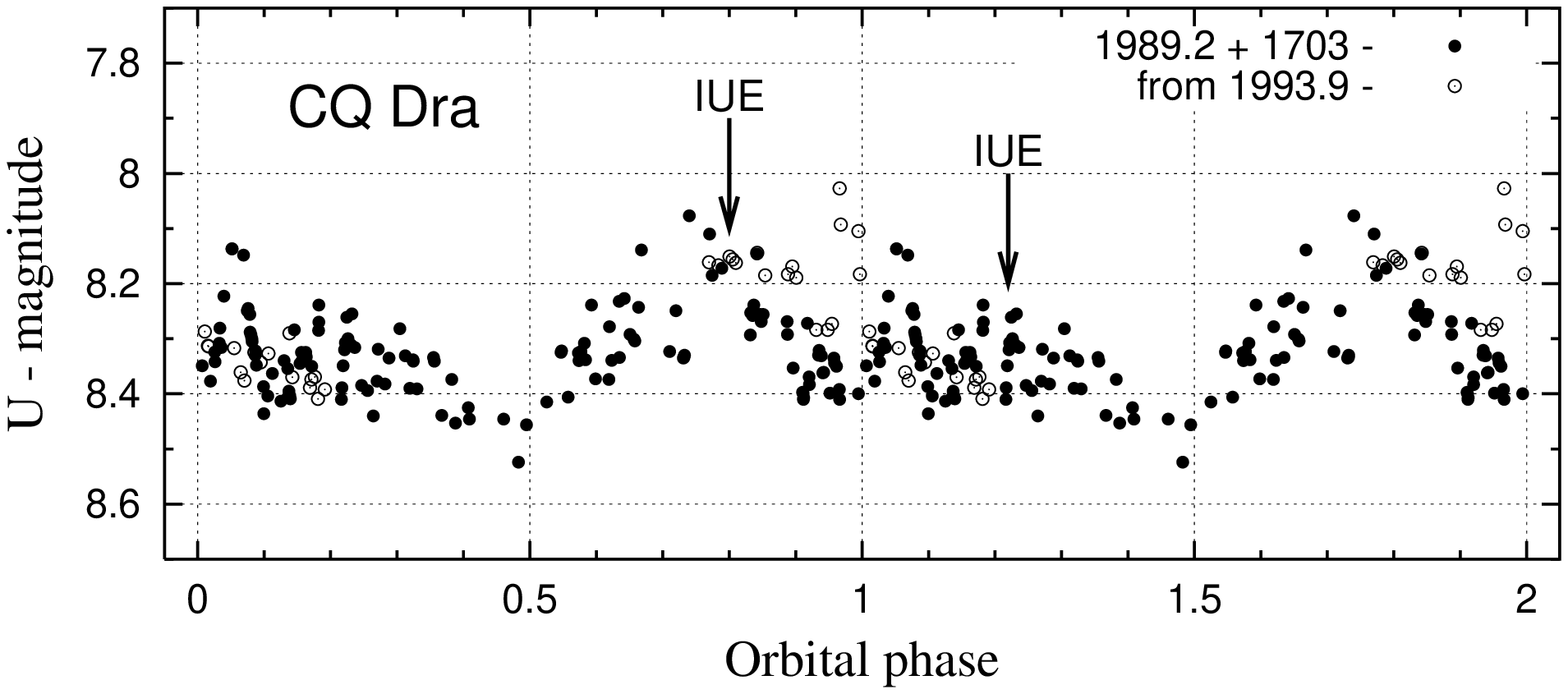}}

\vspace{2mm}

\resizebox{\hsize}{!}{\includegraphics[angle=-90]{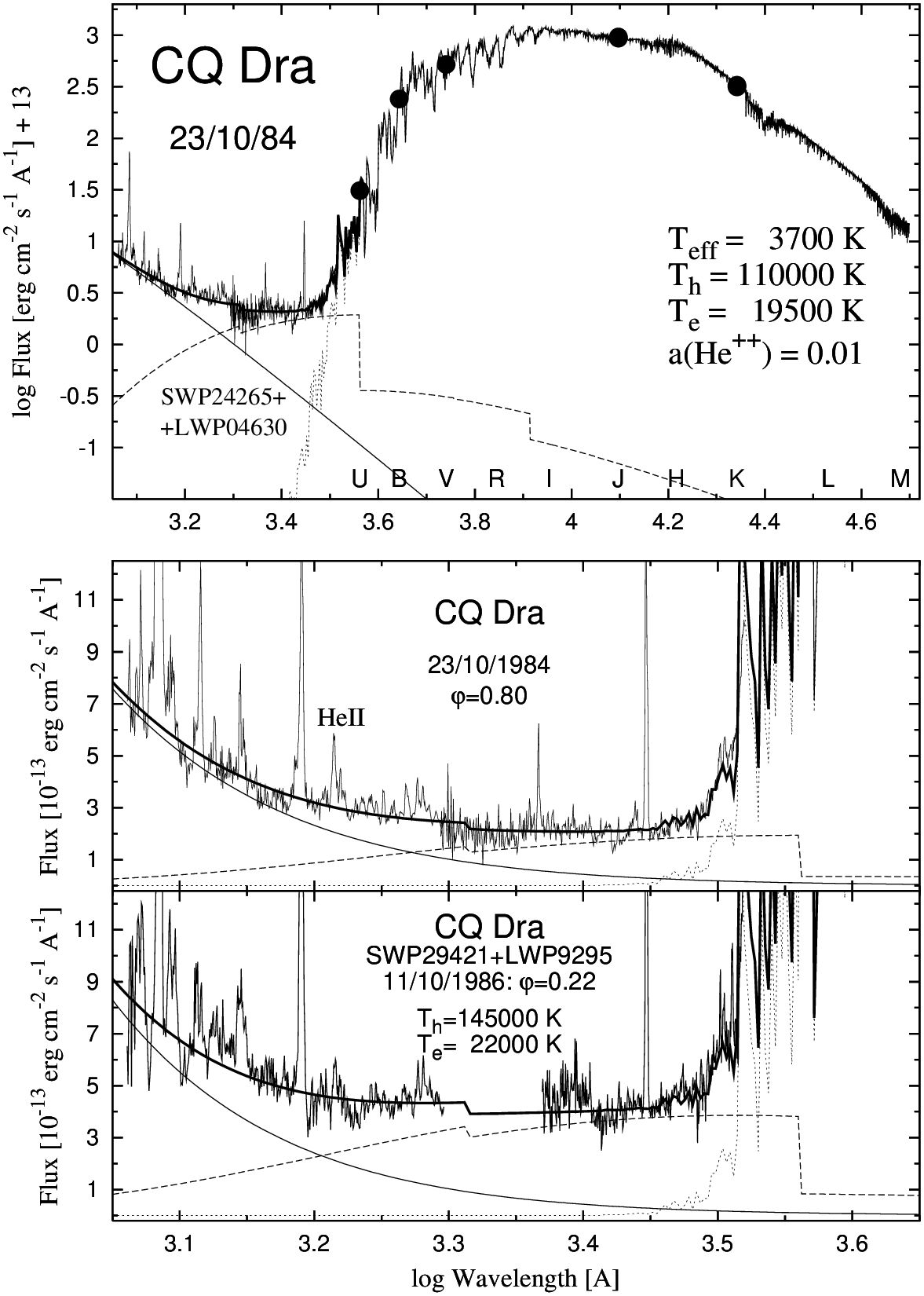}}
\caption[]{
The SED of CQ\,Dra from two shifts of the IUE satellite. 
A strong and variable nebular emission dominates the ultraviolet 
while in the optical/IR we can detect radiation from only 
the red giant. Variation in the $U$ band reflects probably 
that in the nebular radiation. Optical photometry is from 
\cite{hr+94}. 
          }
\end{center}
\end{figure}

\subsection{CQ\,Draconis}
%
%
\cite{rei+88} discovered a UV bright companion of the M3\,\I\I\I\ 
giant 4\,Dra, determined elements of the spectroscopic orbit 
of the giant and suggested that the variation in the UV could 
be produced by a CV of an AM\,Her-type. Recently, \cite{whe+03}, 
based on their X-ray observation, suggested 
that CQ\,Dra is most likely a symbiotic binary containing 
a white dwarf accreting material from the wind of the red giant. 
Here we selected IUE observations from two dates 
(SWP24265 + LWP04630, 23/10/84, $\varphi$\,=\,0.80 
  and 
SWP29421 + LWP09295, 11/10/86, $\varphi$\,=\,0.22)
to verify if the observed properties of the UV continuum 
are consistent with the ionization model of symbiotic binaries. 
%

{\em Radiation from the giant}.  
The red giant in CQ\,Dra is a dominant source of radiation in 
the optical/IR wavelengths (Fig.~14). 
Therefore we used also average values of the $UBV$ measurements 
($U$ = 8.35, $B$ = 6.6 and $V$ = 5.0\,mag) together with 
the $JHK$ photometry of \cite{ka99} to select an appropriate 
synthetic spectrum. In addition, we tried to match an 'emission' 
bump from the giant at $\sim \lambda 3\,200$\,\AA\ observed 
by the IUE. In this manner we selected the synthetic spectrum of 
  $T_{\rm eff} = 3\,700$\,K 
and determined the scaling factor 
  $k_{\rm g} = 1.5\,10^{-16}$, 
which yields the observed bolometric flux 
  $F_{\rm g}^{\rm obs} = 1.6\,10^{-6}$\,\ecs. 
For the HIPPARCOS distance of 178$\pm 15$\,pc \citep{per+97} 
the giant's radius 
 $R_{\rm g} = d \times \theta_{\rm g} = 95\,(d/178\,\pc)\,R_{\sun}$ 
and its luminosity, 
 $L_{\rm g} = 1\,600\,(d/178\,\pc)^2\,L_{\sun}$. 
%

{\em Radiation from the ultraviolet}.  
The profile of the ultraviolet continuum 
with the superposed emission features is typical of a symbiotic 
binary during quiescent phase. The continuum is nearly flat 
between approximately 1\,600 to 3200\,\AA, which suggests 
a dominant contribution from the nebula radiating at a high 
electron temperature. This constrains a high temperature for 
the hot stellar radiation to produce the observed nebular emission. 
In addition, the two selected spectra show a variation in the 
emission measure by a factor larger than 2, while the scaling 
of the hot stellar radiation persists practically unchanged. 
This implies an increase of the hot star temperature from about 
$T_{\rm h}^{\rm min}$ = 110\,000\,K to about 145\,000\,K. 
Parameters of our solution (a low luminosity, 
  $L_{\rm h} = 6.6 - 14\,(d/178\,\pc)^2\,L_{\sun}$ 
and the effective radius, 
  $R_{\rm h}^{\rm eff} = 0.007 - 0.006\,(d/178\,\pc)\,R_{\sun}$) 
suggest that the sole source of the energy detected from 
the ultraviolet is accretion from the giant's wind at the rate of 
  $\dot M_{\rm acc} = {\rm a~few}\times\,10^{-9}$\,\myr\ 
for the mass of a white dwarf, 
  $M_{\rm WD} \equiv 1.0\,M_{\sun}$ \citep{hs61}. 
According to description on wind accretion onto white dwarfs 
\citep{lw84} these accretion rates require the mass-loss rate 
from the giant 
  $\dot M_{\rm W} = \la 10^{-7}$\,\myr\ 
for the separation of the binary components at times 
of observations, 750 and 900\,$R_{\sun}$ 
($M_{\rm T} \equiv 2.5\,M_{\sun}$ and elements from 
\cite{rei+88}). 
Then an increase in $\dot M_{\rm W}$ gives a larger 
$\dot M_{\rm acc}$ and a higher temperature of the ionizing 
source (inner parts of an accretion disk), which causes 
an increase in the emission measure, because a larger 
fraction of the giant's wind can be ionized. 
The first results of studying the accretion process in CQ\,Dra 
are discussed in more detail by \cite{sk05}. 
%
%
\begin{figure}
\centering
\begin{center}
\resizebox{\hsize}{!}{\includegraphics[angle=-90]{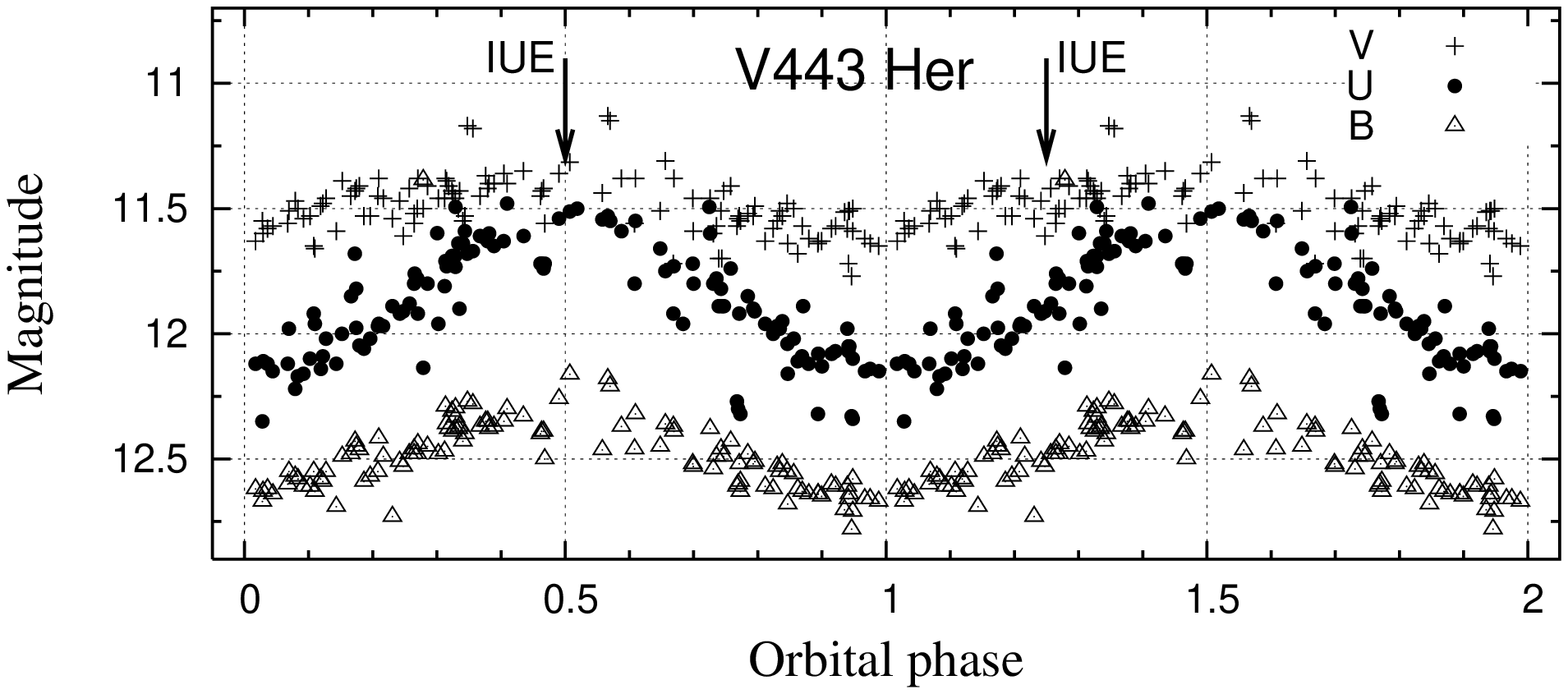}}

\vspace{2mm}

\resizebox{\hsize}{!}{\includegraphics[angle=-90]{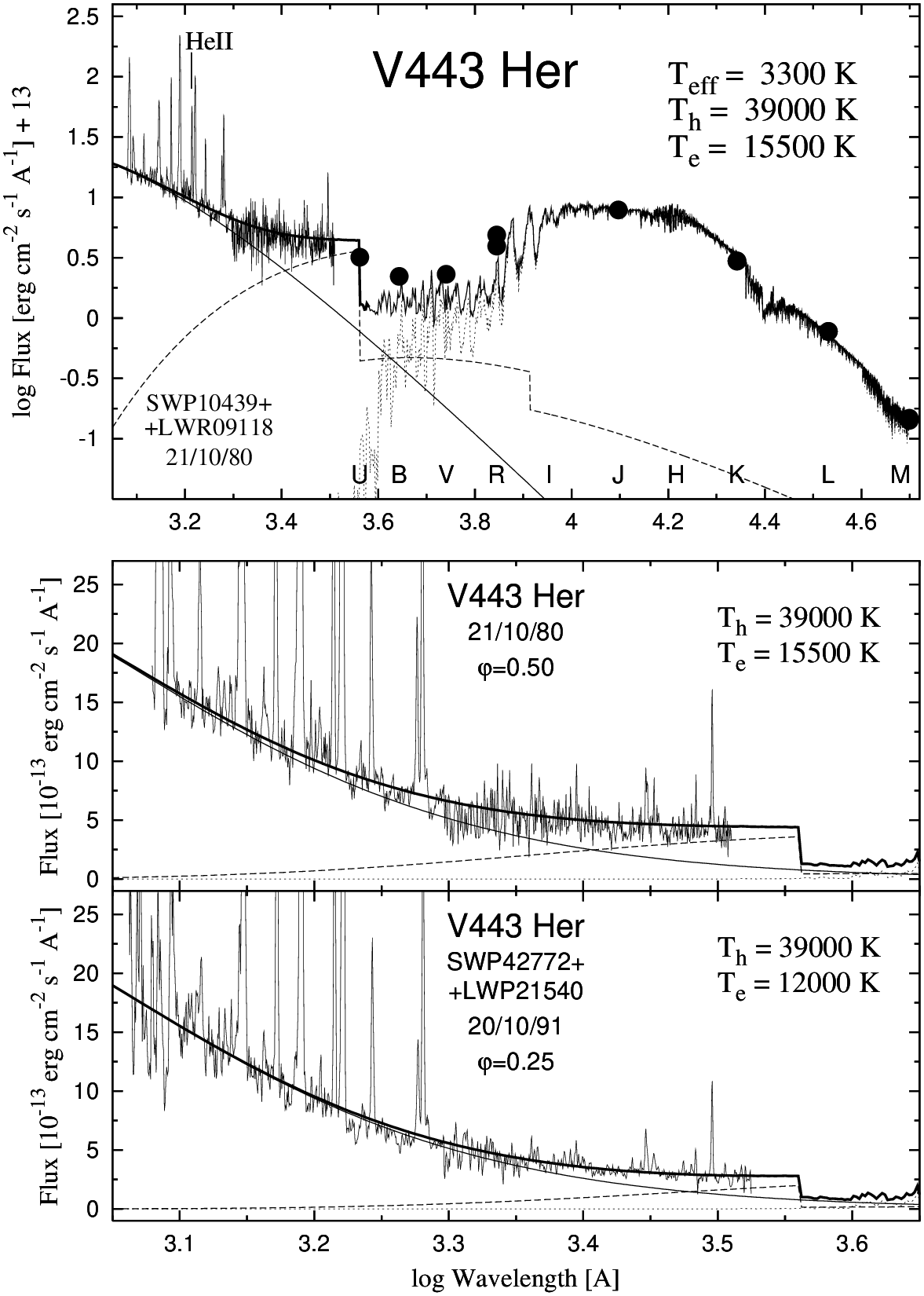}}
\caption[]{
The SED of V443\,Her at two different positions of the binary 
(marked in the LC top). The nebular component of radiation varies 
with the orbital phase, while the hot stellar radiation is stable. 
          }
\end{center}
\end{figure}

\subsection{V443\,Herculis}
%
%
V443~Her is a quiet symbiotic star -- no outburst has been 
recorded to date. The LC displays a marked wave-like variation 
of the optical light along the orbital phase (Fig.~15). 
Recent studies of this symbiotic \citep[][]{dkm93,kmy95,sk96} 
suggested a low orbital inclination ($\approx\,30^{\circ}$, 
and 18$^{\circ}$, respectively). 
We selected two pairs of IUE spectra taken at the maximum of 
the light 
  (SWP10439 + LWR09118, 20/10/80, $\varphi = 0.50$)
and at the quadrature 
  (SWP42772 + LWP21540, 20/10/91, $\varphi = 0.25$). 
No well-exposed spectrum at/around the minimum was available. 
%

{\em Radiation from the giant}.  
The $JKLM$ photometry \citep{ky94} and a flat optical $BVR$ 
continuum suggest a low effective temperature for the giant's 
photosphere (Fig.~15). A synthetic spectrum of 
  $T_{\rm eff} = 3\,300$\,K 
scaled to 
  $F_{\rm g}^{\rm obs} = 1.19\,10^{-8}\,\rm erg\,cm^{-2}\,s^{-1}$
  ($k_{\rm g} = 1.8\,10^{-18}$, $\theta_{\rm g} = 1.33\,10^{-9}$) 
satisfies best the photometric measurements. Also the surface 
brightness relation for M-giants gives a very close value of 
$\theta_{\rm g} = 1.37\,10^{-9}$ for the reddening-free 
magnitudes, $K$ = 5.32 and $J$ = 6.56\,mag.
As there is no reliable estimate of the giant's radius 
in the literature we adopted 
 $R_{\rm g} \sim 130\,R_{\sun}$ 
according to the empirical relations between the radius 
and $T_{\rm eff}$ for M-giants \citep[][]{bel+99}. 
Then the distance 
$d = R_{\rm g}/\theta_{\rm g} 
     \sim 2.2\,(R_{\rm g}/130 R_{\sun})$\,kpc 
and the luminosity
 $L_{\rm g} \sim 1\,800\,(d/2.2\,\kpc)^2\,L_{\sun}$.  
%

{\em Radiation from the ultraviolet.}  
The stellar component of the radiation represents a dominant 
contribution in this spectral region. It has a relatively 
small slope in the far-UV and rivals other components of 
radiation up to the $B$ band (Fig.~15). 
Our model of the 21/10/80 spectrum suggests a low temperature, 
  $T_{\rm h} = 39\,000\,+ 11\,000/\,- 5\,000$\,K, 
where uncertainties were determined in the same way as in 
\cite{sk01b}. 
Both the spectra require the same scaling, 
  $k_{\rm h} = 2.32\,10^{-23}$
  ($\theta_{\rm h} = 4.82\,10^{-12}$), 
which corresponds to the effective radius 
  $R_{\rm h}^{\rm eff} = 0.47(d/2.2\,\kpc)\,R_{\sun}$
and the luminosity 
  $L_{\rm h} = 460(d/2.2\,\kpc)^2\,L_{\sun}$.
On the other hand, the nebular emission is faint and varies 
along the orbit: 
$k_{\rm N} = 
    1.0~{\rm and}~0.4\,10^{15}\,{\rm cm^{-5}}$ 
at $\varphi = 0.5$ ($T_{\rm e} = 15\,500\,\pm\,3\,000$\,K) 
and $\varphi = 0.25$ ($T_{\rm e} = 12\,000\,\pm\,2\,000$\,K), 
respectively. 
However, the corresponding parameter $\delta$ = 3.9 and 2.0 
is not consistent with the fitting parameters 
$T_{\rm h}$, $k_{\rm h}$ and $k_{\rm N}$. The observed hot 
stellar radiation is not capable of producing the nebular 
emission, because of the too low temperature given by the fit 
  ($T_{\rm h} \ll T_{\rm h}^{\rm min} \sim 75\,000$). 
This conflicting situation signals that an unseen part of 
the ionizing source has to radiate at a higher temperature 
to produce a surplus of the nebular flux in addition to that 
generated only by its {\em observed} portion. Qualitatively, 
this could be understandable if the accreted matter has 
a disk-like structure, whose outer rim occults the central 
hot core in the direction to 
the observer. However, this would suggest a rather high 
inclination of the V443\,Her binary orbit. 
Parameters 
  $L_{\rm h}$ and $R_{\rm h}^{\rm eff}$ 
derived from our models represent the lower and upper limits 
of their real values. 
Therefore in Table~3 we also present a formal solution 
for $T_{\rm h} = T_{\rm h}^{\rm min}$ = 75\,000\,K 
(Sect.~4.1, Eq.~23). 
%
%
\begin{figure}[p!th]
\centering
\begin{center}
\resizebox{\hsize}{!}{\includegraphics[angle=-90]{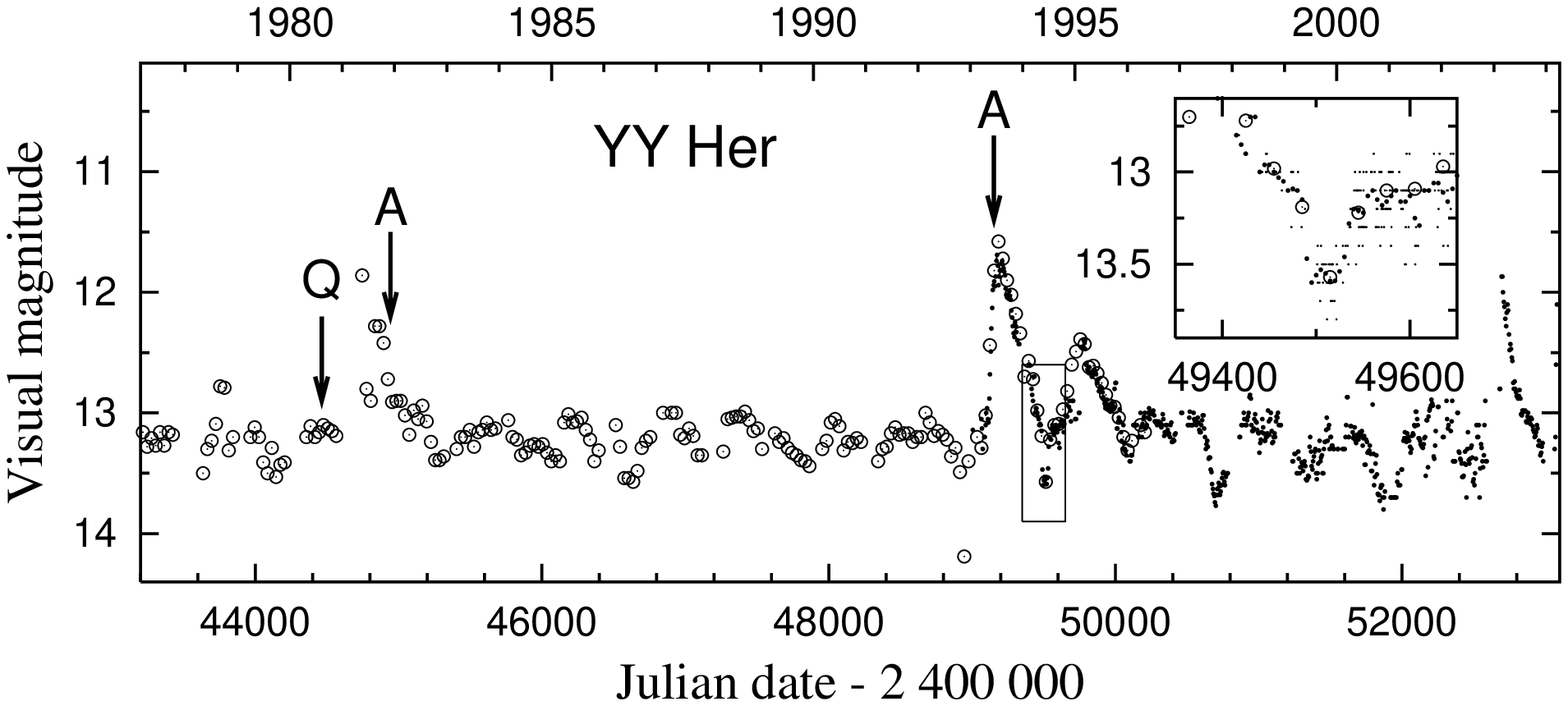}}

\vspace{2mm}

\resizebox{\hsize}{!}{\includegraphics[angle=-90]{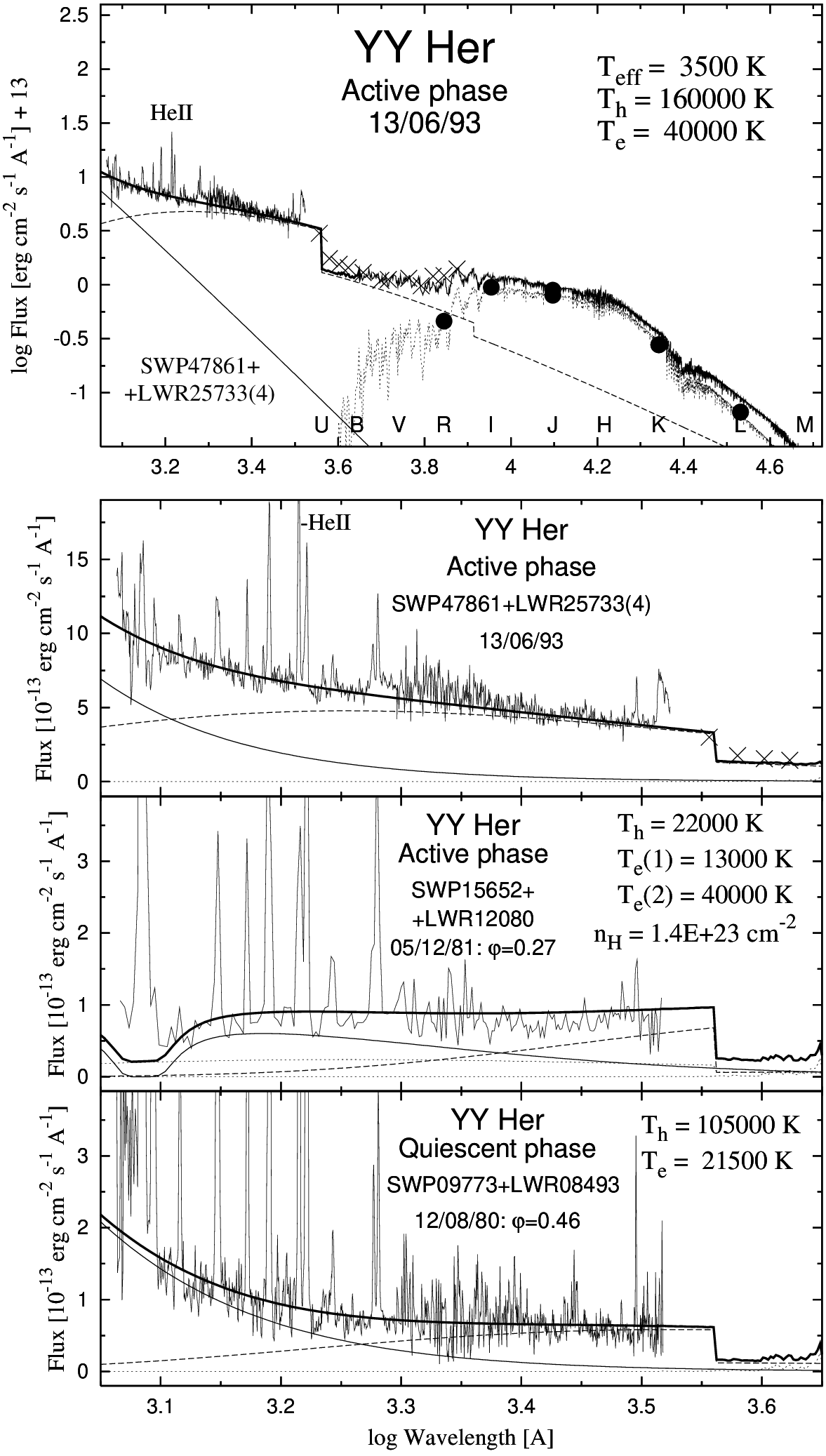}}
\caption[]{
The SED of YY\,Her during quiescence and two different 
phases of activity. Visual LC is composed from the data 
of \cite{mu+97a} ($\circ$) and those available from the CDS 
database ($\bullet$). 
The eclipse effect observed in 1994.43 is shown in detail. 
Crosses ($\times$) in the SED represent continuum flux-points
from the optical spectrum taken on 10/06/93 \citep{mu+97a}.
          }
\end{center}
\end{figure}

\subsection{YY\,Herculis}
%
%
%
\cite{mu+97b} demonstrated that the LC of YY\,Her underwent four 
main outbursts and a few bright stages. Photometric measurements 
made after the main 1993 outburst \citep{hr+01,mika+02} revealed 
a relatively deeper and narrower primary minima than prior to 
the outburst (Fig.~16), which suggests a high inclination of 
the orbital plane. In addition, a secondary minimum in the $VRI$ 
LCs appeared \citep{hr+01}. Such evolution is similar to that of 
AX\,Per during the transition from its 1989 outburst to quiescence 
\citep[Fig.~3 of][Fig.~21 here]{sk+01a}. 
We selected IUE spectra from the maximum of the 1993 outburst 
 (SWP47861 + LWP25733(4), 13/06/93, $\varphi = 0.37$), 
a bright phase in 1981 
 (SWP15652 + LWR12080, 05/12/81, $\varphi = 0.27$) 
and quiescence 
 (SWP09773 + LWR08493, 12/08/80, $\varphi = 0.46$). 
%

{\em Radiation from the giant}. 
Relatively faint $RIJKL$ magnitudes of a normal giant 
\citep[e.g.][]{ms99} suggest that YY\,Her is a distant object. 
The corresponding synthetic spectrum is characterized by 
  $T_{\rm eff} = 3\,500$\,K
and 
  $\theta_{\rm g} = 3.9\,10^{-10}$ 
  (i.e. 
  $F_{\rm g}^{\rm obs} = 1.28\,10^{-9}\,\rm erg\,cm^{-2}\,s^{-1}$). 
The average reddening-free values of $J$\,=8.98 
and $K$\,=7.89 provide nearly identical angular 
radius, $\theta_{\rm g} = 4.0\,10^{-10}$ \citep{ds98}. 
We estimated the radius of the giant from the profile of 
the 1994 minimum in the smoothed visual LC (Fig.~16). 
This can be interpreted in terms of the eclipse of the hot 
star pseudophotosphere by the giant. Its timing 
           ($t_1 \sim JD\,2\,449486$, 
            $t_2 \sim JD\,2\,449492$,
            $t_3 \sim JD\,2\,449522$,
            $t_4 \sim JD\,2\,449537:\,\pm$\,2-3 days)
determines the linear size of the giant's stellar disk that 
eclipses the object, $R_{\rm g}^{\rm E} = 0.21\,A$. This 
parameter can be expressed through the radius of the giant, 
$R_{\rm g}$, and the orbital inclination $i$ as 
\begin{equation}
 R_{\rm g}^{\rm E}/A = \sqrt{(R_{\rm g}/A)^2 - \cos^{2}(i)}.
\end{equation}
With the aid of the empirical relations between the radius
and effective temperature for giant stars \citep[][]{bel+99} 
we adopted $R_{\rm g} \sim 110\,R_{\sun}$, 
which satisfies Eq.~(25) for $i$ = 80$^{\circ}$ 
and $A = 403\,R_{\sun}$ ($M_{\rm T} \equiv 2.5\,M_{\sun}$).
This yields the distance 
  $d = 6.3\,(R_{\rm g}/110 R_{\sun})$\,kpc
and the giant's luminosity 
 $L_{\rm g} = 1\,600\,(d/6.3\,\kpc)^2\,L_{\sun}$.
On the other hand, based on the presence of a secondary minimum 
in the $VRI$ LCs, \cite{mika+02} suggested a Roche-lobe 
filling giant in the system, which corresponds to the distance 
$d = 10\,\pm 3$\,kpc. However, the $I$ LC shows rather narrow 
minima \citep[see Fig.~2 of][]{hr+01} than can be ascribed to 
the wave-like variation due to the tidal distortion of the giant. 
Also a statistical approach of \cite{ms99} suggests the giant 
in YY\,Her to be well inside its Roche lobe. 
%

{\em Radiation from the ultraviolet: Quiescent phase.}  
Radiation of YY\,Her during quiescence comprises a hot 
stellar component of radiation of 
   $T_{\rm h} > 105\,000$\,K = $T_{\rm h}^{\rm min}$, 
the angular radius 
   $\theta_{\rm h} < 4.9\,10^{-13}$ 
and the nebular component of radiation characterized by 
  $T_{\rm e} = 21\,500\,\pm\,3\,000$\,K 
and 
  $k_{\rm N} = 2.4\,10^{14}\,\rm cm^{-5}$. 
Other derived parameters are given in Table~3. 
%

{\em Radiation from the ultraviolet: Active phases.}  
YY\,Her is the only case in our sample of objects that 
shows a very different continuum profile during active phases. 
During the maximum of the star's brightness (1993) the profile 
of the SED is of the same type as observed for AG\,Dra 
(cf. Fig.~13), while during the 1981 bright phase it was similar 
to that of all other active symbiotics with a high orbital 
inclination (e.g. CI\,Cyg). 
In the former case the SED from the ultraviolet to the near-IR 
region is dominated by the nebular radiation at a high electron 
temperature 
            ($T_{\rm e} \sim 40\,000$\,K,
             $k_{\rm N} = 2.7\,10^{15}\,\rm cm^{-5}$), 
which is powered by a strong stellar source radiating at 
             $T_{\rm h} > 160\,000$\,K = $T_{\rm h}^{\rm min}$ 
and scaled with 
             $\theta_{\rm h} < 6.3\,10^{-13}$ (i.e. $\delta < 1$). 
The profile of the simultaneously observed optical spectrum 
\citep{mu+97a} agrees well with the modeled SED (Fig.~16). 
In the latter case the UV continuum results from superposition of 
radiation from a cool stellar pseudophotosphere 
            ($T_{\rm h} = 22\,000$\,K, 
             $\theta_{\rm h} = 3.2\,10^{-12}$) 
attenuated with 
             $n_{\rm H} \sim 1.4\,10^{23}\rm cm^{-2}$, 
a low-temperature nebula 
            ($T_{\rm e} = 13\,000$\,K,
             $k_{\rm N} = 1.5\,10^{14}\,\rm cm^{-5}$)
and a high-temperature nebula 
            ($T_{\rm e} \equiv 40\,000$\,K,
             $k_{\rm N} = 1.3\,10^{14}\,\rm cm^{-5}$).
%
%
%
\begin{figure}
\centering
\begin{center}
\resizebox{\hsize}{!}{\includegraphics[angle=-90]{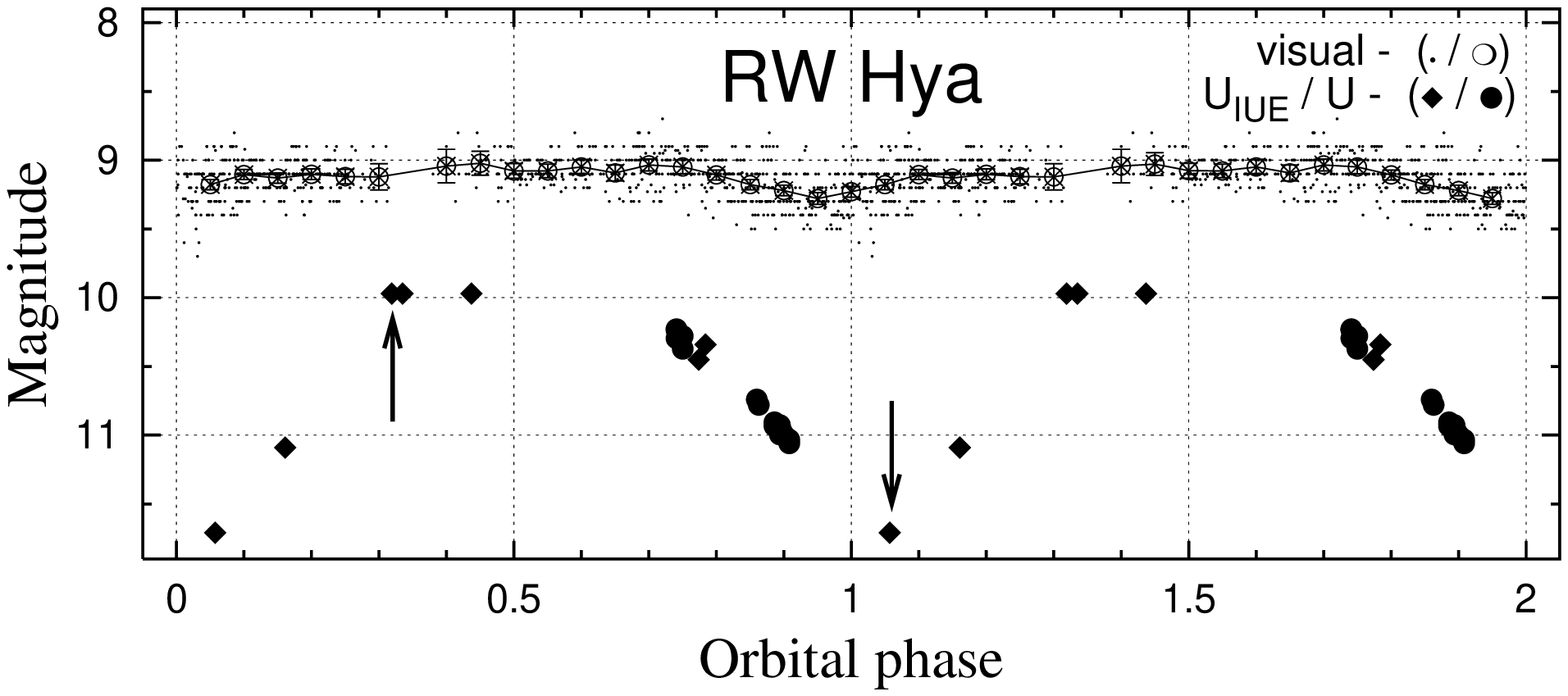}}

\vspace{2mm}

\resizebox{\hsize}{!}{\includegraphics[angle=-90]{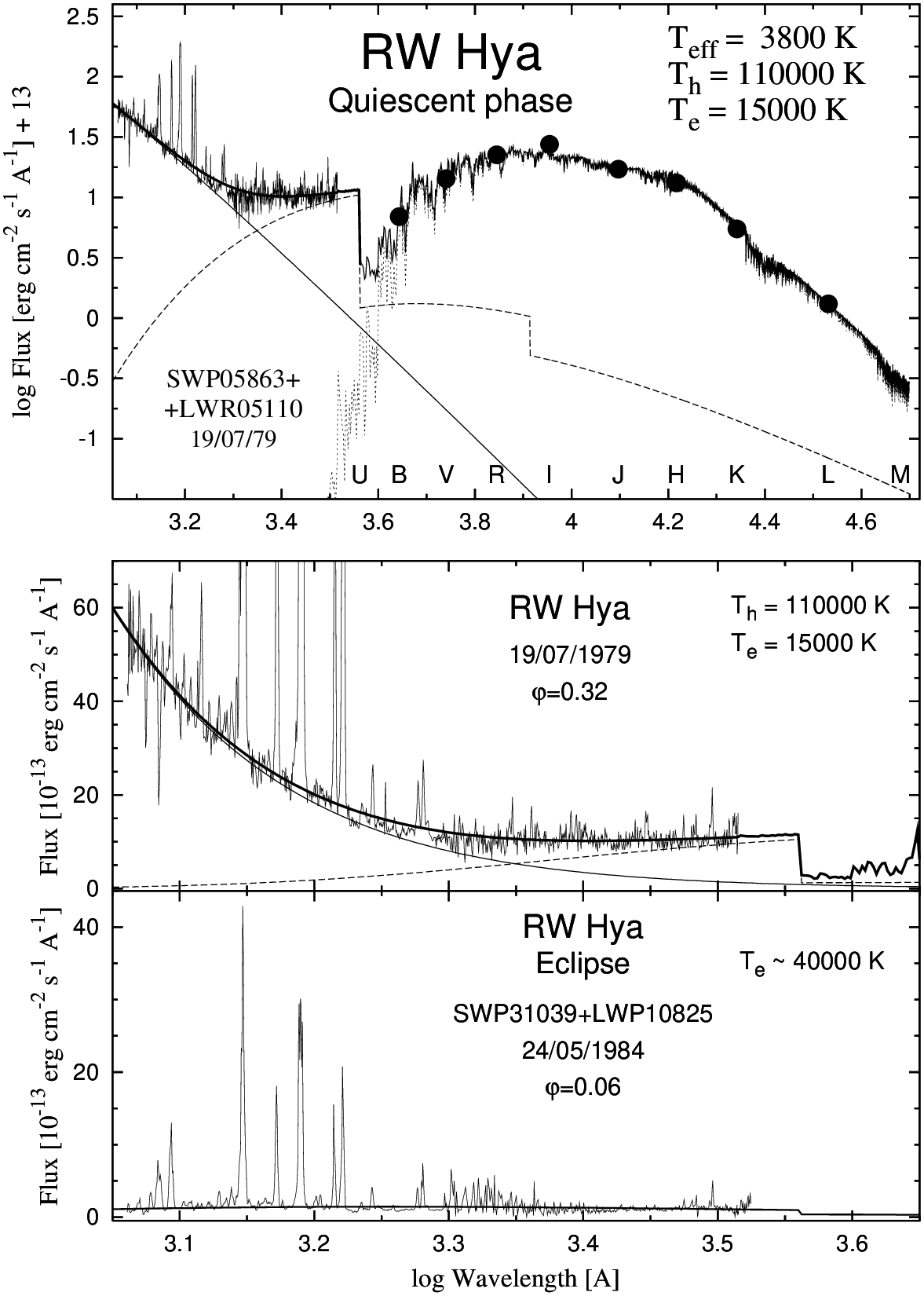}}
\caption[]{
Top: Visual and $U$ LCs of RW\,Hya \citep{sk+02b} with added 
$U$-magnitudes estimated from available IUE observations. 
The arrow marks position of the used spectrum. 
Middle and bottom: 
The SED near to the optical maximum shows typical characteristics 
of a quiescent phase with a high luminosity of the hot object, while 
during the eclipse only a very faint high-temperature nebular emission 
can be recognized. 
          }
\end{center}
\end{figure}

\subsection{RW\,Hydrae}
%
%
RW\,Hya is a stable symbiotic system for which no optical 
eruption has been observed. Its LC displays a wave-like 
variation as a function of the orbital phase (Fig.~17). 
A high inclination of the orbital plane is indicated by 
the eclipse effect due to attenuation of the far-UV continuum 
by Rayleigh scattering \citep[e.g.][]{sch+96}. IUE spectra 
of RW\,Hya have been studied in detail by many authors 
\citep{km95,sch+96,d+99,sion+02}. We selected one 
observation around the maximum 
(SWP05863 + LWR05110, 19/07/79, $\varphi = 0.32$) 
  and one around the minimum of the optical light 
(SWP31039 + LWP10825, 24/05/87, $\varphi = 0.06$)
to demonstrate our approach for this case. \\
\hspace*{5mm}{\em Radiation from the giant}. 
A synthetic spectrum of 
  $T_{\rm eff} = 3\,800$\,K
with scaling of 
  $k_{\rm g} = 2.6\,10^{-18}$ 
  ($\theta_{\rm g} = 1.61\,10^{-9}$, 
  $F_{\rm g}^{\rm obs} = 3.10\,10^{-8}$\,\ecs) 
matches well the flux-points corresponding to the $BVRIJHKL$ 
photometry. Reddening-free magnitudes, $K$ = 4.65 and 
$J$ = 5.72\,mag, also give a similar value of 
$\theta_{\rm g} = 1.77\,10^{-9}$ \citep{ds98}. The giant 
dominates the spectrum from the $B$ band to the infrared. 
This results in a large amplitude difference of the wave-like 
variation at different passbands ($\Delta U \gg \Delta V$, 
Fig.~17). Thus, the nebular emission, which is responsible 
for this type of variability, is not able to rival the giant's 
contribution from $B$ to longer wavelengths. 
Assuming the synchronous rotation of the giant with the orbital 
revolution, \cite{sch+96} determined its radius to 
  $R_{\rm g} = 58.5\,\pm\,8\,R_{\sun}$. 
This quantity then yields the distance
  $d = R_{\rm g}/\theta_{\rm g}
     = 820\,\pm\,112\,(R_{\rm g}/58.5 R_{\sun})$\,pc
and the luminosity, 
  $L_{\rm g} = 650\,\pm\,180\,(d/0.82\,\kpc)^2\,L_{\sun}$, 
where uncertainty in only $R_{\rm g}$ is included. 
These parameters are similar to those determined 
by \cite{sch+96}. \\
\hspace*{5mm}{\em Radiation from the ultraviolet.}  
A steep far-UV continuum (practically within the SWP 
spectrum) reflects a dominant contribution from the hot 
stellar source in this region. On the other hand, a flat 
near-UV continuum of the LWR spectrum signals a significant 
nebular radiation here -- an illustrative example of 
a quiescent symbiotic's UV spectrum. 
Our best fit of the 19/07/79 spectrum corresponds to 
   $T_{\rm h} = 110\,000\,\pm\,30\,000$\,K,
   $k_{\rm h} = 6.3\,10^{-24}$ 
and 
  $T_{\rm e} = 15\,000\,\pm\,2\,000$\,K with scaling, 
  $k_{\rm N} = 2.8\,10^{15}\,\rm cm^{-5}$. 
This solution gives the parameter $\delta$ = 0.54, which 
means that a fraction of ionizing photons is not converted 
to the nebular radiation. In the sense of the STB model 
the ionized zone is open. The situation in which all 
the $L_{\rm ph}$ photons are involved in the 
ionization process corresponds to 
  $T_{\rm h}^{\rm min} = 80\,000$\,K, 
which represents the lower limit of the hot source 
temperature in RW\,Hya. It is also close to the Zanstra 
temperature of 75\,000\,K derived by MNSV. 
The spectra of RW\,Hya taken close to the giant's inferior 
conjunction suffer an additional (to the Rayleigh scattering) 
wavelength-independent attenuation \citep{d+99} as observed 
for other symbiotics with a high orbital inclination. 
During the totality only a high-temperature nebula can be 
recognized. 
%
%
\begin{figure}
\centering
\begin{center}
\resizebox{\hsize}{!}{\includegraphics[angle=-90]{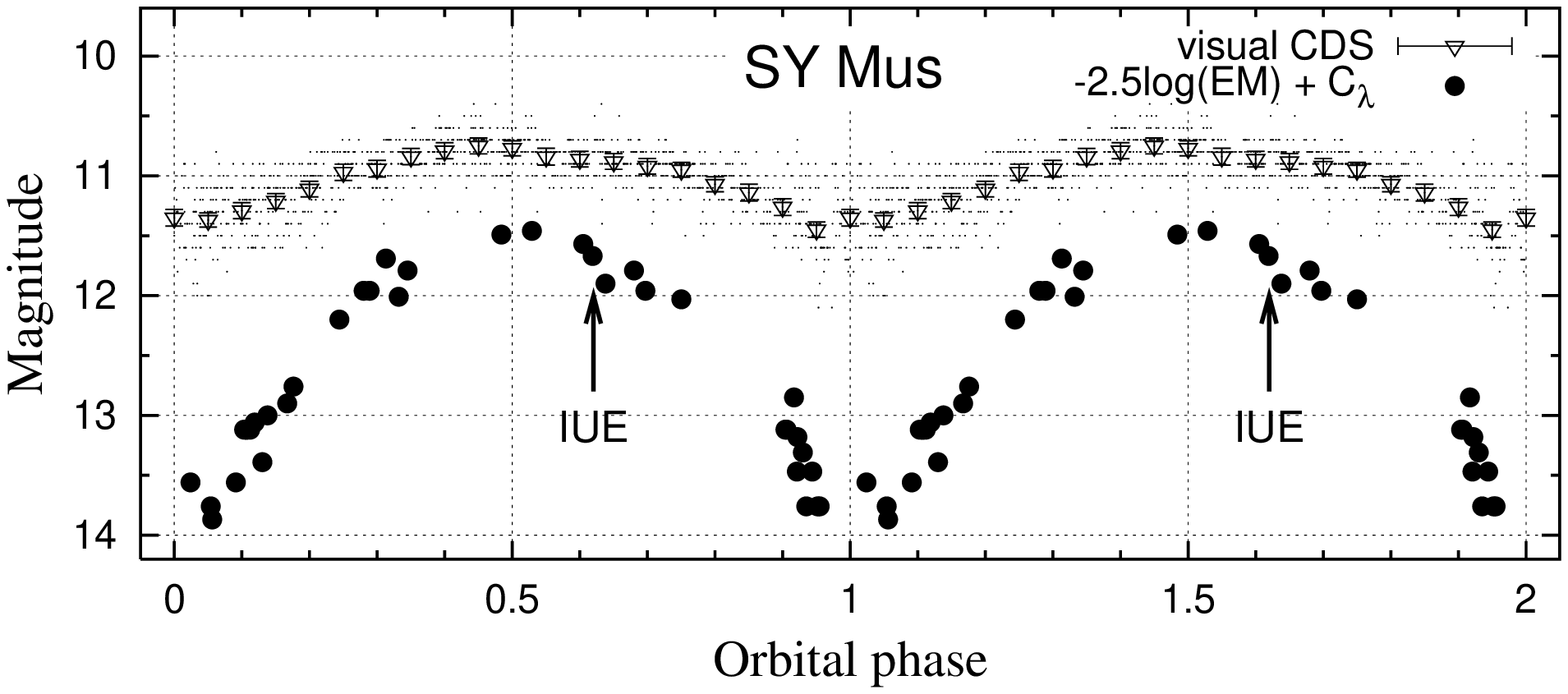}}

\vspace{2mm}

\resizebox{\hsize}{!}{\includegraphics[angle=-90]{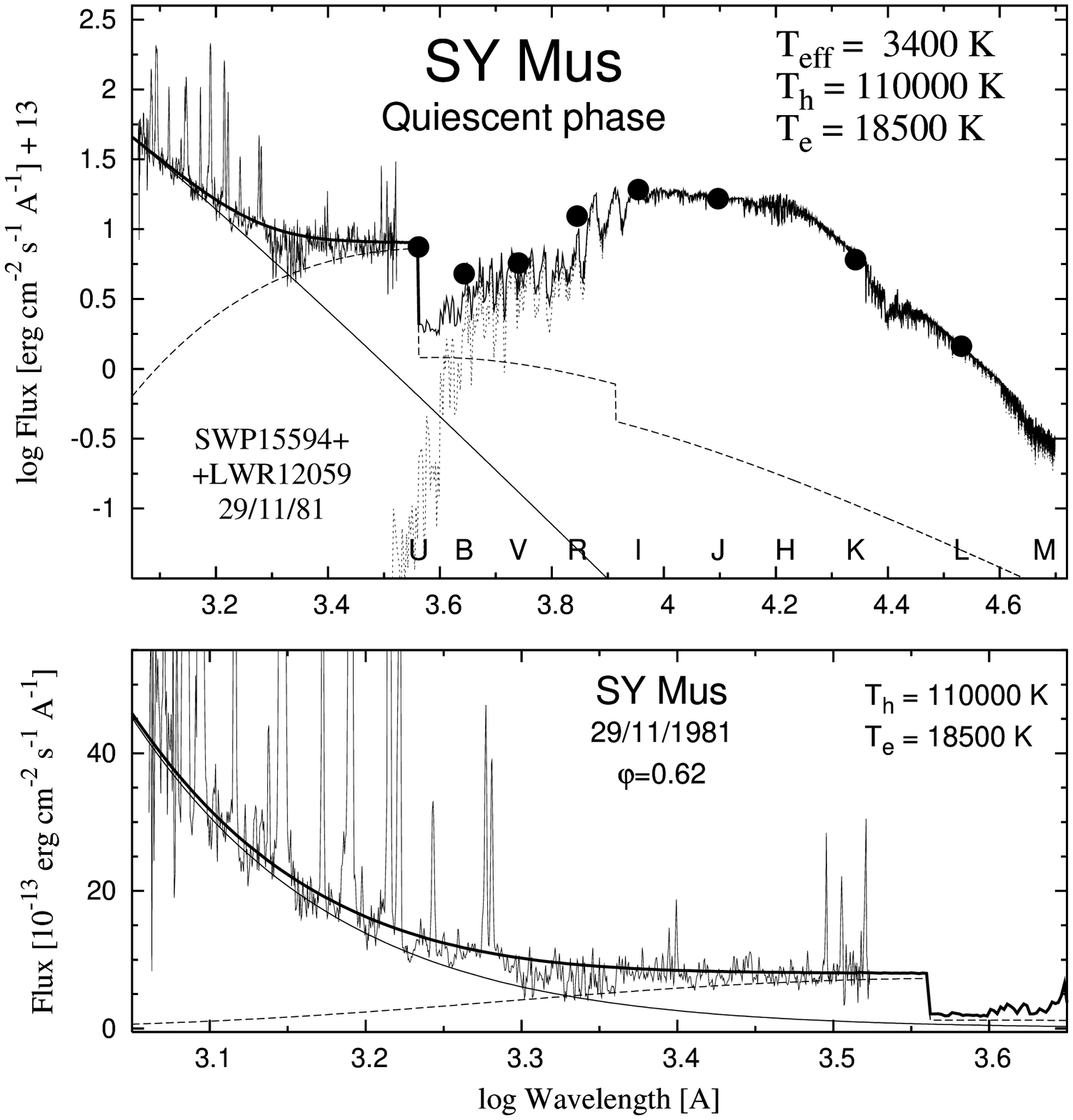}}
\caption[]{
Top: 
Visual and $U$ LC of SY\,Mus. The former was compiled 
from visual estimates available at the CDS database. 
The latter was reconstructed from the IUE low-resolution 
spectra by the same way as suggested by \cite{sk01a}.
Middle and bottom: 
Observed and reconstructed SEDs at the position near 
to its optical maximum. 
          }
\end{center}
\end{figure}

\subsection{SY\,Muscae}
%
%
SY\,Mus is another stable system with a high luminosity 
of the hot component. Its visual LC displays a strictly 
periodic wave-like variation with the orbital phase 
\citep[e.g.][]{p+95}. Figure~18 also shows a variation in 
the emission measure along the orbital phase scaled to the 
effective wavelength of the $U$ band as suggested by 
\cite{sk01a}. The emission measure was estimated from the near-UV, 
where the nebula represents a dominant source of radiation 
(Eq.~18, $T_{\rm e} \equiv 18\,500$\,K). 
The system is eclipsing. When the binary approaches the 
position of the inferior giant's conjunction, the Rayleigh 
scattering significantly affects the far-UV radiation 
\citep[e.g.][]{p+95}. 
Near to the conjunction 
($\varphi = 0\,\pm\sim 0.05$) the UV continuum is practically 
flat and at the level of 
1 -- 2\,$10^{-13}\rm\,erg\,cm^{-2}\,s^{-1}\,\AA^{-1}$. 
Also in this case an additional wavelength-independent 
continuum attenuation can be recognized 
\citep[Table~2 of][]{d+99}. As a result we selected only one 
observation exposed at the position near to the maximum, where 
the effect of the additional opacity source can be assumed 
to be minimal 
(SWP15594 + LWR12059, 29/11/81, $\varphi = 0.62$). \\
\hspace*{5mm}{\em Radiation from the giant}.  
In good agreement with the spectral classification of the 
giant in SY\,Mus \citep[M4.5\,\I\I\I,][]{ms99} a synthetic 
spectrum of
  $T_{\rm eff} = 3\,400$\,K 
matches well photometric measurements in the IR. 
Its scaling, 
  $k_{\rm g} = 3.6\,10^{-18}$, 
 ($\theta_{\rm g} = 1.9\,10^{-9}$) 
corresponds to the bolometric flux, 
  $F_{\rm g}^{\rm obs} = 2.6\,10^{-8}\,\rm erg\,cm^{-2}\,s^{-1}$. 
According to the surface brightness relation for M-giants 
\citep{ds98}, the reddening-free magnitudes, 
$K$ = 4.55 and $J$ = 5.75\,mag, give the same value of 
$\theta_{\rm g}$. 
These fitting parameters define the distance 
  $d = 1.0\,\pm\,0.15\,(R_{\rm g}/86 R_{\sun})$\,kpc
and the luminosity
 $L_{\rm g} = 850\,\pm\,250\,(d/1\,\kpc)^2\,L_{\sun}$, 
where the radius 
 $R_{\rm g} = 86\,\pm\,13\,R_{\sun}$ 
was determined by \cite{schmutz+94} in the same way 
as for RW\,Hya. \\
\hspace*{5mm}{\em Radiation from the ultraviolet.}  
There is a striking similarity in the SEDs of SY\,Mus and 
RW\,Hya (Figs.~17 and 18), which suggests very similar 
ionization conditions in these systems. 
The evolution in the H$\alpha$ line profile along the orbital 
motion is qualitatively of the same type in both systems 
\citep{schmutz+94,sch+96}. 
Our solution for the selected spectrum corresponds to 
   $T_{\rm h} = 110\,000$\,K ($\equiv$ Zanstra temperature), 
   $k_{\rm h} = 4.8\,10^{-24}$
and
  $T_{\rm e} = 18\,500\,\pm\,2\,500$\,K with scaling,
  $k_{\rm N} = 2.5\,10^{15}\,\rm cm^{-5}$, 
which gives the parameter $\delta$ = 0.55. 
These parameters can be discussed in the same way as those 
for RW\,Hya. Here the lower limit of the hot source 
temperature is $T_{\rm h}^{\rm min} = 80\,000$\,K. The only 
larger difference is in the relative contributions from 
the giant and those produced by the hot star in the $U$ band, 
which is due to the difference in $T_{\rm eff}$. 
This situation makes the amplitude 
$\Delta U$(SY\,Mus) $>$ $\Delta U$(RW\,Hya) despite that 
orbital inclinations of both systems are comparable. 
As a by-product of fitting the UV spectra, we determined 
a new value of reddening to SY\,Mus, $E_{\rm B-V}$ = 0.35. 
A larger quantity of $E_{\rm B-V}$ = 0.5 -- 0.45, previously 
suggested by MNSV and \cite{p+95}, yields too steep a far-UV 
continuum (from $\approx 1\,350$\,\AA), which is not 
possible to match by any reasonable black-body radiation. 
The 2\,100 -- 2\,300\,\AA\ spectral region displays 
a higher continuum level than that at surrounding wavelengths 
if one consider the influence of the iron curtain. We note that 
this region is rich in strong absorption features 
(see in detail the well-exposed spectra, e.g. LWR12059). 
\subsection{AR\,Pavonis}

%
%
\cite{sk+01b} recently summarized the historical, 1889 - 2001, 
LC of AR\,Pav. It is characterized by about 2\,mag deep 
minima -- eclipses -- and strong out-of-eclipse 
variations of between about 12 and 10\,mag. This behaviour 
suggests that AR\,Pav persists in an active phase. 
The UV continuum shows the same type of profile during 
the whole IUE mission \citep[e.g. Fig~1 of][]{sk03b}. 
Therefore we selected observations well outside the giant's 
inferior conjunction 
  (SWP13956 + LWR10570, 10/05/1981, $\varphi = 0.37$) 
and just in the eclipse 
  (SWP16949 + LWR13235, 13/05/1982, $\varphi = 0.98$). \\
\hspace*{5mm}{\em Radiation from the giant}. 
We compared the $RIJHKL$ fluxes with a synthetic spectrum of 
  $T_{\rm eff} = 3\,400$\,K
and 
  $\theta_{\rm g} = (6.4\,\pm\,0.5)\,10^{-10}$, 
which corresponds to the bolometric flux 
  $F_{\rm g}^{\rm obs} = 
           (3.1\,\pm\,0.3)\,10^{-9}\,\rm erg\,cm^{-2}\,s^{-1}$. 
The average reddening-free values of 
   $J$ = 7.96 and $K$ = 6.86\,mag 
yield $\theta_{\rm g} = 6.5\,10^{-10}$ \citep{ds98}. 
\cite{sch+01} derived the distance 
  $d$ = 4.9\,kpc 
from the 'rotational' giant's radius, 
 $R_{\rm g} = (130\,\pm\,25)\,R_{\sun}$ 
and the brightness surface relation for M-giants. This value 
is in particular agreement with that derived from a 106-day 
period of pulsations of the giant seen in some parts of the 
visual LC \citep{sk+00}. Our value of $\theta_{\rm g}$ then 
gives the radius of the giant 
  $R_{\rm g} = (139\,\pm\,10)\,R_{\sun}$, 
which agrees with that determined by \cite{sch+01}. 
The uncertainty results only from that in $\theta_{\rm g}$. 
This quantity corresponds to the giant's luminosity, 
  $L_{\rm g} = (2\,300\,\pm\,400)\,(d/4.9\,\kpc)^2\,L_{\sun}$. 
%
%
%
\begin{figure}
\centering
\begin{center}
\resizebox{\hsize}{!}{\includegraphics[angle=-90]{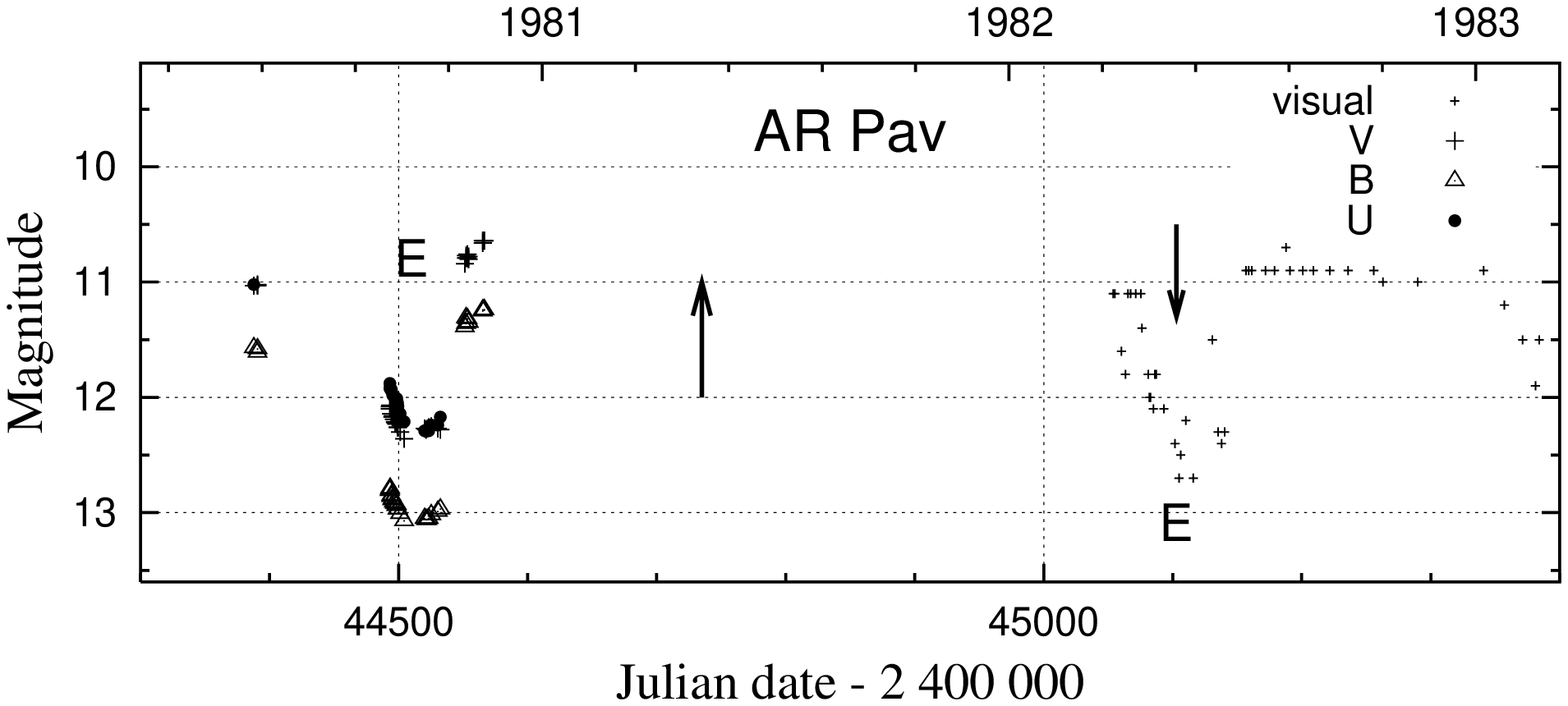}}

\vspace{2mm}

\resizebox{\hsize}{!}{\includegraphics[angle=-90]{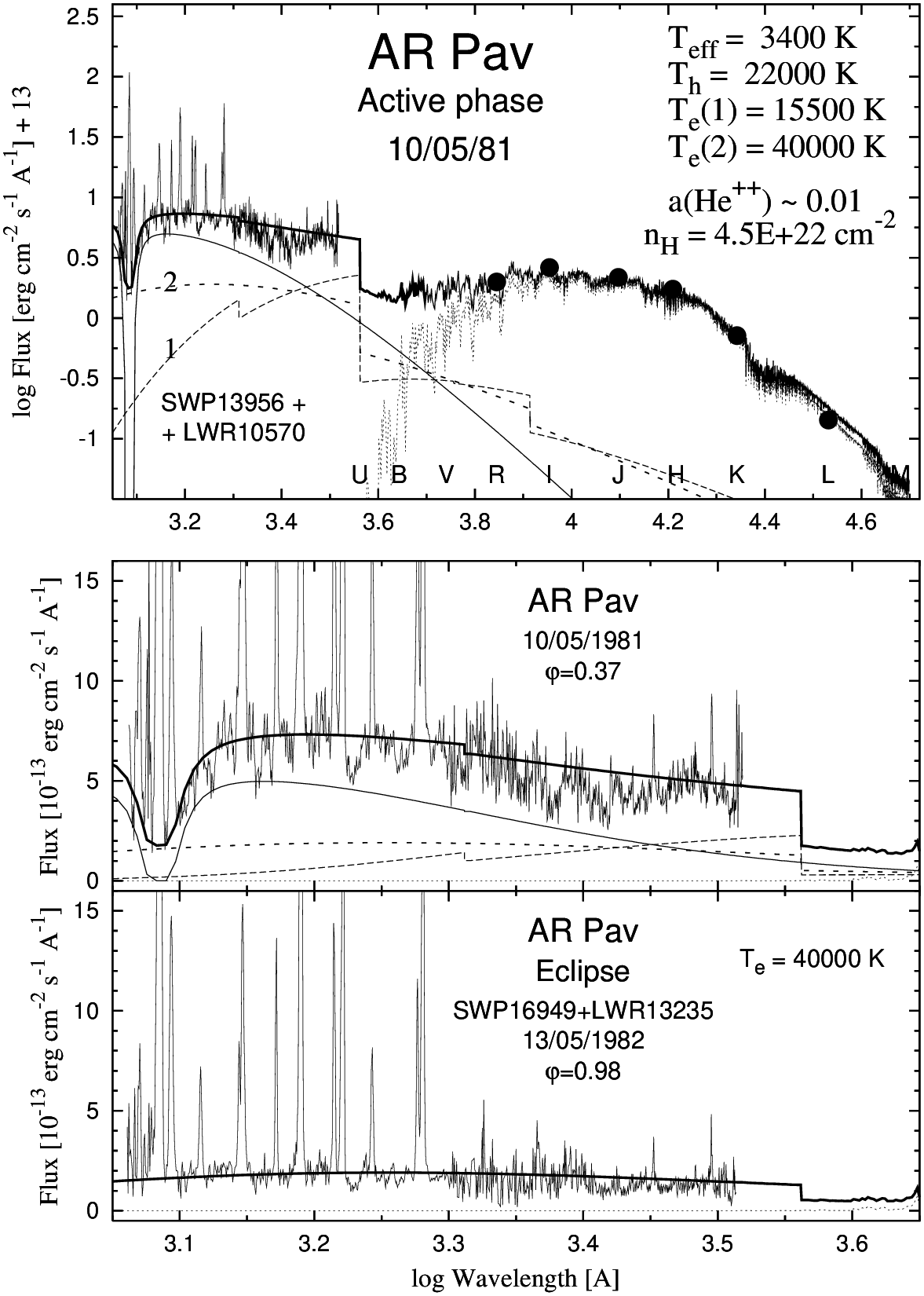}}
\caption[]{
The SED for AR\,Pav out and in the eclipse (positions of the 
spectra used are marked by arrows in the relevant part of the 
LC). Its out-of-eclipse profile in the ultraviolet resembles 
that observed in other eclipsing systems during active phases. 
          }
\end{center}
\end{figure}
According to the orbital solution, the separation between 
the components, 
  $A\sin(i) = 457\,\pm\,35\,R_{\sun}$ \citep{q+02} 
and the linear size of the giant's stellar disk that 
eclipses the object, 
  $R_{\rm g}^{\rm E}/A = 0.30\,\pm\,0.02$ 
\citep[][]{sk+00} limit the $R_{\rm g}$ radius to between 
  207\,$\pm\,25\,R_{\sun}$ ($R_{\rm g} \sim R_{\rm L}$, $i = 72^{\circ}$) 
and 
  137\,$\pm\,20\,R_{\sun}$ ($i = 90^{\circ}$) 
\citep[Eq.~25, see also][]{q+02}. 
From this point of view, $R_{\rm g} = (139\,\pm\,10)\,R_{\sun}$ 
corresponds to the orbital inclination 
  $i = 87\,\pm\,3^{\circ}$. 
%

{\em Radiation from the ultraviolet}.  
The profile of the UV continuum on the out-of-eclipse spectra 
is of the same type as observed for other active symbiotics 
with a high orbital inclination: 
A Rayleigh attenuated far-UV continuum 
   (here $n_{\rm H}\,\sim\,4.5\,10^{22}\,\rm cm^{-2}$), 
a rather cool radiation from the hot stellar source 
   ($T_{\rm h} \sim 22\,000$\,K, $k_{\rm h} = 7.9\,10^{-23}$), 
which is markedly affected by the iron curtain absorptions, 
and the nebular components of radiation from both the 
   LTN ($T_{\rm e} = 15\,500$\,K) 
and 
   HTN ($T_{\rm e} \sim 40\,000$\,K). 
Radiation from the HTN is identified in all spectra. 
During eclipses, as the only light visible in the ultraviolet 
and on the out-of-eclipse spectra it is indicated by 
a non-zero level of the Rayleigh attenuated continuum near 
to the $Ly\alpha$ line. Its level is about 
  1 -- 2\,$\times\,10^{-13}$\,\ecsa, 
which corresponds to a large emission measure, 
   $EM_{\rm HTN} = 2.9\,10^{60}(d/4.9\,\kpc)^2$\,\cmtri\
and/or the luminosity
   $L_{\rm HTN} = 530\,(d/4.9\,\kpc)^2\,L_{\sun}$. 
Our solution is plotted in Fig.~19. 

\subsection{AG\,Pegasi}
%
%

AG\,Peg is the slowest symbiotic nova. In the mid-1850 
it rose in brightness from $\sim 9$\,mag to $\sim 6$\,mag 
and afterwards followed a gradual decline to the present 
brightness of $\sim 8.7$\,mag in $V$. Evolution in its spectrum 
during the last century was described by many authors 
\citep[e.g.][]{k+93}. The orbitally related wave-like 
variation in the LC developed around 1940 
\citep[Fig.~6 of][and references therein]{sk98}. From that 
time to the present AG\,Peg faded from about 7.5 to 10\,mag 
in the $B$ band. Therefore we selected two earlier observations 
within one orbital cycle, but at/around different conjunctions 
of the components 
 (SWP03830 + LWR03376, 05/01/79, $\varphi = 0.96$; 
  SWP07407 + LWR06390, 14/12/79, $\varphi = 0.38$) 
to show the orbitally-dependent nebular emission at 
a constant hot stellar radiation, and one observation 
taken later when the star's brightness significantly 
faded 
 (Y1JO0308T + Y1JO0403T + Y1JO0309T + Y1JO0406T, 13/11/93, 
$\varphi = 0.63$). This was made by the Faint Object 
Spectrograph (FOS) on the board the Hubble Space Telescope. 
%

{\em Radiation from the giant}. 
The cool component in AG\,Peg was classified as a normal 
M3\,\I\I\I\ red giant \citep{ms99,k+93}. Its rather bright 
IR magnitudes suggest it cannot be very distant. The flux 
points corresponding to the $RIJKLM$ measurements can be 
matched by a synthetic spectrum with 
  $T_{\rm eff} = 3\,600$\,K, 
  $\log(g) = 1.0$ and 
  $\theta_{\rm g} = 2.4\,10^{-9}$, 
which yields the flux 
$F_{\rm g}^{\rm obs} = 5.6\,10^{-8}$\,\ecs. 
Our mean dereddened magnitudes, $K$ = 3.85 and $J$ = 4.98\,mag, 
also give a similar value of $\theta_{\rm g} = 2.6\,10^{-9}$. 
\cite{k+93} determined the distance modulus for bolometric 
magnitudes under the assumption that the giant in AG\,Peg is 
similar to other red giants. They obtained 
  $d$ = 800\,pc, 
which corresponds to 
  $R_{\rm g} = d \times \theta_{\rm g} = 85\,R_{\sun}$
and its luminosity 
  $L_{\rm g} = 1\,100\,(d/800\,\pc)^2\,L_{\sun}$. 
These quantities are identical to those determined by 
\cite{k+93}. Other estimates in the literature also rely 
on common properties of red giants as a class. 
For example, MNSV employed the same method, but for $K$ 
magnitudes, and obtained $d$ = 650\,pc. As there is no 
independent determination of the giant's radius in AG\,Peg 
we adopted that published by \cite{k+93}, which conforms 
better to our solution. 
%

{\em Radiation from the ultraviolet.}  
Properties of the UV spectrum and optical LCs during the 
investigated period resemble those of classical symbiotic 
stars during quiescent phases (Fig.~20). In addition, the 
overall fading of the star's brightness results from that 
in $L_{\rm h}$ \citep[see Fig.~4 of][and Fig.~20 here]{mn94}. 
According to \cite{mn94} a constant hot star temperature 
  $T_{\rm h} = 95\,000$\,K can be assumed during the period 
1978 -- 1993. We adopted this quantity in our modeling the SED. 
The following points are relevant. 

(i) 
The 1979 spectra show the phase-dependence of the nebular 
radiation, which follows that in the LCs. The observed $EM$ 
increased by a factor of about 2 from 05/01/79 ($\varphi$ = 0.96) 
to 14/12/79 ($\varphi$ = 0.38). In contrast to, e.g. RW\,Hya 
and SY\,Mus, the far-UV continuum is not subject to 
orbitally-related variation, which suggests that the AG\,Peg 
orbit is not highly inclined to the observer. 

(ii) 
We estimated the lower limit of the hot star temperature, 
$T_{\rm h}^{\rm min} = 70\,000$\,K for the 14/12/79 spectrum. 
The  H\,\I\I\ region in AG\,Peg is very open 
\citep[Fig.~11 of][]{k+93}, which means that a significant 
fraction of the $L_{\rm ph}$ photons escapes the system. 
This implies $T_{\rm h} \gg T_{\rm h}^{\rm min}$ to produce 
the observed $EM$ for an appropriate fit. 
For example, if only one half of the ionizing photons gives 
rise to the observed nebular emission then the hot star 
temperature has to increase to $T_{\rm h} = 100\,000$\,K 
to fit the far-UV continuum (Eq.~23 for $\delta$ = 0.5). 

(iii) 
The situation for AG\,Peg is complicated by its hot star wind 
  ($v_{\infty} \sim $\,900 -- 1\,000\,\kms, 
   $\dot M \sim 3\,10^{-7}$\,\myr, \cite{vn94}), 
which represents an additional source of emitters 
contributing to the observed nebula. However, the emission 
measure from the hot star wind, $EM_{\rm W}$, is negligible with 
respect to that observed. We estimated its quantity to be 
  $EM_{\rm W}\,\dot =\,8.3\,10^{58}$\,\cmtri\ 
by using relation (8) of \cite{sk+02a} for 
  $R_{\star} \equiv R_{\rm h}^{\rm eff} = 0.18\,R_{\sun}$, 
the radius $r_{\rm min} \sim 1.5\,R_{\star}$, from which the wind 
becomes optically thin \citep{l88} and the parameter $\beta = 1$ 
as assumed for hot star winds. Thus $EM_{\rm W} \ll EM^{\rm obs}$. 
This suggests that a major part of the nebular emission has 
to come from a more dense region in the system, which could 
be associated with the colliding region of the two winds 
located probably at the vicinity of the giant's hemisphere 
facing the hot star \citep{k+93}. 

(iv) 
Finally, the 1993-spectrum shows a higher efficiency in 
producing the nebular component of radiation ($\delta$ = 0.73). 
A decrease of the mass-loss rate from the hot star by a factor 
of $\sim$\,5 \citep{vn94} results in a larger expansion of 
the massive giant's wind towards the hot star and thus in 
a larger opening of the colliding zone. This makes it possible 
for a larger fraction of the $L_{\rm ph}$ photons to be involved 
in the ionization/recombination process. 
A quantitative model should support this view. 
%
%
\begin{figure}[p!t]
\centering
\begin{center}
\resizebox{\hsize}{!}{\includegraphics[angle=-90]{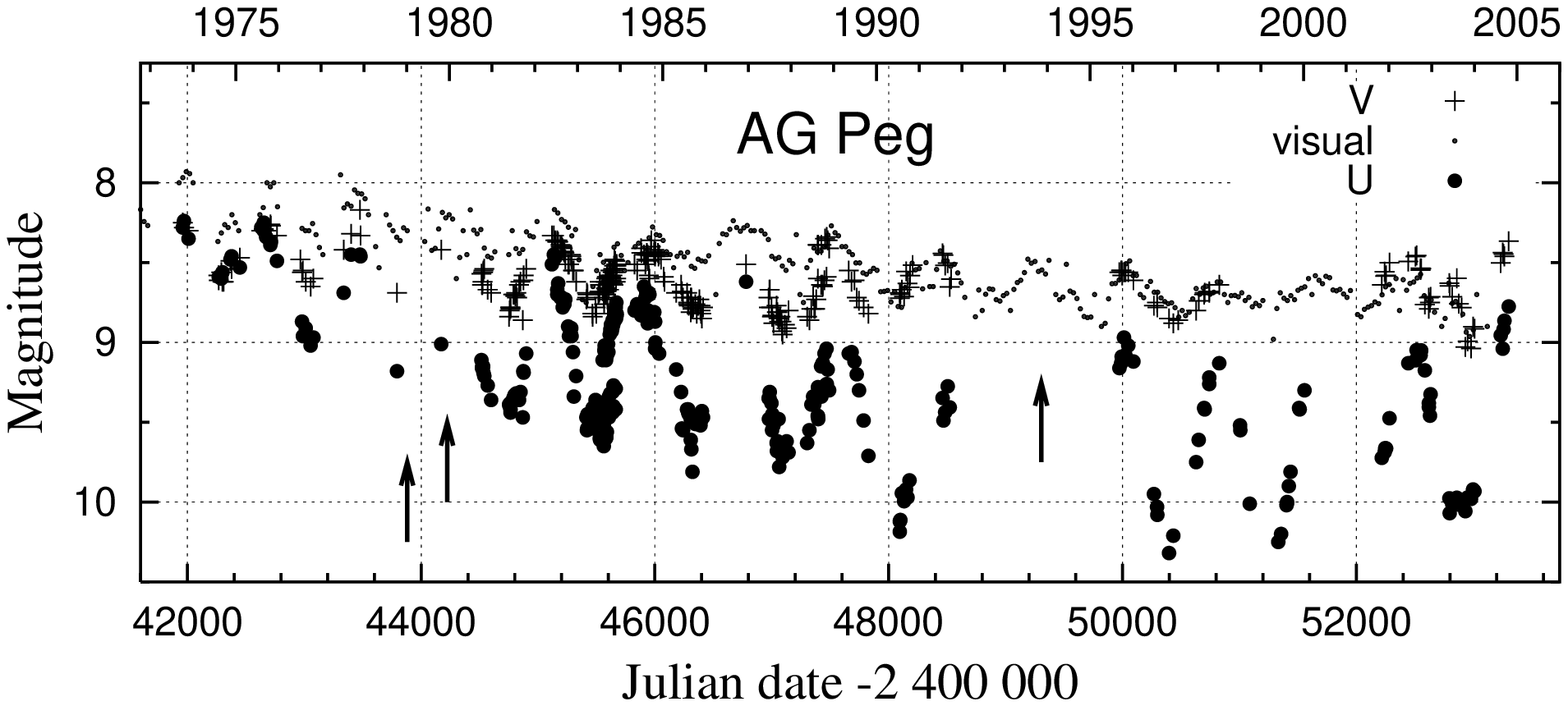}}

\vspace{2mm}

\resizebox{\hsize}{!}{\includegraphics[angle=-90]{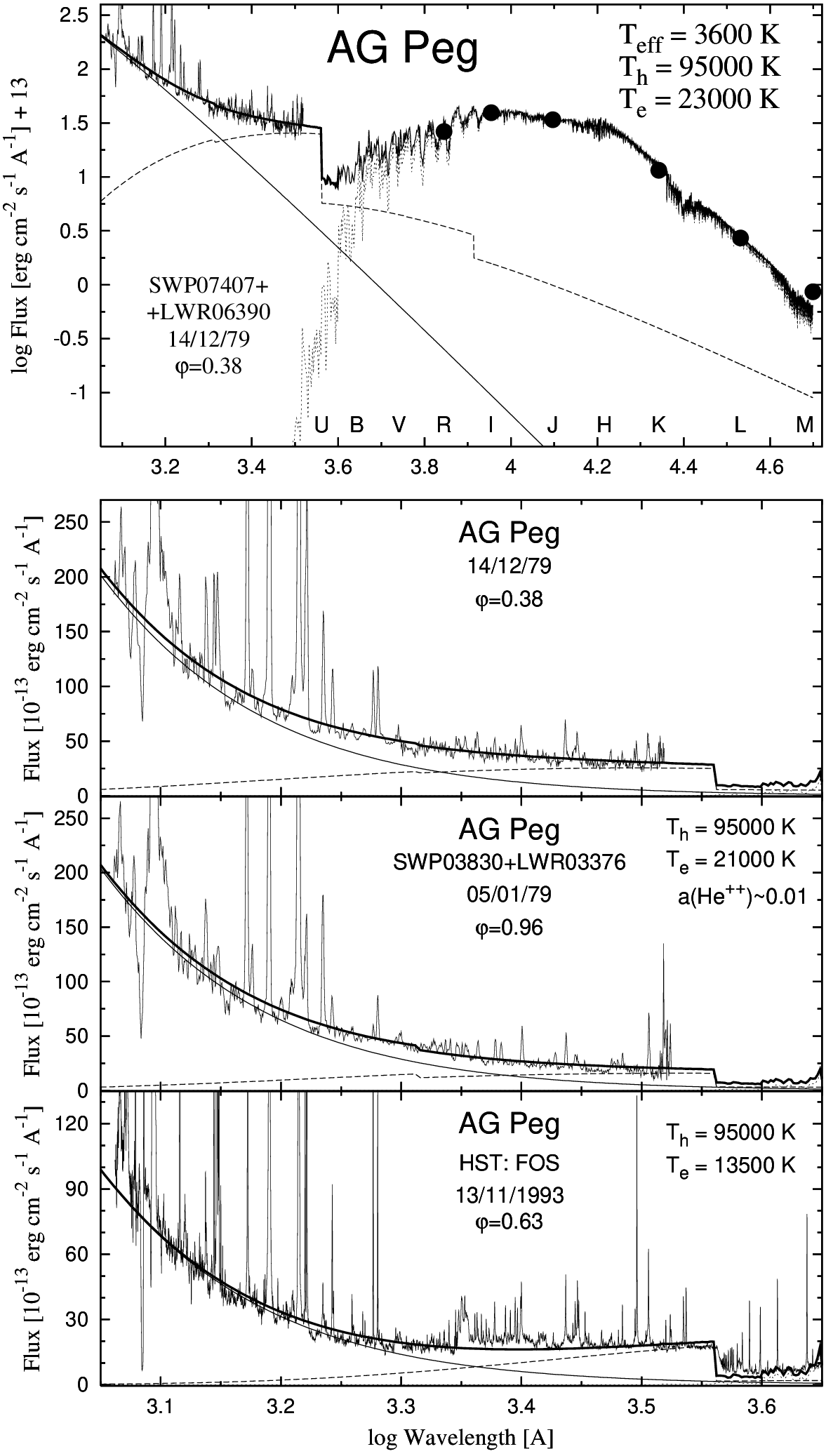}}
\caption[]{
The SED for AG\,Peg for three selected dates at different 
brightness of the system. Positions of the used spectra are 
marked by arrows in the LC (top). 
Data are from 
\cite{lu84,lu89}, 
\cite{ko90}, 
\cite{bel92}, 
\cite{tt98,tt01}, 
\cite{sk+04} 
and from 2003 our unpublished observations. 
          }
\end{center}
\end{figure}
%
%
%
\begin{figure}
\centering
\begin{center}
\vspace*{-3mm}
\resizebox{\hsize}{!}{\includegraphics[angle=-90]{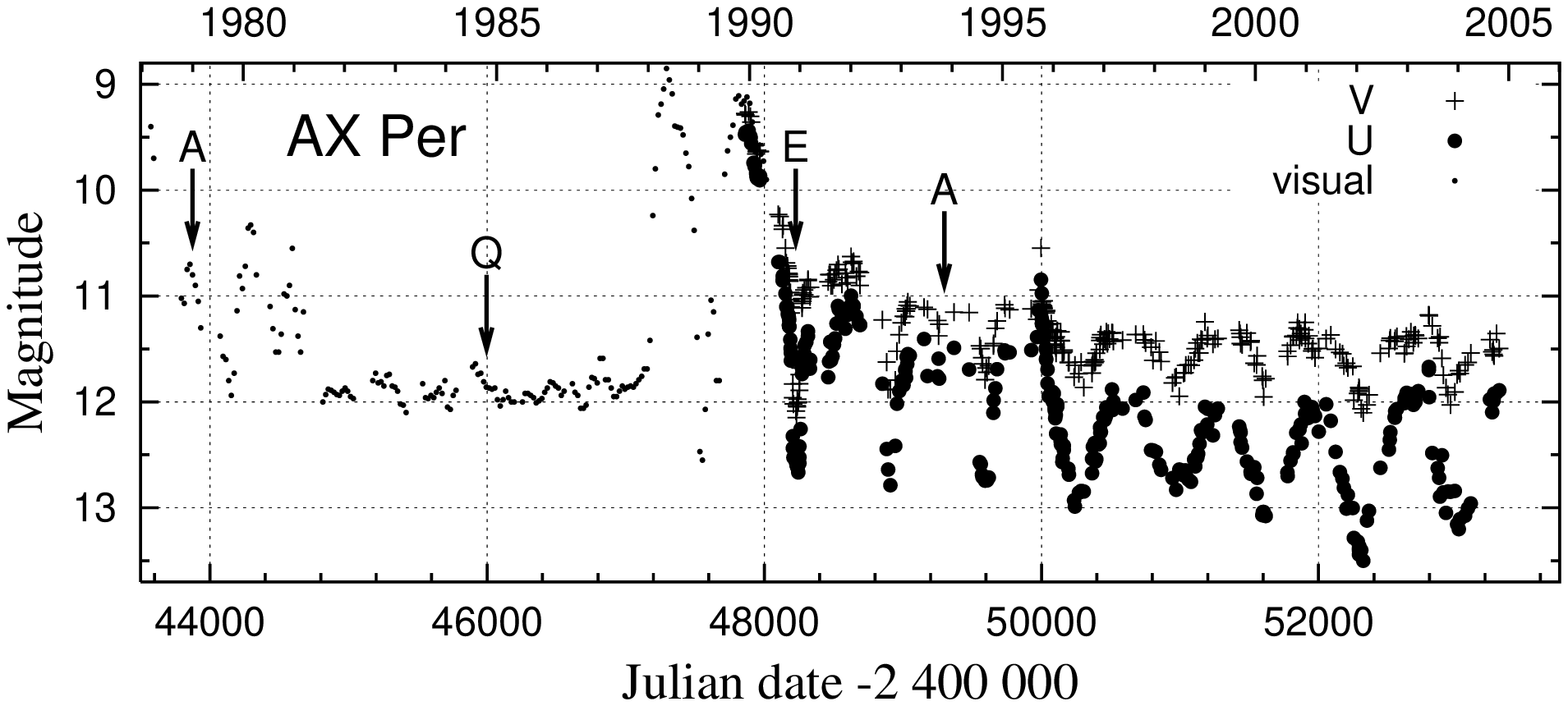}}

\vspace{2mm}

\resizebox{\hsize}{!}{\includegraphics[angle=-90]{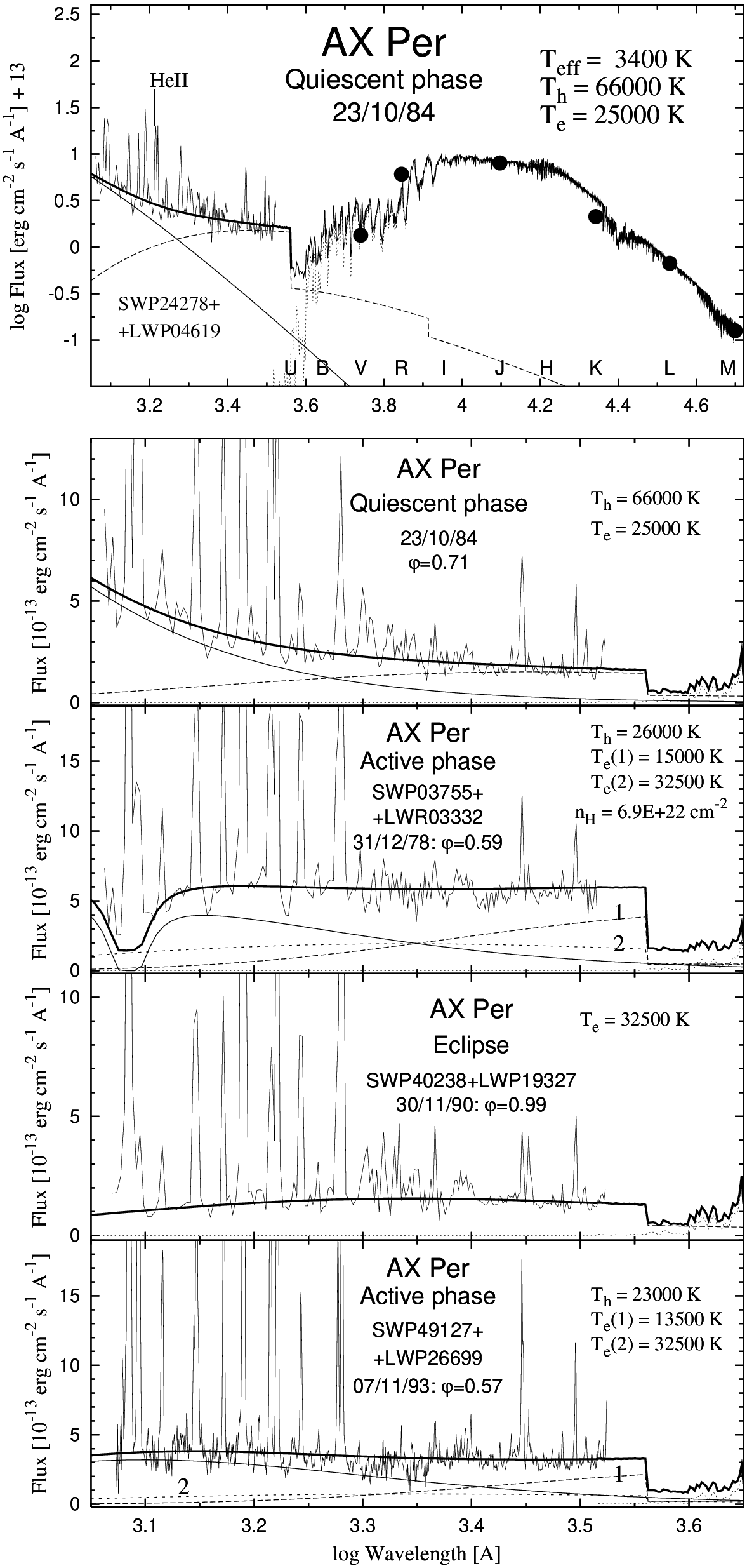}}
\caption[]{
The SED for AX\,Per during the activity, quiescence and eclipse. 
Positions of the spectra used are marked by arrows in the LC. 
Small dots represent means of the CDS visual data in 
the 30-day bins. 
          }
\end{center}
\end{figure}

\subsection{AX\,Persei}
%
%
The eclipsing nature of this classical symbiotic star was 
unambiguously revealed by photometric observations of the 
$\sim 1$\,mag deep narrow minimum -- eclipse -- in the 
$UBV$ LCs during its recent 1989 major outburst \citep{sk91}. 
The historical LC of AX\,Per is characterized by long-lasting 
periods of quiescence with superposition of a few bright 
stages \citep{sk+01a}. 
Accordingly we selected four IUE spectra taken at very 
different brightness/activity levels: 
During quiescence 
  (SWP24278 + LWP04619, 23/10/84, $\varphi = 0.71$), 
during the 1978 outburst 
  (SWP03755 + LWR03332, 31/12/78, $\varphi = 0.59$), 
during the 1990 eclipse 
  (SWP40238 + LWP19327, 30/11/90, $\varphi = 0.99$) 
and during the transition to quiescence 
  (SWP49127 + LWP26699, 07/11/93, $\varphi = 0.57$). 
%

{\em Radiation from the giant}.  
Infrared photometry of AX\,Per was summarized by 
\cite{sk00}, who derived the following parameters 
of the giant in the same way as in this paper: 
 $T_{\rm eff} = 3\,400 \pm 150$\,K, 
 $\theta_{\rm g} = (1.3 \pm 0.2)\,10^{-9}$, 
 $F_{\rm g}^{\rm obs} = 
      (1.4 \pm 0.3)\,10^{-8}\,\rm erg\,cm^{-2}\,s^{-1}$. 
The angular radius then gives the distance 
  $d = 1\,700\,\pm\,200\,(R_{\rm g}/102\,R_{\sun})$\,pc, 
where the giant's radius was derived from the eclipse 
timing under the assumption that $i = 90^{\circ}$, which 
is supported by a rectangular profile of the minimum 
\citep{sk94}. 
The giant's corresponding luminosity 
 $L_{\rm g} = 1\,200\,\pm\,400\,(d/1.7\,\kpc)^2\,L_{\sun}$. \\
\hspace*{5mm}{\em Radiation from the ultraviolet: Quiescent phase.}  
The best solution of the 23/10/84 spectrum yields a low 
temperature of the observed hot stellar source, 
   $T_{\rm h} = 66\,000\,+ 15\,000/\,- 7\,000$\,K 
   ($k_{\rm h} = 1.7\,10^{-24}$), 
but a strong nebular emission characterized by 
   $T_{\rm e} = 25\,000\,\pm\,3\,000$\,K
and scaled with 
   $k_{\rm N} = 7.2\,10^{14}\,{\rm cm^{-5}}$. 
These parameters give $\delta = 2.3$ (Eq.~21), i.e. 
the corresponding amount of the $L_{\rm ph}$ photons is not 
capable of producing the observed quantity of the nebular 
emission. This situation signals the presence of a material 
around the accretor of a disk-like structure, which makes it 
optically thick to a larger distance from the hot 
surface of the central star 
  ($\ge R_{\rm h}^{\rm eff}$) 
in directions on the orbital plane 
(i.e. on the lines of sight; $i \sim 90^{\circ}$). 
This means that in AX\,Per the accreted matter has 
a disk-like structure even during quiescent phases.
For comparison, we also present a formal solution for 
$\delta = 1$, i.e. 
$T_{\rm h} = T_{\rm h}^{\rm min}\dot = 100\,000$\,K (Table~3).  
The solution for this case gives 
  $R_{\rm h}^{\rm eff} = 0.065(d/1.7\,kpc)\,R_{\sun}$
and 
  $L_{\rm h} = 380(d/1.7\,kpc)^2\,L_{\sun}$, 
which are similar to those determined by MNSV. 
Our somewhat larger value of $L_{\rm h}$ is probably caused by 
different scaling, because we considered the influence of the 
iron curtain absorptions. \\
\hspace*{5mm}{\em Radiation from the ultraviolet: Active phases.}  
It is characterized by a flat profile with signatures of 
the Rayleigh attenuated far-UV continuum and the influence 
of the iron curtain absorptions. An appropriate discussion 
of such a spectrum was introduced, for example, in Sect.~4.10 
for CI\,Cyg. 
Also here the three-component model, composed of the HSS, LTN 
and HTN radiative contributions, has to be applied to get 
a satisfactory fit to observations. Fitting and derived 
parameters of our solutions for both selected spectra 
(31/12/78, 07/11/93) are given in Table~4 and Fig.~21. \\
\hspace*{5mm}{\em Radiation from the ultraviolet: Eclipse.}  
The UV spectrum during the 1990-eclipse can be matched 
by the HTN radiating at 
 $T_{\rm e} = 32\,500\,\pm\,3\,000$\,K 
and scaled with 
 $k_{\rm N} = 8.0\,10^{14}\,\rm cm^{-5}$. 
The same component was used in fitting the 31/12/78 spectrum 
to fill in the Rayleigh attenuated far-UV continuum. 
However, during the 1994-eclipse 
(SWP52027 + LWP29094, 04/09/94, $\varphi = 0.01$), 
at transition to quiescence, the level of the HTN 
emission was a factor of about 2.5 fainter. This suggests 
that the amount of the HTN 
contribution \mbox{is~a~function~of~the~activity.} 
%
%
%
\begin{figure}[p!t]
\centering
\begin{center}
\resizebox{\hsize}{!}{\includegraphics[angle=-90]{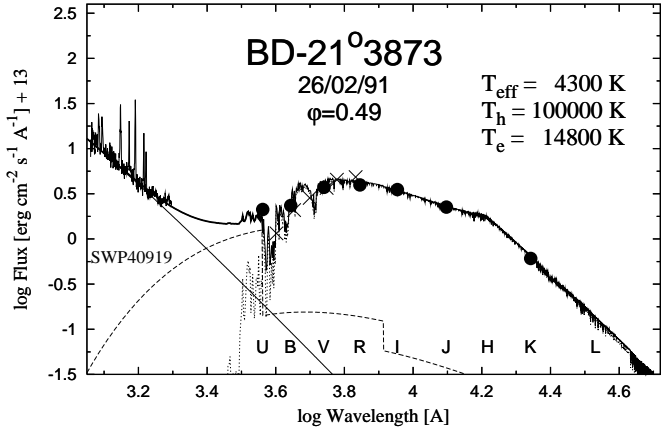}}
\caption[]{
Approximate SED for BD-21$^{\circ}$3873 given by its SWP spectrum in 
the ultraviolet and $UBVRIJK$ fluxes from photometric measurements 
($\bullet$). The value of $U(\varphi = 0.31$) from \cite{mu+92} was
extrapolated to $U(\varphi = 0.5$) according to the profile of the 
$u$-LC. Crosses ($\times$) represents flux-points from the 
spectrum taken on 24/02/88 ($\varphi = 0.61$) by \cite{nbd}. 
          }
\end{center}
\end{figure}

\subsection{BD-21$^{\circ}$3873}
%
\cite{smith+97} presented the largest set of $uvby$ photometry 
of this symbiotic star. The LCs show a pronounced orbitally-related 
variation with superposed double-sine patterns in $y$, $b$ and $v$. 
There are no signatures of an active phase 
\citep[also Fig.~9 of][]{sk+02b}. 
As for LT\,Del only SWP spectrum was available in the IUE archive 
(SWP40919, 26/02/91, $\varphi = 0.49$). 
We determined a new value of $E_{\rm B-V}$ = 0.20 by the same 
way as for LT\,Del (Sect.~4.12). \\
\hspace*{5mm}{\em Radiation from the giant.}  
Fluxes of the $BVRIJK$ photometric measurements can be compared 
well with a synthetic spectrum calculated for 
  $T_{\rm eff} = 4\,300$\,K 
with agreement with the atmospheric analysis of \cite{per+mello} 
and \cite{smith+97}. The scaling, 
  $k_{\rm g} = 2.4\,10^{-19}$, 
then gives the bolometric flux
  $F_{\rm g}^{\rm obs} = 4.7\,10^{-9}$\,\ecs. 
From the surface gravity, $\log(g) = 1.69 \pm 0.17$ 
\citep{per+mello}, one can derive the giant's radius as 
  $R_{\rm g} = (29 \pm 6)(M_{\rm g}/1.5M_{\sun})^{1/2}\,R_{\sun}$
and the distance
  $d = R_{\rm g}/\theta_{\rm g} = 
       (1.3 \pm 0.3)(M_{\rm g}/1.5M_{\sun})^{1/2}$\,kpc,
where the giant's mass was adopted as for AG\,Dra, because of 
their very similar atmospheric parameters \citep{smith+97}. 
The giant's luminosity 
  $L_{\rm g} = 260\,(d/1.3\,\kpc)^2\,L_{\sun}$. 
%

{\em Radiation from the ultraviolet.}  
Observed and approximate modeled SEDs are shown in Fig.~22. 
As for LT\,Del we adopted 
  $T_{\rm h} \equiv 100\,000$\,K, 
which requires a scaling of 
  $k_{\rm h} = 1.7\,10^{-24}$ 
and suggests parameters for the nebular emission, 
  $T_{\rm e} \sim 15\,000$\,K 
and 
  $k_{\rm N} \sim 3.3\,10^{14}\,{\rm cm^{-5}}$. 
This solution gives $\delta = 0.36$. 
Also $T_{\rm h} \equiv 75\,000$\,K \citep{schn93} 
is above $T_{\rm h}^{\rm min}$, because the corresponding 
$\delta = 0.63$. 
This results suggest an open ionized zone ($X > 1$), which 
is important in calculating the mass-loss rate from the giant 
(Sect.~5.2.1). 
%
%
%
%
\begin{table}
\begin{center}
\caption{Fitting and derived parameters: Cool components}
\begin{tabular}{ccccc}
\hline
\hline
Object            &
$T_{\rm eff}$     &
$\theta_{\rm g}$  &
$R_{\rm g}$       &
$L_{\rm g}$       \\
                  &
[$K$]             &
                  &
[$R_{\sun}$]      &
[$L_{\sun}$]      \\
\hline
EG\,And               & 3\,500 & 4.7\,10$^{-9}$  & 124 & 2\,070 \\
Z\,And                & 3\,400 & 1.6\,10$^{-9}$  & 106 & 1\,400 \\
AE\,Ara               & 3\,200 & 9.0\,10$^{-10}$ & 140 & 1\,800 \\
CD-43$^{\circ}$14304  & 4\,000 & 4.3\,10$^{-10}$ &~~40 & ~~~320 \\
TX\,CVn               & 4\,100 & 6.7\,10$^{-10}$ &~~30 & ~~~230 \\
T\,CrB                & 3\,400 & 1.8\,10$^{-9}$  &~~75 & ~~~680 \\
BF\,Cyg               & 3\,400 & 8.8\,10$^{-10}$ & 150 & 2\,700 \\
CH\,Cyg               & 2\,600 & 3.1\,10$^{-8}$  & 370 & 5\,600 \\
CI\,Cyg               & 3\,300 & 2.0\,10$^{-9}$  & 180 & 3\,400 \\
V1329\,Cyg            & 3\,300 & 7.2\,10$^{-10}$ & 132 & 1\,900 \\
LT\,Del               & 4\,100 & 1.7\,10$^{-10}$ &~~30 & ~~~230 \\
AG\,Dra               & 4\,300 & 7.0\,10$^{-10}$ &~~33 & ~~~360 \\
CQ\,Dra               & 3\,700 & 1.2\,10$^{-8}$  &~~96 & 1\,600 \\
V443\,Her             & 3\,300 & 1.3\,10$^{-9}$  & 130 & 1\,800 \\
YY\,Her               & 3\,500 & 3.9\,10$^{-10}$ & 110 & 1\,600 \\
RW\,Hya               & 3\,800 & 1.6\,10$^{-9}$  &~~59 & ~~~650 \\
SY\,Mus               & 3\,400 & 1.9\,10$^{-9}$  &~~86 & ~~~850 \\
AR\,Pav               & 3\,400 & 6.4\,10$^{-10}$ & 139 & 2\,300 \\
AG\,Peg               & 3\,600 & 2.4\,10$^{-9}$  &~~85 & 1\,100 \\
AX\,Per               & 3\,400 & 1.3\,10$^{-9}$  & 102 & 1\,200 \\
BD-21$^{\circ}$3873   & 4\,300 & 4.9\,10$^{-10}$ &~~29 & ~~~260 \\
%
%
\hline
\end{tabular}
\end{center}
Useful conversions according to the section 3.1: $k_{\rm g} = 
\theta_{\rm g}^2 = F_{\rm g}^{\rm obs}/\sigma T_{\rm eff}^{4}$, 
~~$R_{\rm g} = d \times \theta_{\rm g}$, 
~~$L_{\rm g} = 4\pi d^2 F_{\rm g}^{\rm obs}$ and $d$ in Table~1. 
\end{table}
%
%
%
\begin{table*}
\begin{center}
\caption{Fitting and derived parameters: Hot components during 
                                         quiescent phases}
\begin{tabular}{cccrccccccrc}
\hline
\hline
Object            &
Date              &
Phase             &
$T_{\rm h}$       &
$\theta_{\rm h}$  &
$R_{\rm h}^{\rm eff}$ &
$L_{\rm h}$       &
$T_{\rm e}$       &
$EM$              &
$L_{\rm N}$       &
$\delta$          &
$\log(\dot M_{\rm W})$     \\
                  &
                  &   
                  &
[$K$]             &   
                  &
[$R_{\sun}$]      &
[$L_{\sun}$]      &
[$K$]             &
[cm$^{-3}$]       &  
$L_{\sun}$        &
                  &
[\myr]   \\
%
\hline
EG\,And & 10/09/95 & 0.52 & 95\,000 & 1.2\,10$^{-12}$ &  0.032  
                          &  77     & 25\,000         
                          & 5.2\,10$^{58}$ &  12   & 1.0  &   -6.46$^a$\\
        & 28/08/91 & 0.48 &   "     & 1.1\,10$^{-12}$ &  0.029  
                          &  60     & 13\,500 
                          & 1.0\,10$^{58}$ &   2.6 & 0.46 &   -6.80$^b$\\
        & 13/01/83 & 0.95 &   "     &       "         &   "     
                          &   "     & 14\,500 
                          & 7.5\,10$^{57}$ &   1.9 & 0.33 &       \\
\hline
Z\,And  & 19/11/82 & 0.49&$>$120\,000&$<$1.7\,10$^{-12}$  &$<$0.11  
                          &$>$2\,300& 20\,500 
                          & 9.8\,10$^{59}$ & 220   &$<$0.92&  -6.10\\
        & 03/02/88 & 0.99 &   "     &$<$1.0\,10$^{-12}$   &$<$0.068 
                          &$>$860   & 22\,000 
                          & 1.6\,10$^{59}$ &  35   &$<$0.20&       \\
        & 29/06/95 & 0.57 &   "     &$<$1.7\,10$^{-12}$&  $<$0.11 
                          &$>$2\,300& 11\,500 
                          & 7.3\,10$^{59}$ & 200   &$<$1.1&   -6.16\\
\hline
BF\,Cyg & 10/04/92 & 0.46 & 95\,000 & 3.0\,10$^{-12}$ &  0.51   
                          & 18\,800 & 18\,000 
                          & 6.2\,10$^{60}$ &1\,400 & 0.72 &   -6.20\\
        &          &      &         &                 &          
                          &         & 42\,000 
                          & 5.7\,10$^{60}$ &1\,050 & 0.31 &        \\
        & 26/09/96 & 0.61 & 95\,000 & 2.9\,10$^{-12}$ &  0.50   
                          &18\,100  & 22\,800 
                          & 3.1\,10$^{60}$ & 670   & 0.24 &   -6.34\\
\hline
CI\,Cyg & 22/05/89 & 0.82 &115\,000 & 1.2\,10$^{-12}$ &  0.11   
                          & 1\,700  & 24\,000 
                          & 4.8\,10$^{59}$ & 110   & 0.48 &   -6.88\\
\hline
V1329\,Cyg & 29/03/83 & 0.57&$>$170\,000 &$<$5.4\,10$^{-13}$ &$<$0.10  
                          &$>$7\,500& 18\,000 
                          & 2.3\,10$^{60}$ & 530   &$<$0.95&  -5.87\\
           & 07/08/81 &0.9&    "    & 3.6\,10$^{-13}$&  0.068  
                          & 3\,400  & 18\,000 
                          & 4.8\,10$^{59}$ & 110   & 0.43 &        \\
\hline
LT\,Del & 16/06/90 & 0.49 &100\,000$^g$&   6.7\,10$^{-13}$  &   0.12
                          &   1\,200 & 17\,000
                          & 1.5\,10$^{59}$ &  35   &  0.28&   -6.8\\
\hline
AG\,Dra & 27/10/89 & 0.62 &$>$110\,000&$<$7.3\,10$^{-13}$&$<$0.034   
                          & $>$153  & 21\,500 
                          & 8.0\,10$^{58}$ &  18   &$<$0.93&  -6.49\\
        & 25/09/79 & 0.91 &     "     &       "          &    "  
                          &     "   & 21\,800 
                          & 5.8\,10$^{58}$ &  13   &  0.68 &       \\
\hline
CQ\,Dra & 23/10/84 & 0.80 &110\,000 & 8.9\,10$^{-13}$&0.007 
                          &   6.6   & 19\,500 
                          & 2.6\,10$^{57}$&    0.63&  0.84 &  -6.92$^c$\\
        & 11/10/86 & 0.22 &$>$145\,000&$<$7.5\,10$^{-13}$&$<$0.006 
                          &$>$14    & 22\,000 
                          & 5.8\,10$^{57}$&    1.4 &$<$0.94&  -6.60$^d$\\
\hline
V443\,Her& 20/10/80& 0.47 &$>$39\,000$^\dagger$&$<$4.8\,10$^{-12}$&$<$0.47
                          & $>$460  & 15\,500 
                          & 5.8\,10$^{59}$&  140   &$<$3.9 &       \\
        &          &      &  75\,000$^\ddagger$&  2.2\,10$^{-12}$&0.21
                          & 1\,300  &    "    
                          &      "        &    "   &   1.0 &  -6.42\\
        & 21/10/91& 0.23 &$>$39\,000$^\dagger$&$<$4.8\,10$^{-12}$&$<$0.47
                          & $>$460  & 12\,000 
                          & 2.3\,10$^{59}$&   63   &$<$2.0 &       \\
\hline
YY\,Her & 12/08/80& 0.46 &$>$105\,000&$<$4.9\,10$^{-13}$&$<$0.14 
                         &$>$2\,100  & 21\,500 
                         & 1.1\,10$^{60}$&  250    &$<$1.0 &  -6.17\\
\hline
RW\,Hya & 19/07/79& 0.32 &  110\,000& 2.5\,10$^{-12}$&  0.091   
                         &  1\,100  & 15\,000  
                         & 2.3\,10$^{59}$&   57    &  0.54 &  -6.50\\
        & 24/05/87& 0.06 &          &              &      
                         &          & 40\,000
                         & 6.0\,10$^{58}$&   11    &       &       \\
\hline
SY\,Mus & 29/11/81& 0.62 &  110\,000& 2.2\,10$^{-12}$&  0.097   
                         &  1\,200  & 18\,500  
                         & 3.0\,10$^{59}$&   69    &  0.55 &  -6.39\\

\hline
AG\,Peg & 14/12/79& 0.38 &  95\,000 & 5.2\,10$^{-12}$&  0.18    
                         &  2\,400  & 23\,000  
                         & 8.3\,10$^{59}$&  173    &  0.58 &  -6.19\\
        & 05/01/79& 0.96 &  95\,000 & 5.2\,10$^{-12}$&  0.18    
                         &  2\,400  & 21\,000  
                         & 4.4\,10$^{59}$&   94    &  0.33 &       \\
        & 13/11/93& 0.63 &  95\,000 & 3.6\,10$^{-12}$&  0.13    
                         &  1\,180  & 13\,500  
                         & 3.2\,10$^{59}$&   79    &  0.73 &  -6.38\\
\hline
AX\,Per & 23/10/84& 0.71 &$>$66\,000$^\dagger$&$<$1.3\,10$^{-12}$&$<$0.098
                         &  $>$170  & 25\,000  
                         & 2.5\,10$^{59}$&   53    &$<$2.3 &       \\
        &         &      & 100\,000$^\ddagger$& 8.7\,10$^{-13}$&  0.065
                         &     400  &     "    
                         &      "        &    "  &     1.0 &  -6.14$^e$\\
        &         &      & 140\,000 & 6.7\,10$^{-13}$&  0.051
                         &     900  &     "
                         &      "        &    "  &    0.54 &  -6.26$^h$\\
\hline
IV\,Vir$^f$&26/02/91&0.49&100\,000$^g$& 1.3\,10$^{-12}$&$<$0.072
                         &  480       & 14\,800
                         & 6.7\,10$^{58}$&   16     &  0.36 & -7.05    \\
%
\hline
\end{tabular}
\end{center}
$^\dagger$  $T_{\rm h}$ given by the fit,~~
$^\ddagger$  $T_{\rm h}$ = $T_{\rm h}^{\rm min}$,~~ 
$^a$  $X$ = 0.73,~~
$^b$  $X$ = 0.65,~~
$^c$  $X$ = 1.5,~~
$^d$  $X$ = 0.51,~~
$^e$  $X$ = 1.0, ~~
$^h$  $X$ = 20 \\
$^f$ = BD-21$^{\circ}$3873, ~~ 
$^g$ adopted value 
\end{table*}
%
%
%
\begin{table*}
\begin{center}
\caption{Fitting and derived parameters: Hot components 
                                         during active phases}
\begin{tabular}{ccccccccccccc}
\hline
\hline
Object            &
Date              &
Phase             &
$T_{\rm h}$       &
$\theta_{\rm h}$  &
$R_{\rm h}^{\rm eff}$&
$L_{\rm h}$       &
$\log(n_{\rm H})$ &
$T_{\rm e}$       &
$EM_{\rm LTN}$    &
$L_{\rm LTN}$     &
$EM_{\rm HTN}$    &
$L_{\rm HTN}$     \\
                  &
                  &   
                  &
[$K$]             &
                  &   
[$R_{\sun}$]      &
[$L_{\sun}$]      &
[cm$^{-2}$]       &
[$K$]             &
[cm$^{-3}$]       &
[$L_{\sun}$]      &
[cm$^{-3}$]       &
[$L_{\sun}$]      \\
%
%
\hline
Z\,And  & 06/04/84& 0.15& 24\,000& 1.4\,10$^{-11}$& 0.93   &   250 
                        &   22.85&     11\,000$^d$& 6.7\,10$^{59}$
                        &     190&  1.2\,10$^{60}$&            225  \\
        & 24/12/85& 0.98& 19\,000& 3.4\,10$^{-11}$& 2.25   &   590
                        &   23.34&     18\,000$^d$& 8.6\,10$^{59}$
                        &     200&  1.2\,10$^{60}$&            225  \\
\hline
AE\,Ara & 08/10/79& 0.53& 23\,000& 1.2\,10$^{-11}$& 1.9    &   850
                        &        &         17\,000& 5.3\,10$^{59}$
                        &     125&                &                 \\
\hline
CD-43$^{c}$
        & 15/05/91& 0.95& 20\,000& 4.9\,10$^{-12}$& 0.46   &    30
                        &    23.40 &  12\,000$^d$& 2.6\,10$^{58}$ 
                        &$\sim\,7$&1.0\,10$^{59}$&             19  \\
\hline
TX\,CVn & 05/12/81& 0.06&  17\,000 & 1.6\,10$^{-11}$& 0.70   &    37
                        &    22.36 &    10\,000   & 1.2\,10$^{58}$
                        & $\sim\,4$&              &                 \\
\hline
T\,CrB$^a$&03/06/87&0.99& 60\,000$^b$&            &   0.07 &   380
                        &    22.60 &    10\,000   & 1.3\,10$^{58}$
                        & $\sim\,4$&              &                 \\
        & 05-06/86& 0.52& 50\,500$^b$&            &   0.09 &   270
                        &    22.00 &    16\,000   & 4.4\,10$^{58}$
                        &       11 &              &                 \\
\hline
BF\,Cyg & 30/06/90& 0.59&  21\,500 & 4.2\,10$^{-11}$&  7.1 & 9\,600 
                        &    22.68 &    42\,000     & $\sim\, 0$ 
                        &$\sim\, 0$& 1.2\,10$^{61}$&         2\,200 \\
        & 01/05/91& 0.0 &\multicolumn{5}{c}{E~ c~ l~ i~ p~ s~ e}
                                   &    42\,000     &
                        &          & 1.1\,10$^{61}$ &        2\,100 \\
\hline
CH\,Cyg & 30/05/93& 0.30&  14\,500 & 7.5\,10$^{-11}$& 0.89 &     32
                        &    22.11 &
                     \multicolumn{5}{c}{nebular emission $\sim\, 0$} \\
        & 05/12/94& 0.03&\multicolumn{5}{c}{E~ c~ l~ i~ p~ s~ e}
                                   &  $10^4/10^5$   & 7.8\,10$^{57}$
                        &      2.3 & 1.1\,10$^{58}$ &            1.7 \\
        & 30/03/95& 0.18&  14\,200 &5.2\,10$^{-11}$ & 0.62  &     14
                        &    22.47 &  $10^4/10^5$   & 7.2\,10$^{57}$ 
                        &      2.1 & 9.2\,10$^{57}$ &            1.3 \\
\hline
CI\,Cyg & 05/01/79& 0.39&  28\,000 &7.6\,10$^{-12}$ & 0.67  &    250
                        &    23.00 &     12\,000$^d$& 5.3\,10$^{59}$
                        &      150 & 4.8\,10$^{59}$ &            90  \\
        & 10/06/80& 0.0 &\multicolumn{5}{c}{E~ c~ l~ i~ p~ s~ e}
                                   &    40\,000     &
                        &          & 5.3\,10$^{59}$ &            100 \\
\hline
AG\,Dra & 08/01/81& 0.78&$>$155\,000 &1.6\,10$^{-12}$ & 0.079&$>$3\,300
                        &          &    30\,000     &  $\sim\, 0$
                        &$\sim\, 0$& 1.8\,10$^{60}$ &            360 \\
        & 28/07/94& 0.79&$>$150\,000 &2.1\,10$^{-12}$ & 0.10 &$>$4\,900
                        &          &    35\,000     &  $\sim\, 0$             
                        &$\sim\, 0$& 3.1\,10$^{60}$ &            600 \\
        & 13/03/85& 0.56&$>$180\,000&7.8\,10$^{-13}$ & 0.038 & 1\,400
                        &          &    18\,600     & 4.6\,10$^{59}$
                        &   106    &  $\sim\, 0$    &  $\sim\, 0$   \\
        & 30/04/86& 0.31&$>$185\,000&7.3\,10$^{-13}$ & 0.036 & 1\,350
                        &          &    18\,000     & 4.2\,10$^{59}$
                        &    96    &  $\sim\, 0$    &  $\sim\, 0$   \\
\hline
YY\,Her & 05/12/81& 0.27&  22\,000 &3.2\,10$^{-12}$ & 0.89   &   170
                        &  23.14   &    13\,000$^d$ & 7.2\,10$^{59}$
                        &   190    & 5.9\,10$^{59}$ &            110 \\
        & 13/06/93& 0.37&$>$160\,000&$<$6.3\,10$^{-13}$ &$<$0.18&$>$18\,400
                        &          &    40\,000     & $\sim\, 0$
                        &$\sim\, 0$&  1.3\,10$^{61}$&          2\,400\\
\hline
AR\,Pav & 10/05/81& 0.37&  22\,000 &8.9\,10$^{-12}$ & 1.9    &   800
                        &  22.65   &    15\,500$^d$& 1.7\,10$^{60}$
                        &   415    & 2.9\,10$^{60}$ &            530 \\
        & 13/05/82& 0.98&\multicolumn{5}{c}{E~ c~ l~ i~ p~ s~ e}
                                   &    40\,000     &
                        &          &  2.9\,10$^{60}$&            530 \\
\hline
AX\,Per & 31/12/78& 0.59&  26\,000 &5.6\,10$^{-12}$ & 0.42    &    75
                        &  22.84   &    15\,000$^e$ & 3.6\,10$^{59}$
                        &    89    & 3.5\,10$^{59}$ &           68$^d$\\
        & 30/11/90& 0.99&\multicolumn{5}{c}{E~ c~ l~ i~ p~ s~ e}
                                   &    32\,500     &          
                        &          & 2.8\,10$^{59}$ &              55 \\
        & 07/11/93& 0.57&  23\,000 &6.2\,10$^{-12}$ & 0.47    &    55
                        &  $<$21   &    13\,500$^e$ & 1.7\,10$^{59}$
                        &    45    & 1.3\,10$^{59}$ &           26$^d$\\
%
%
\hline
\end{tabular}
\end{center}
$a$ -- accretion disk model, see Sect.~4.7,~ 
$b$ -- $T_{\rm max}$ of the AD-model,~ 
$c$ -- CD-43$^{\circ}$14304,~ 
$d$ -- $T_{\rm e}$(HTN) = 40\,000\,K,~ 
$e$ -- $T_{\rm e}$(HTN) = 32\,500\,K
\end{table*}
%
%
%

\section{Summary}

By disentangling the composite spectrum of 21 S-type symbiotic 
stars we isolated four basic components of radiation 
contributing to their UV/optical/IR continuum: 
Radiation from the 
(i) cool giant, 
(ii) hot stellar source (HSS), 
(iii) low-temperature nebula (LTN) 
and 
(iv) high-temperature nebula (HTN). 
Physical parameters of these components are introduced in 
Tables 2 -- 4 and corresponding solutions are shown in the 
relevant figures. 
During quiescent phases, components from the hot object show 
orbitally-related variations and unpredictable changes in 
$T_{\rm e}$, $L_{\rm h}$ and $EM$. 
Transition to active phases is followed by a dramatic change 
in $T_{\rm h}$ and by the appearance of the HTN component 
in the spectrum. 
Below we summarize and discuss their common properties 
and observed effects. 

\subsection{Radiation from the giant}

This component of radiation in the spectra of S-type symbiotic 
stars is determined by two fitting parameters -- the angular 
radius, $\theta_{\rm g}$ and the effective temperature, 
$T_{\rm eff}$ (Sect.~3.1). Both are connected throughout 
the observed bolometric flux, 
$F_{\rm g}^{\rm obs} = \theta_{\rm g}^2\sigma T_{\rm eff}^{4}$, 
which defines the giant's luminosity as 
\begin{equation}
  L_{\rm g} = 4\pi d^2 F_{\rm g}^{\rm obs}. 
\end{equation}
To test common properties of giants we examined the quantity 
of $F_{\rm g}^{\rm obs}$ as a function of the distance $d$ in 
the sense of the last relation. Fitting the relevant quantities 
from Table~1 and 2 determines this function as 
%
\begin{equation}
  \log(F_{\rm g}^{\rm obs}) = -2\log(d) - (7.30 \pm 0.05)~~~
                             {\rm for~red~giants}
\end{equation}
and
\begin{equation}
  \log(F_{\rm g}^{\rm obs}) = -2\log(d) - (8.03 \pm 0.05)~~~
                             {\rm for~yellow~giants}.
\end{equation}
The flux is in \ecs\ and the distance in kpc. 
The constant corresponds to a characteristic luminosity 
$L_{\rm g} = 1\,600 \pm 200$ and $290 \pm 30\,L_{\sun}$ 
for red and yellow giants, respectively. 
Figure~23 plots the observed data and fits. 
In the figure we distinguish giants having independently 
determined radii (from eclipses, co-rotation, atmospheric 
analysis, lobe-filling giant; 11 objects), which then provide 
the distance from their $\theta_{\rm g}$. 
For CQ\,Dra and CH\,Cyg we adopted their HIPPARCOS distances. 
This group of giants confirms that cool components in S-type
systems are normal giants. 
Otherwise the distances were estimated from $ST/R_{\rm g}$ 
dependencies \citep{bel+99} and the measured angular radii. 
CH\,Cyg, which is suspected to contain two red giants 
in accordance with its triple-star model, was not included 
in the analysis. 
Finally, having the SED in the infrared, these relations can 
be used to estimate distances of S-type symbiotic stars. 

\subsection{The quiescent phase}

\subsubsection{Nebular continuum radiation} 

{\em General properties:} 
Our solutions for 15 symbiotic stars during the quiescent phase 
showed that the nebular component of radiation can be 
characterized by a unique electron temperature. For individual 
objects $T_{\rm e}$ runs between about 12\,000 and 25\,000\,K 
with an average value of 19\,000\,K (Table~3). 
For Z\,And and AG\,Peg we observed a large decrease in 
$T_{\rm e}$ by about 10$^4$\,K. In the case of Z\,And, 
$T_{\rm e}$ decreased at approximately the same $L_{\rm N}$ 
and probably also $L_{\rm h}$, while for AG\,Peg this change 
was followed by a significant decrease in $L_{\rm N}$, 
obviously caused by that in $L_{\rm h}$ and thus $L_{\rm ph}$ 
(Table~3). A decrease in $T_{\rm e}$ could be caused by 
a decrease of the hot star wind, which causes removal of 
a fraction of hot electrons from the plasma created 
by collisions. 
For EG\,And and CQ\,Dra a transient enhancement of the emission 
measure observed at the same orbital phase can result from 
a relevant increase in the mass-loss rate from the giant, which 
thus supplies more emitters into the ionized zone. 
This change has a counterpart in the increase of the star's 
brightness in $U$. 

{\em Mass-loss rate from the giant:} 
The nebula during quiescence arises from ionization of 
the cool giant wind. 
\cite{sk01a} showed that the observed emission measure 
is consistent with that produced by the ionization model. 
Therefore we can calculate $EM$ directly by integrating 
emission contributions throughout the volume of the fully 
ionized zone as 
\begin{equation}
  EM = \int_{\ion{H}{ii}} \!n^2(r)\,{\rm d}V, 
\end{equation}
which can be determined through parameters of the giant's wind. 
The particle density $n(r)$ satisfies the equation 
of continuity, 
  $\dot M_{\rm W} = 4\pi r^{2}\mu m_{\rm H}n(r)v_{\rm W}(r)$, 
in which the velocity structure of the wind, 
  $v_{\rm W}(r) = v_{\infty}(1- R_{\rm g}/r)^{\beta}$ 
\citep{cak}, $v_{\infty}$ is the terminal velocity 
of the wind and the parameter ${\beta}$ characterizes its 
acceleration; ${\beta}$ = 2.5 satisfies a wind from 
a late-type giant \citep{schroder}. Under these conditions 
the upper limit of the $EM$ calculated in accordance with 
the relation (29) can be expressed analytically as 
\begin{equation}
  EM = \frac{-1}{16\pi (\mu m_{\rm H})^2}
       \Big(\frac{\dot M_{\rm W}}{v_{\infty}}\Big)^2 
       \frac{1}{R_{\rm g}}
       \Big[1 - \Big(1 - \frac{R_{\rm g}}{Q}\Big)^{-4}\Big]
\end{equation}
\citep[][]{nv89,sk01a}, which is applicable for extended 
ionized zones in symbiotic binaries (i.e. very open in 
the sense of the STB model; parameter $X \gg$ 1). This 
solution corresponds to an upper limit of the $EM$ 
assuming the sphere around the cool star to be fully ionized 
from $r = Q$ to $r = \infty$. The parameter $Q$ is the location 
of the \ion{H}{i}/\ion{H}{ii} boundary on the binary axis and 
the wind starts at the giant's surface. 
We comment on our calculations of $\dot M_{\rm W}$ as follows: 

(i) The mass-loss rate estimated from relation (30) represents 
its lower limit. In the real case the ionized volume is smaller. 
It represents only a fraction of the whole sphere, which 
requires a higher $\dot M_{\rm W}$ to give rise the observed 
quantity of $EM$. 
For example, mass-loss rates calculated according to Eq.~(29) 
for the open $\ion{H}{ii}$ regions characterized with $X$ = 20 
are factor of about 2.5 larger than those estimated from Eq.~(30). 
Most of the objects we investigated here contain extended 
nebulae which is mainly due to the high number of the ionizing 
photons they produce and a large separation between the binary 
components. Therefore for these objects we calculated 
$\dot M_{\rm W}$ according to Eq.~(29) for $X$ = 20. 
In the cases of a less 
opened ionized zone (EG\,And, CQ\,Dra and possibly AX\,Per) 
we integrated Eq.~(29) throughout its volume given by the 
parameter $X$. For EG\,And and CQ\,Dra, \cite{sk05} obtained 
$\dot M_{\rm W}$ and the corresponding parameter $X$ from 
accretion processes in these binaries. For AX\,Per \cite{sk+01a} 
determined $X$ = 1.0 from timing of the broad minimum observed 
during the 1958-72 quiescencent phase. However, as this value 
could change significantly (following the change of the LC 
profile), we present also a solution for $X$ = 20. Results of our 
analysis are given in the last column of Table~3 and Fig.~24 shows 
a comparison of our values with those determined by MNSV 
and \cite{mio02} from radio observations as a function of the 
hot star luminosity. The graph confirms well the relationship between 
these two parameters. \cite{mio02} discussed this correlation 
in terms of the illumination heating of the outer red giant's 
atmoshere, which enhances the mass-loss rate and consequently 
can induce a larger $L_{\rm h}$. 

(ii) Production of the nebular emission depends on the terminal 
velocity of the wind. A low-velocity wind corresponds to a higher 
concentration of emitters in the $\ion{H}{ii}$ zone, which thus 
gives rise to the observed $EM$ for a lower mass-loss rate.
We adopted $v_{\infty}$ = 20\,\kms. Note that MNSV 
adopted $\mu \cdot v_{\infty}$ = 10\,\kms\ 
(i.e. $v_{\infty} \dot = $ 7\,\kms), which provides a factor of 
about 3 lower values of $\dot M_{\rm W}$. 

(iii) There are other sources of the nebular radiation -- e.g. 
emitters from the hot star wind and ionizations by collisions. 
The former accounts for about 10\% of the total $EM$ for well 
developed winds (e.g. AG\,Peg) and the latter can be important 
for nebulae with $T_{\rm e} \ga 20\,000$\,K. Therefore 
the emission measure produced solely by emitters of the 
giant's wind is lower than what we measure. From this point 
of view our estimates of $\dot M_{\rm W}$ represent upper 
limits as they were derived from the total $EM$. 
For example, the mass-loss rate derived independently 
from accretion process in CQ\,Dra 
\citep[$\dot M_{\rm W} = 8.2\,10^{-8}$\,\myr,][]{sk05}
is by factor of $\sim$\,1.5 lower than that derived here 
($\dot M_{\rm W} = 1.2\,10^{-7}$\,\myr, Table~3). 
%
%
\begin{figure}[p!t]
\centering
\begin{center}
\vspace*{-3mm}
\resizebox{\hsize}{!}{\includegraphics[angle=-90]{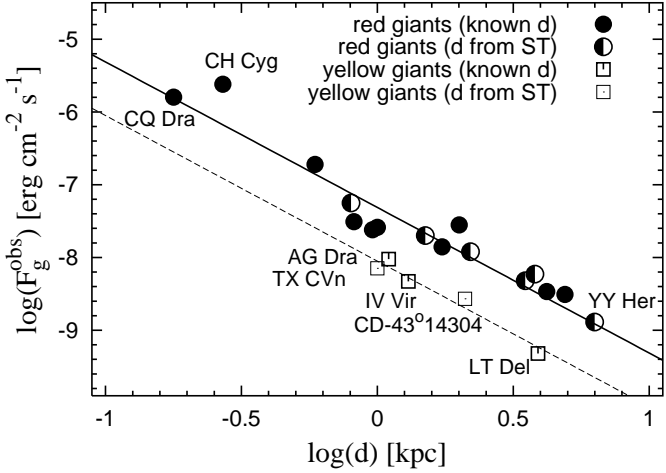}}
\caption[]{
The relationship between the observed bolometric flux of 
giants, $F_{\rm g}^{\rm obs}$ and the distance $d$. 
Giants with independently known radii or distances are 
distinguished (see the text). Data are from Tables~1 and 2. 
Solid and dashed line correspond to their fit given by 
relations (27) and (28). CH\,Cyg was omitted in the fitting 
(see the text). 
        }
\end{center}
\end{figure}
%
%
%
\begin{figure}[p!t]
\centering
\begin{center}
\vspace*{-3mm}
\resizebox{\hsize}{!}{\includegraphics[angle=-90]{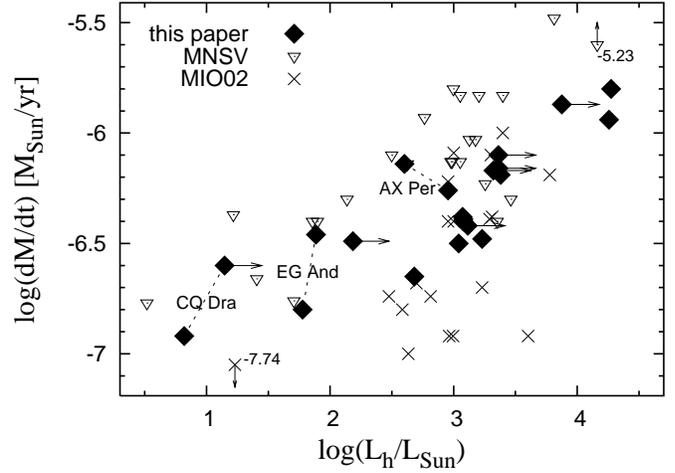}}
\caption[]{
Comparison of our mass-loss rates from Table~3 with those 
published by MNSV ($\triangledown$, recalculated to 
our $d$ and $v_{\infty}$ = 20\,\kms) and by \cite{mio02} 
($\times$, obtained from radio observations and scaled 
with $v_{\infty}$ = 20\,\kms). Horizontal arrows indicate 
a shift in the luminosity by a factor of 2 for values of 
$T_{\rm h} > T_{\rm h}^{\rm min}$. 
          }
\end{center}
\end{figure}

{\em Variability:}
The reconstructed SED at very different binary positions 
confirmed the orbitally-related variation in the emission 
measure which causes the well-known wave-like variation in 
the LCs as a function of the orbital phase. The nature of 
this type of variability was recently investigated 
by \cite{sk01a}. 
Here we used his conversion between the observed $EM$ and 
the magnitude to reconstruct the $U$ LCs of RW\,Hya and 
SY\,Mus from the IUE spectra (Figs.~17 and 18). 
The observed amplitude of the wave-like variation is 
proportional to the ratio of fluxes from the nebula and 
the giant, which is a function of the wavelength -- flux 
from the giant/nebula increases/decreses with increasing 
lambda. 
The contribution from the giant at the considered passband depends 
mainly on its spectral type, while the quantity of the nebular 
flux results from ionization/recombination processes and can 
be very different for individual objects. 
Figure~25 shows that the contribution from the giant in 
the $UBV$ bands is a function of its effective temperature, 
which determines the wavelength-dependence of the wave-like 
variation in LCs: Yellow objects show 
             $\Delta U / \Delta V \gg 1$, 
while those containing late-type M-giants have 
             $\Delta U / \Delta V \sim 1 - 3$. 
In the column of the same $T_{\rm eff}$, objects with 
a strong nebula are located at the bottom 
(i.e. $\Delta U / \Delta V \sim 1$). This is the case of 
V1329\,Cyg at the column of $T_{\rm eff}$ = 3\,300\,K 
and BF\,Cyg among objects at $T_{\rm eff}$ = 3\,400\,K. 

\subsubsection{Stellar continuum radiation}

{\em The hot star temperature, $T_{\rm h}$}:
It was not possible to determine directly this parameter with 
our fitting procedure for $T_{\rm h} \ga 10^5$\,K. As a result 
we selected appropriate values of $T_{\rm h}$ according to 
the ionizing capacity of the hot source and the observed 
emission measure: 

(i) 
If the parameter $\delta < 1$ for $T_{\rm h} \ga 10^5$\,K 
we adopted the Zanstra temperature for the ionizing object 
according to MNSV. This was the case of CI\,Cyg, SY\,Mus, 
and AG\,Peg. 
For two limiting cases, EG\,And(10/09/95) and BF\,Cyg(10/04/92), 
we adopted the corresponding minimum temperatures, 
$T^{\rm min}_{\rm h}$, because the referred Zanstra temperatures 
were below this lower limit. Note that for 
$T_{\rm h} < T^{\rm min}_{\rm h}$ the ionizing capacity is not 
sufficient to produce $EM_{\rm obs}$. 
For RW\,Hya we adopted $T_{\rm h}$ = 110\,000\,K, because its 
ionized zone is very open even for 
$T_{\rm h} = T^{\rm min}_{\rm h}$ = 80\,000\,K 
and the suggested Zanstra temperature is below
$T^{\rm min}_{\rm h}$. In this special case it was also 
possible to get this temperature directly by fitting 
the well exposed spectrum from 19/07/79. 

(ii) 
If the parameter $\delta \sim 1$ for a certain temperature 
$T_{\rm h} > 10^5$\,K, i.e. $EM_{\rm obs}$ is very large, 
we adopted $T_{\rm h} = T^{\rm min}_{\rm h}$. Then the 
parameters derived from the SED, $L_{\rm h}$ and $\theta_{\rm h}$ 
represent their lower and upper limit, respectively 
(Z\,And, V1329\,Cyg, AG\,Dra, CQ\,Dra; Table~3). 

(iii) 
For two cases, V443\,Her and AX\,Per, it was possible 
to determine the $T_{\rm h}$ temperature directly as the fitting 
parameter of our procedure, because of its low value. 
However, a relatively strong nebular radiation in these systems 
yields the parameter $\delta \gg 1$. 
Therefore, in Table~3 we also give parameters for the case 
$\delta = 1$ ($T_{\rm h} = T^{\rm min}_{\rm h}$). 

{\em The effective radius, $R_{\rm h}^{\rm eff}$}: 
According to Eq.~(6), $R_{\rm h}^{\rm eff}$ is the radius of 
a sphere which has the same luminosity as the observed hot 
object seen under the angle $2\theta_{\rm h}$. 
According to present knowledge, accretors in symbiotic 
binaries are white dwarfs of a mean mass 
$M_{\rm WD} \sim 0.6\,M_{\sun}$ \citep{cmm03}. 
However, only in the case of CQ\,Dra, from our sample of 
the objects, is the $R_{\rm h}^{\rm eff}$ radius comparable 
to that of a white dwarf (Table~3). Here the very small 
effective radius limits the white dwarf radius to 
$R_{\rm WD} \la 0.006\,R_{\sun}$, which implies 
a rather massive white dwarf, $M_{\rm WD} \sim 1\,M_{\sun}$ 
(Hamada \& Salpeter 1961).
Effective radii derived from the SED of other objects are 
a factor of about 10 (or more) larger than a typical white 
dwarf's radius. Qualitatively, this can be understood as 
a result of accretion process that creates an optically 
thick photosphere around the mass core. Below we investigate 
this possibility: 
%
%
\begin{figure}[p!t]
\centering
\begin{center}
\resizebox{\hsize}{!}{\includegraphics[angle=-90]{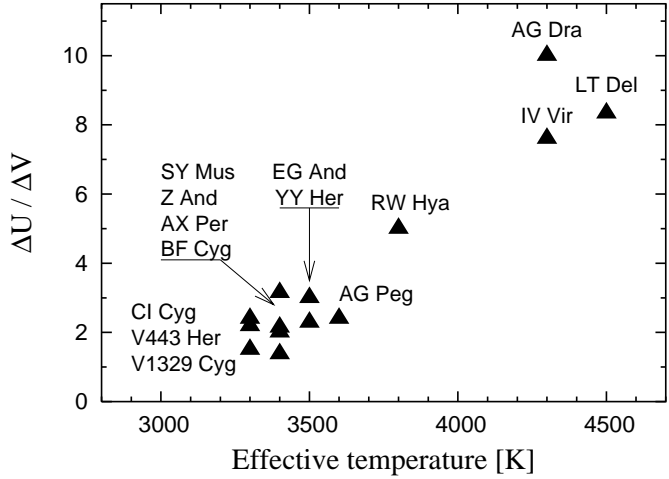}}
\caption[]{
Ratio of amplitudes in $U$ and $V$ of the orbitally-related 
variation in quiescent LCs as a function of the effective 
temperature of the giant. The ratio of fluxes from the giant 
and the nebula depends largely on the giant's effective 
temperature, which determines the overall relationship. 
Systems with a higher nebular emission have lower ratio of 
$\Delta U /\Delta V$ for the given $T_{\rm eff}$. 
          }
\end{center}
\end{figure}
%
%
%
\begin{figure}[p!t]
\centering
\begin{center}
\resizebox{\hsize}{!}{\includegraphics[angle=-90]{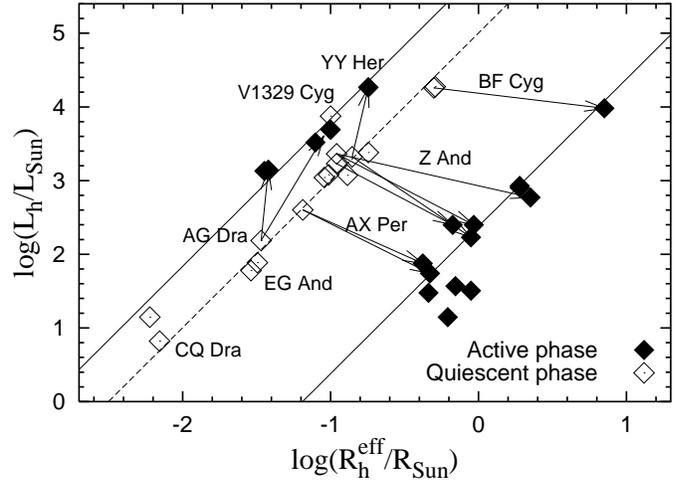}}
\caption[]{
The $L-R$ relationship for the hot stellar component 
of radiation. It satisfies well radiation of stellar 
photospheres (Eq.~31). During quiescence (open squares, 
dashed line) stellar components of investigated objects 
radiate at a characteristic temperature of $\sim 103\,000$\,K, 
while during outbursts (filled squares, solid lines) 
this temperature decreases significantly to $\sim 22\,000$\,K 
(1st-type of outbursts) or increases to $\sim 165\,000$\,K 
(2nd-type of outbursts). 
          }
\end{center}
\end{figure}

(i)
The extreme cases of SEDs with $\delta \ga 1$ 
(i.e. $T_{\rm h} < T_{\rm h}^{\rm min}$) suggest 
a disk-like structure of the hot source. The {\em observed} 
hot stellar radiation is not capable of giving rise to the strong 
nebula in the system which constrains that an unseen part of 
the hot source radiates at a higher temperature to produce 
the necessary surplus of the ionizing photons. 
Thus, the inner ionizing source of the hot object 
has to be obscured by the accreted matter more in directions 
forwards the observer than at its poles. 
The effect is significant for highly inclined systems. 

(ii)
Models corresponding to $\delta < 1$. 
If the hot stellar source radiation mimics that of a stellar 
photosphere, then it has to satisfy the $LRT$ relation, 
which can be expressed in solar units as 
\begin{equation}
 \log\Bigl(\frac{L_{\rm h}}{L_{\sun}}\Bigr) = 
    2\log\Bigl(\frac{R_{\rm h}^{\rm eff}}{R_{\sun}}\Bigr) + 
    4\log\Bigl(\frac{T_{\rm h}}{T_{\sun}}\Bigr).
\end{equation}
A good agreement between this relation and our quantities 
of $L_{\rm h}$ and $R_{\rm h}^{\rm eff}$ from 
Table~3 confirms the abovementioned suggestion (Fig.~26). 
The best fit gives 
4$\log(T_{\rm h}/T_{\sun}) = 5.0 \pm 0.04$, i.e. they 
radiate at a characteristic temperature of 
   $103\,000 \pm 2\,500$\,K 
for $T_{\sun} = 5\,800$\,K. Here, to get a better solution, 
we omitted the cases of V1329\,Cyg and CQ\,Dra(11/10/86) 
(see Fig.~26).
Thus we cannot observe just the accretor's surface. 

{\em Variability}:
Observations at/around opposite conjunctions of the binary 
components confirmed the presence of an additional source 
of extinction, which attenuates the continuum (seemingly) 
at all wavelengths and is more pronounced at positions 
around the inferior conjunction of the giant 
(here Z\,And, V1329\,Cyg). 
MNSV first noticed this extinction process. The effect 
is known for RW\,Hya, SY\,Mus, BF\,Cyg \citep[e.g.][]{d+99}. 
\cite{ho+94} modeled the influence of the absorbing medium 
to the ultraviolet, optical and near-infrared spectrum of 
of the white dwarf in the eclipsing dwarf nova OY\,Car. They 
found that besides deep features around 1\,600 and 2\,000\,\AA\ 
the absorbing gas also produces a modest optical depth in 
the Balmer and Pashen continuum which makes it lower by about 
20\% in the ultraviolet (see their Fig.~8). The magnitude 
of this effect is probably a function of the orbital 
inclination. We did not find it in the spectra of AG\,Dra 
and AG\,Peg. 

\subsection{The active phase}

\subsubsection{The high-temperature nebula}

The HTN is seen directly during optical eclipses of all 
investigated stars 
  (BF\,Cyg, CH\,Cyg, CI\,Cyg, AR\,Pav and AX\,Per) 
as the only component of the hot object radiation in the UV 
spectrum. Out of eclipses it is indicated by a non-zero 
level of the Rayleigh attenuated far-UV continuum. The nebular 
nature of this component was proved directly by the optical 
spectrum containing the Balmer jump in emission taken 
simultaneously with the ultraviolet measurements of 
the strong 1993-outburst of YY\,Her (Fig.~16). 
Also simultaneous UV/optical/radio observations of BF\,Cyg 
taken at the totality of the optical eclipse during 
its 1989-92 active phase suggest the nebular origin of 
this component of radiation. In the UV it is given by its 
flat profile (Fig.~8), in the optical by the very negative 
colour index, $U-B \sim -0.7$ \citep{sk92}, and in the radio 
the observed flux is increased to 0.84\,mJy with respect 
to precedent values \citep{sea+93}. The only counterpart 
to the radio emission is the HTN in the ultraviolet, 
which thus confirms its nebular nature.
Its profile corresponds to high electron temperatures, 
$T_{\rm e} > 3\,10^4$\,K (Table 4) and it is complementary 
to other components of radiation throughout the UV/optical/IR 
wavelengths to match the observed SED. 
These characteristics imply that the HTN-radiation is not 
subject to the Rayleigh scattering process and the corresponding 
ionized medium is not eclipsed by the giant. Thus 
a major fraction of the HTN region is located 
at least $\sim 1\,R_{\rm g}$ above/below the orbital plane. 
The HTN appears to be very strong in the spectra from active 
phases. During transitions to quiescence its emission measure 
gradually decreases (see figures for Z\,And, CH\,Cyg, BF\,Cyg). 
Other IUE observations also support such evolution. 
For example, the HTN emission measure from AX\,Per spectra 
on 30/11/90 (Fig.~21) decreased by a factor of about 
2.5 on its transition to quiescence on 04/09/94 
(SWP52027 + LWP29094, $\varphi$ = 0.01 -- not presented here). 
During quiescent phases it can be recognized as a very faint 
emission in UV (e.g. RW\,Hya, Fig.~17). \cite{p+95} noted 
the presence of such the remaining light also during the eclipse 
of SY\,Mus. They assigned it most likely to nebular emission. 
This suggests that creation of a strong HTN region is connected 
with the mass-outflow from the central star during outbursts. 
Below we discuss its origin. 

(i)
Photoionization of a rather massive hot star wind can be excluded, 
because this would produce a significant emission within about 
100\,$R_{\sun}$ from its source. The emissivity of such the wind 
is proportional to the square of the particle concentration, 
which is diluted in the wind with the radial distance $r$ as 
$1/r^2$. In addition, a small wind velocity at the vicinity of 
its source makes the particle density even higher, which 
significantly enhances the emissivity of the inner parts of 
the ionized wind \citep[cf. Fig.~C1 of][]{sk+02a}. 
If this were the case, we would have observed a significant 
decrease in the emission from the HTN during the total eclipse. 
This, however, is not the case. 

(ii)
Photoionization of the neutral particles from the giant's wind 
would produce nebular emission at $T_{\rm e} \sim 1 - 2 \times 
10^4$\,K as we observe during {\em quiescent phases}. 
However, this does not correspond to the HTN properties. 

(iii)
The ejected material can make the surrounding nebular medium
more opaque, which could lead to a significant heating of
the outer layers of the symbiotic nebula, because the
ionizing radiation is modified by absorption in the nebula -- 
the higher-energy photons penetrate into the gas at larger
(i.e. more opaque) distances from the central star, and thus 
the mean energy of the produced electrons is higher there 
\citep{ost74}.
The result probably depends strongly on the structure of
the surrounding ionized material.

(iv)
In the analogy to the colliding wind model 
\citep[e.g.][]{gw87,nw93}, a high temperature 
collisionally-ionized shock layer, which can be ionized also 
by the radiation 
field of the hot star, can be created well above the accretion 
disk. The collisions are between the high-velocity hot star 
wind and the slowly moving particles of the cool giant wind 
and/or circumbinary material. The temperature of this region 
can reach well above 10$^6$\,K. Problems here are connected 
mainly with the efficiency of cooling 
the hot plasma and properties of the driven hot star wind. 
Nevertheless, physical separation of the HTN from the central 
hot object, its very high electron temperature and production 
of the emission lines with very high ionization potential 
could be produced by this type of interaction. 

\subsubsection{The low-temperature nebula}

The LTN is subject to eclipses, which determines its location
in the vicinity of the central star and limits its linear size 
to 2\,R$_{\rm g}$. This case is reproduced by CI\,Cyg, 
AR\,Pav and AX\,Per, for which there are available ultraviolet 
observations taken {\em in} and {\em out} of their optical 
eclipses during activity. 
The radiative properties of the LTN are characterized by its
emission measure and luminosity. They are determined by
the fitting parameters $k_{\rm N}$, $T_{\rm e}$ and
$\tilde a$ in relations (13) and (14). 
Its electron temperature is between 10\,000 and 18\,000\,K, 
which means that this region is predominately due 
to photoionization. 
Geometrical properties of the LTN -- mainly its size and
location -- suggest its very high ionization degree. 
According to the ionization formula \citep[e.g.][]{gur}, 
  $N_{+}/N_{1} \sim $ a few $\times\,10^3$ 
even for very high electron concentrations of 
$N_{\rm e} \sim $ a few $\times\,10^{10}$\,\cmtri, 
because of a high value of the dilution factor (see below). 
Therefore and due to a small or negligible contribution
from the doubly ionized helium ($\tilde a \sim 0$) 
we can assume 
  $N_{\rm e} = N_{\rm H^{+}}$. 
Further, for the temperature of the ionizing source 
  $T_{\star} \equiv 10^5$\,K, 
the average electron temperature
  $T_{\rm e} = 14\,000$\,K, 
a dilution factor
  $W\, \dot =\,(R_{\star}/R_{\rm LTN})^2 
             \sim (0.01/100)^2 = 10^{-8}$
and the optical depth of the nebula
  $\tau_{\rm c} = \kappa_{\rm c}\,N_1\,R_{\rm LTN} \sim 1$ 
($\kappa_{\rm c} = 6.3\,10^{-18}$\,cm$^2$ is the hydrogen 
absorption coefficient at the Lyman limit, $N_1$ is 
the concentration of neutral H atoms and for $R_{\rm LTN}$ 
we adopt 100\,$R_{\sun}$ as a maximum), the ionization 
formula provides the electron density
\begin{equation}
  N_{\rm e} \sim 10^{9} ~~~ \rm cm^{-3}.
\end{equation}
An independent estimate of $N_{\rm e}$ can be made from 
the size of the LTN (it is subject to eclipse)
and its emission measure. 
If we approximate its volume by a sphere of radius 
  $R_{\rm LTN} \equiv 100\,R_{\sun}$ then 
its emitting volume 
  $V_{\rm LTN} = 1.4\,10^{39}\,\rm cm^{3}$. 
Consequently, the emission measure obtained from the spectra 
out of eclipses, 
  $EM_{\rm LTN} = N_{\rm e}^2\,V_{\rm LTN} 
                = 10^{58} - 10^{60}$\,\cmtri, 
yields the average electron concentration 
\begin{equation}
  N_{\rm e} \sim {\rm a~ few}\,\times 10^{9} ~~~ \rm cm^{-3},
\end{equation}
which is comparable with the estimate made above using 
the ionization formula. This agreement supports that the LTN 
is mainly due to photoionization. \cite{k+91}, based on 
an analysis of the line emitting regions in CI\,Cyg, also 
suggested the presence of two nebular regions in this 
system -- a compact nebula located within the Roche lobe of 
the accretor and a low density, collisionally-ionized nebula 
having a form of bipolar streams above/below the disk's surface. 
%
%
%
\begin{figure*}
\centering
\begin{center}
\resizebox{\hsize}{!}{\includegraphics[]{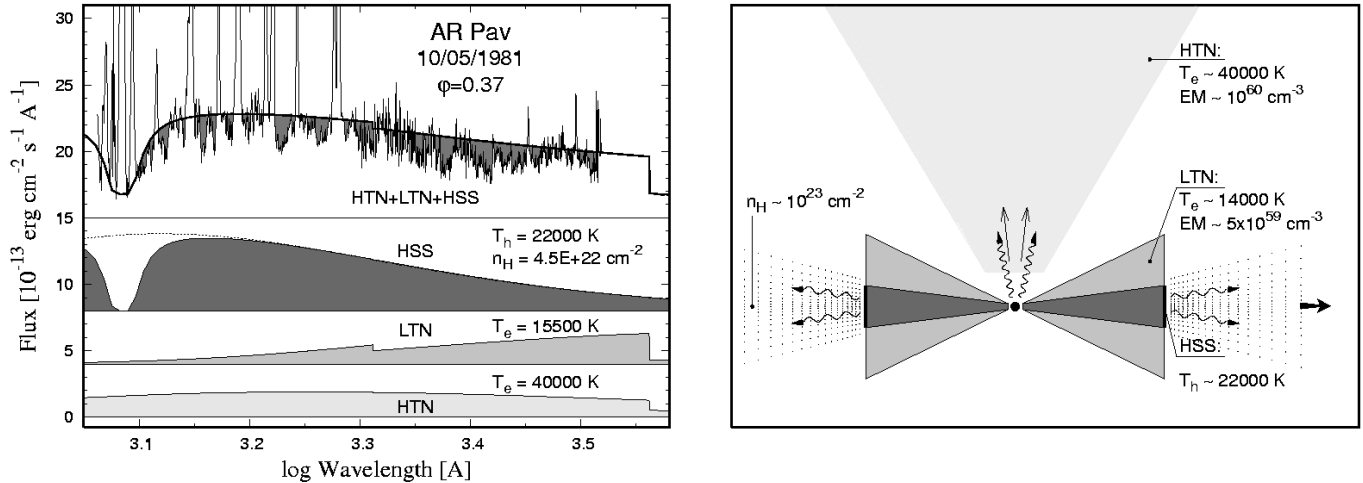}}
\caption[]{
Schematic representation of a basic structure of hot objects 
during outbursts of the 1st-type. 
The left panel shows example of individual components of 
radiation we isolated from the AR\,Pav spectrum observed 
on 10/05/81. Note that the HSS radiation is Rayleigh attenuated 
at the far-UV region and veiled by a forest of blended 
\ion{Fe}{ii} absorptions. 
The right panel shows a sketch of the corresponding emitting 
regions as seen on a cut perpendicular to the orbital plane 
containing the accretor. During outbursts the disk's outer rim 
has to be flared to occult permanently the central source of 
the hot ionizing radiation. 
The model is described in Sect.~5.3.5. 
          }
\end{center}
\end{figure*}

\subsubsection{The hot stellar source}

The region of the HSS is subject to eclipses 
(Figs.~8, 9, 10, 19 and 21). 
Its radiation can be approximated by that of a black body 
at low temperatures. Below we summarize its radiative and 
geometrical properties as follows: 

1. Figure~26 shows: (i) The HSS radiation of most of objects 
satisfies that of a stellar photosphere at a characteristic 
temperature of $\sim$\,22\,800\,K. The HSS of CH\,Cyg and 
TX\,CVn radiates even at lower temperatures; we did not 
include them in the fit in Fig.~26. 
 (ii) There is a significant enlargement of 
the $R_{\rm h}^{\rm eff}$ radius by a factor of $\sim$\,10 
with respect to quantities from quiescence. This and 
the simultaneous presence of the two-temperature-type of 
the UV spectrum (see bellow, Sect.~5.3.4) suggest that 
the optically thick medium expands at the orbital plane in 
the form of a disk. 
  (iii) The {\em observed} luminosity of the HSS during the 
1st-type of outbursts (see below; Fig.~26) decreases by a factor 
of about 2 -- 10 with respect to the hot object luminosity 
during quiescence (Table~3 and 4). This is because the HSS, 
which develops during these outbursts, blocks a fraction of 
radiation from the centre in the direction to the observer. 
We discuss this situation in Sect.~5.3.6 in more detail. 

2. The HSS radiation is attenuated by the Rayleigh scattering 
process for {\em all} the well-known eclipsing systems 
presented here 
  (BF\,Cyg, CH\,Cyg, CI\,Cyg, AR\,Pav and AX\,Per; Table~4). 
The effect is measured at any orbital phase. We therefore 
believe that the absorbing gas is located at the outer rim of 
the disk. On the other hand the presence of 
the Rayleigh-attenuated far-UV continuum in Z\,And and TX\,CVn 
confirms the recent suggestion of their high orbital 
inclination \citep{sk03a,sk+04}, and thus suggests it also for 
CD-43$^{\circ}$14304 (Fig.~5). Generally, all the orbits of 
the objects showing the two-temperature-type of UV spectrum 
during activity should be highly inclined to the observer. 
Here this applies to AE\,Ara (Fig.~4). 

3. Outbursts of AG\,Dra and the major 1993-outburst of YY\,Her 
differ significantly from those described above. A strong 
nebular and very hot stellar components of radiation 
developed immediately at the beginning of these outbursts. 
The hot stellar radiation is characterized by a higher 
temperature $T_{\rm h} \sim 165\,000$\,K, a larger $L_{\rm h}$ 
and smaller $R_{\rm h}^{\rm eff}$ with respect to values from 
quiescence (Fig.~26). 
The case of AG\,Dra could be caused mainly by an inclination 
effect: This system is seen more from its pole 
\citep[$i=60\,\pm 8^{\circ}$,][]{ss97ag} and thus the central 
ionizing source can be directly observed. As a result both 
a very hot and luminous stellar and strong nebular component 
of radiation can be measured directly. The observed large difference 
in the HSS temperature during the minor 1981 and the major 
1993 outburst of YY\,Her (Fig.~16) is probably related to 
different properties of the mass outflow. Material during 
the major outburst was ejected probably at very high velocity, 
which precluded formation of a cool stellar pseudophotosphere. 
As a result the simultaneous massive injection of new particles 
(emitters) by the active star and the increase of its ionizing 
capacity enhanced the nebular emission by a factor of 
$\sim$\,20 (Table~4). 

Accordingly, these observational properties of the HSS during 
active phases suggest two types of outbursts: 
The 1st type is characterized by 
         a larger $R_{\rm h}^{\rm eff}$ and lower $T_{\rm h}$
         (points 1 and 2 above) 
and 
the 2nd type is determined by a smaller $R_{\rm h}^{\rm eff}$ 
                       and higher $T_{\rm h}$ (the point 3 above) 
with respect to their quiescent values (Fig.~26). 
All the active objects with a high orbital inclination show 
characteristics of the 1st type of outbursts. 
\cite{g-r+99} identified 'cool' and 'hot' outbursts of AG\,Dra. 
They lie within the 2nd type of outbursts. 
According to the point 1.\,(iii) above one can speculate that 
the 1st-type of outbursts represent a special case of 
the 2nd-type due to a high orbital inclination (Sect.~5.3.6). 
Finally, we note that a large number of objects with highly 
inclined orbits, among those intensively investigated, 
reflects a systematic selection, because the symbiotic 
phenomenon is best recognizable for such systems. 

\subsubsection{The two-temperature UV spectrum}

The presence of the HTN, LTN and a warm HSS components 
of radiation determines the two-temperature-type of 
the ultraviolet spectrum. In our approach, the cool spectrum 
is produced by the warm HSS radiation ($\sim 22\,000$\,K) 
and the hot spectrum is represented by the emission from 
the HTN+LTN powered by an unseen hot ionizing source 
($\ga 10^5$\,K) in the system. 
Superposition of these components results in the observed, 
more or less flat ultraviolet continuum. The far-UV region 
is dominated by the HSS radiation, while the near-UV continuum 
often contains a large contribution from the LTN. The slope 
of the HTN radiation throughout the ultraviolet is very small. 
This type of the UV spectrum represents a common feature of 
active phases classified here as the 1st type. 

\subsubsection{Basic structure of the hot active object}

Radiative and geometrical properties of individual components 
of radiation observed during the 1st type of outbursts allow 
us to reconstruct a basic configuration of the hot active object. 
The following points are relevant. 

(i) The two-temperature-type of the UV spectrum suggests the 
presence of an optically thick disk-like torus around the hot 
star, which is seen approximately edge-on. 
The disk's outer rim has to be flared sufficiently to occult
permanently the central source of the hot ionizing radiation, 
which is not seen directly in the spectrum. 
Under such conditions, the energetic radiation from 
the central source is absorbed and diffused by the disk body 
in the direction to the observer. At the distance $R_{\rm HSS}$ 
from the center, it releases the disk at much lower temperature 
($\sim$\,22\,000\,K). We identify this warmer pseudophotosphere 
as the HSS. 
The HSS radiation then continues throughout the neutral 
gas surrounding the accretor beyond the disk, which attenuates 
its amount by the Rayleigh scattering process and affects 
(often drastically) its profile by numerous absorptions of 
mainly \ion{Fe}{ii} lines, the so-called 'iron curtain'. 

(ii) The geometrical and physical properties of the LTN 
localize this region 
above/below the disk plane approximately within the linear 
size of the HSS (it is also subject to eclipse). It is 
probably also flared at the outer edge of the disk, because 
directions from the hot inner parts of the disk to poles 
have to be rather free for high-velocity particles and 
ionizing photons, which give rise the uneclipsed emission 
from the HTN. In addition, the LTN has to 'see' well 
the ionizing radiation from the hot center. This view is 
supported by theoretical modeling of \cite{ph94}, who found 
that the vertical outer edge of accretion disks in symbiotic 
binaries can be extremely extended. 
In the real case the vertical extension is limited by 
the eclipsing nature of the LTN. 
For the sake of simplicity, we draw schematically the LTN
vertical dimension as a linear function of the disk radius
(cf. Fig.~27).

(iii) 
The HTN region has to be located high enough above/below 
the disk ($> 1\,R_{\rm g}$) not to be eclipsed by the giant 
(e.g. Fig.~10 and 19). 
However, our observations were not sufficient to determine 
its geometry more accurately.
The HTN can be in part collisionaly ionized, which suggests that 
a mechanisms similar to that of the colliding wind model could 
be responsible for its rise (Sect.~5.3.1). 

A sketch of the above described geometrical structure of 
the hot object in symbiotic binaries during the 1st-type 
of outbursts is shown in Fig.~27. 

\subsubsection{Luminosity of the hot active object}

In this section we discuss the total luminosity of hot objects 
undergoing the 1st-type of outbursts. 
The edge-on disk/torus, which develops during these outbursts, 
blocks a fraction of radiation from the accretor in the 
direction to the observer. We measure directly only its part 
in the form of the HSS radiation. The remaining part of 
the accretor's luminosity can become visible due to its 
interaction with the medium surrounding the hot object. 
We detect only its fraction converted into the nebular 
emission in the form of the LTN and HTN radiation. However, 
a part of the accretor's luminosity, $L_{\rm out}$, can escape 
the system. Therefore we write the bolometric luminosity, 
$L_{\rm T}$, produced at/around the white dwarf as 
\begin{equation}
 L_{\rm T} = L_{\rm HSS} + L_{\rm HTN} + L_{\rm LTN} + L_{\rm out}.
\end{equation}
The not-detectable light includes the radiation, which is not 
capable of ionizing the surrounding medium 
(for $T_{\rm h} = 10^{5}$\,K and hydrogen it is about 10\% of 
the total radiation \citep[Fig.~2 of][]{sk03b}), and a part of 
ionizing photons, which are not converted into the nebular 
emission. Thus we, generally, observe  
\begin{equation}
   L_{\rm h}^{\rm obs} \ll L_{\rm T}.
\end{equation}
This is the case of all systems showing signatures of the 
1st type of outbursts 
(compare data from Tables~3, 4 and Fig.~26). 
On the other hand when viewing the system more pole-on we can 
detect directly all the radiation released during outbursts. 
An example here is AG\,Dra. During its 'hot' outbursts 
(Sect.~4.13) we have 
\begin{equation}
   L_{\rm h}^{\rm obs} \approx L_{\rm T} \equiv L_{\rm Q},
\end{equation}
if one adopts the total luminosity as that from quiescence, 
$L_{\rm Q}$. Note 
that for $T_{\rm h} > T_{\rm h}^{\rm min}$ one can easily get 
equivalence between the luminosities in the above relation. 
This suggests an accretion-powered nature of these outbursts. 
However, for 'cool' outbursts we measure 
\begin{equation}
   L_{\rm h}^{\rm obs} \gg L_{\rm Q},
\end{equation}
which signals an additional source of the energy during these 
events, probably from a TNR. Some details on the nova-like 
eruptions of symbiotic stars can be found in \cite{mk92b}. 

\subsection{Comparison with previous models}

A modeling of the composite continuum for a large sample 
of well-studied symbiotic stars similar to the present 
one has been carried out by MNSV and KW. The basic conceptional 
differences between these and our approach were introduced 
in the section 1. 
Here we present a concise summary of differences between 
our models and those suggested by other authors. 
The following points are relevant: 

(i) 
Our SED model (Eq.~13) is less sophisticated than that suggested 
by MNSV. We did not include a geometry of radiating regions and 
consider just basic sources of radiation and effects that can 
be recognized by observations. This simplifies significantly 
modeling the SED and makes it more accessible for a wider community. 
On the other hand, this does not result in decisive differencies 
in $R_{\rm h}^{\rm eff}$ and $L_{\rm h}$ from quiescent phases. 

(ii)
In our approach the electron temperature represents the fitting 
parameter, while MNSV considered it to be within a small range 
(Sect.~1). This revealed drastic variations in $T_{\rm e}$ 
even for the same object (e.g. Z\,And, AG\,Peg). This allowed us 
to determine corresponding $EM$s more accurately. 
We note that an adoption of a 'typical' value of $T_{\rm e}$ 
can result in an incorrect SED \citep[e.g. Fig.~4 in][]{mu+97a}. 

(iii)
In our approach we used temperatures required by the basic 
photoionization model (i.e. $T_{\rm h}^{\rm min}$ (Eq.~23) 
or $T_{\rm Zanstra}$, see Sect.~5.2.2) as nebulae during 
quiescent phases are largely radiatively ionized 
($T_{\rm e} \la 20\,000$\,K). In cases of AX\,Per and 
V443\,Her a large temperature difference, 
   $\Delta T \equiv T_{\rm h}^{\rm min}$ - $T_{\rm fit} 
              \sim$ 35\,000\,K (Table~3), 
suggests the presence of a disk-like structured material 
around their accretors even during quiescent phases 
(Sects.~4.15 and 4.21). MNSV discarded the presence of 
accretion disks in their solutions, whereas KW considered 
them as a reasonable possibility for some cases. 
Most recently \cite{sion+02}, \cite{sion03} and \cite{ko+04} 
applied model atmospheres in spectral analyses of the hot 
stellar component of radiation from RW\,Hya, AG\,Dra and EG\,And. 
They also indicated significantly lower effective temperatures 
than those given by conditions of photoionization in these 
systems. Effective temperatures of their models and our 
$T_{\rm h}^{\rm min}$ for EG\,And and RW\,Hya (Table~3) 
give a similar $\Delta T$ to what we found for V443\,Her and 
AX\,Per. This suggests that we detect temperatures 
of different regions in the hot object. 
Within our interpretation we measure {\em directly} 
a cooler pseudophotosphere ($\sim$\,edge-on disk-like 
structured accreting material), while the underlying 
hotter accretor's surface can be 'seen' only by the surrounding 
medium, which converts its radiation to the nebular emission, 
based on whose properties we can derive {\em indirectly} 
the temperature of the hottest part of the ionizing source 
as $T_{\rm Zanstra}$ and/or $T_{\rm h}^{\rm min}$. Such 
a structure of hot components can explain the apparent 
inconsistence between effective temperatures from model 
atmospheres analyses and those required by the observed 
nebular emission. 
In accordance with this view the temperature difference 
$\Delta T$ should be larger for objects with a high orbital 
inclination and large mass-loss rates, and vice versa. This 
idea could be tested by applying the model atmosphere 
spectral analysis to objects seen rather pole-on and with 
a small $\dot M_{\rm W}$ (e.g. CQ\,Dra in our sample). 
Such testing is beyond the possibilities of our simple 
black-body fitting. 

(iv)
The mass-loss rate was calculated for a more realistic structure 
of the giant's wind. We considered a $\beta$-law wind, whereas MNSV 
adopted a constant wind. In addition, we used a more accurately 
detemined $EM$s (the point (ii) above). This difference, 
however, did not yield a large quantitative effect. \cite{mio02} 
estimated $\dot M_{\rm W}$ from their radio mm/sub-mm observations 
assuming a partially ionized red giant wind determined by opened 
ionized zones. This approach is, in principle, similar to our 
calculations. A comparison of our $\dot M_{\rm W}$ quantities 
with those mentioned above is satisfactory (Fig.~24). 

(v)
We quantified contributions from cool components by comparing 
an appropriate synthetic spectrum to IR flux-points given by 
photometric measurements. KW reconstructed the spectrum of cool 
components according to energy distributions of \cite{j66} and 
\cite{lee70} and scaled it to absolute $V$ magnitudes 
of K3, M2 and M4 giants. MNSV restricted their analysis 
only to UV spectra. Our approach allowed us to estimate 
effective temperatures and angular radii of giants 
independently from their spectral types. Application of 
synthetic spectra of red/yellow giants is of a particular 
importance to model SEDs for objects with a considerable cool 
component contribution at the near-UV (yellow symbiotics and 
those with a small nebular emission, e.g. CQ\,Dra) to get 
a correct $T_{\rm e}$ and $EM$. For comparison, a simple 
blackbody fit gives a large contribution at these wavelengths 
\citep[e.g. Fig.~7 in][]{sk+00}. 

(vi)
A major qualitative difference is that we modeled spectra from 
all the quiescent phases, active phases and eclipses, whereas 
MNSV restricted their analysis exclusively to quiescent phases
at binary positions well outside eclipses. Their trials to model 
active phases failed (Fig.~8 and Table~6 in MNSV). Application 
of our method to spectra taken during active phases revealed 
their {\em common} properties (Sect.~5.3; Fig.~27). 
This represents the main novelty of the present paper. 
For comparison, in our model of the active hot object of 
the 1st-type the flared disk permanently occults its hotter 
inner parts in the line-of-sight, whereas according to previous 
suggestions \citep[e.g. KW;][]{k+91,mk92a} we observe radiation 
from the inclined disk's surface including its central hot 
parts. However, fits to the observed continuum by using parameters 
of the latter models are poor (e.g. Fig.~9 in MNSV). A quantitative 
application of standard disk models to the UV/optical/IR continuum 
was carried out by \cite{sk03b} for AR\,Pav. This approach failed 
to explain the UV continuum with the disk and its boundary layer. 

(vii)
To select the continuum flux-points we considered the influence 
of the iron curtain. This led to determination of more realistic
model parameters for spectra from mainly active phases. 
Particularly, we obtained solution also for CH\,Cyg, on which
approaches of both the groups (KW and MNSV) failed. 
A first suggestion was made by \cite{sa93} for the recurrent 
nova T\,CrB. However, ignoring the effect of 
the iron curtain for this case, \cite{s+04} obtained an 
ineffectual SED corresponding to false parameters, which are 
not consistent with the observed hot object luminosity. 
Also \cite{sk03a} obtained false quantities of He$^{++}$ 
abundance and $T_{\rm e}$ of the LTN in his modeling the 
Z\,And spectra from the active phase. 

(viii)
Finally, the presented comprehensive atlas of modeled and 
observed SEDs for intensively studied S-type symbiotic stars 
accompanied by LCs represents a large source of information 
for future works. 

\section{Conclusions}

We introduced a method of disentangling the composite spectrum 
consisting of two stellar components of radiation (from cool 
giants and hot objects in the binary) and nebular radiation. 
We applied the method to 21 S-type symbiotic binaries to isolate 
individual radiative components from their 0.12 -- 5\,$\mu$m SED 
in the continuum during quiescent as well as active phases. 
The modeled continuum is determined by 8 fitting parameters: 
Temperatures, $T_{\rm eff}$, $T_{\rm h}$ and $T_{\rm e}$, 
angular radii, $\theta_{\rm g}$ and $\theta_{\rm h}$, 
scaling of the nebular emission, $k_{\rm N}$, 
hydrogen column density, $n_{\rm H}$ and average abundance 
of doubly ionized helium, $\tilde{a}$ (Sect.~3.1). 
These parameters then determine luminosities and radii of the 
stellar components and the emission measure of the nebula.
Common properties of individual radiative components and their 
sources can be summarized as follows:  

1. Radiation from the giant: 
(i) The SED of giants is defined by the fitting parameters, 
$T_{\rm eff}$ and $\theta_{\rm g}$. Those having 
independently-determined radii (11 objects from our sample, Fig.~23) 
then provide the distance from their $\theta_{\rm g}$. 
Relations (27) and (28) show that their characteristics satisfy 
best those of normal giants. This confirms a generally accepted 
view that cool components in S-type systems are normal giants 
as given by their spectral types \citep[e.g.][]{ms99}. 
(ii) The effective temperature of giants is critical for 
the amplitudes of the wave-like orbitally-related variation 
in the LCs (Fig.~25). 

2. Hot object radiation during quiescence: 
(i) The nebulae radiate at a mean temperature of 19\,000\,K. 
By comparing the measured and calculated emission measure 
we derived the mass-loss rate from giants via their wind as 
$\dot M_{\rm W}$ = a few $\times\,10^{-7}$\,\myr (Table~3, Fig.~24). 
(ii) Stellar continuum radiation determines the effective radii 
of the hot objects to be a factor of $\sim$\,10 larger than 
the accretors. We thus observe pseudophotospheres due to 
the accretion process (Fig.~26). 
The extreme cases, for which the SED suggests 
$T_{\rm h} < T_{\rm h}^{\rm min}$ (the {\em observed} hot 
source is not capable of giving rise to the nebular emission), 
signal the presence of a disk-like structure of the circumstellar 
matter around the accretor. This is in agreement with theoretical 
modeling of accretion from the giant's wind that predicts 
formation of a stable, permanent disk around the accretor 
\citep{mm98}. 

3. Hot object radiation during activity: 
(i) The stellar radiation satisfies that of a black-body 
photosphere. There are two types of outbursts: 
The 1st-type with the HSS radiating at significantly lower 
temperatures than during quiescence and the 2nd-type with 
higher characteristic temperatures (cf. Fig.~26). 
All the active objects with a high orbital inclination show 
features of the 1st type of outbursts: Rayleigh attenuation 
at any orbital phase, a strong influence from the iron curtain 
absorptions, a low characteristic temperature of 
$\sim 22\,000$\,K, strong HTN emission, two-temperature-type 
UV spectrum and enlargement of $R_{\rm h}^{\rm eff}$ 
by a factor of $\sim$\,10 with respect to the quiescent values. 
The last two characteristics suggest an expansion of 
an optically thick medium at the orbital plane in the form 
of a disk. 
(ii) The low-temperature nebula (LTN) radiates at a mean 
electron temperature of 14\,000\,K. This region is subject 
to eclipses, which restricts its maximum linear radius to 
2\,$R_{\rm g}$. Its emission measure, $10^{58} - 10^{60}$\,\cmtri, 
suggests the electron concentration 
$N_{\rm e} \sim {\rm a~ few}\,\times 10^{9}$\,\cmtri. 
%
%
(iii) The high-temperature nebula (HTN) is characterized by 
$T_{\rm e} > 30\,000$\,K and its major fraction is located 
well away from the hot object, not subject to eclipses. 
Its emission measure is between $10^{58} - 10^{61}$\,\cmtri. 
Its origin could be within a scenario of the colliding wind 
model. 

The characteristics of the UV-continuum and its components 
we observe during the 1st type of outbursts can be interpreted 
in terms of an edge-on flared disk surrounded by the neutral 
material at the orbital plane and the nebulae located 
above/below its surface. The model is described in Sect.~5.3.5 
and in Fig.~27. 

\begin{acknowledgements}
The author is grateful to the anonymous referees for several 
helpful comments. Special thanks are due to Ivan Huben\'y for 
his constructive comments on the iron curtain and providing 
his code {\scriptsize CIRCUS}. 
The author would also like to thank Albert Jones for 
providing his visual light curves of AE\,Ara, 
CD-43$^{\circ}$14304 and RW\,Hya and Nikolai Tomov for 
his unpublished photometry of AG\,Peg in the $U$ band 
for this paper. 
The first parts of this research were done within the project 
No. SLA/1039115 of the Alexander von Humboldt foundation 
(from 1998), throughout the bilateral research project between 
the Royal Society of Great Britain and the Slovak Academy of 
Sciences. Early versions of this work were presented at workshops 
in G\"ottingen (2001), Los Cancajos (2002), Dubrovn\'{\i}k (2003) 
and Strasbourg (2004). 
The final work on this project was partly supported by the Slovak 
Academy of Sciences under grant No.~2/4014/4. 
\end{acknowledgements}

\end{document}